\documentclass[11pt, twoside, a4paper]{article}

\usepackage[left=2.5cm,top=2.cm,right=2.cm,bottom=2.5cm]{geometry}  
\usepackage[english]{babel} 
\usepackage{amsmath, amssymb, mathtools}
\usepackage{txfonts}
\usepackage{natbib} 
\usepackage{graphicx, pgfplots} 
\usepackage[colorlinks=false,citecolor=black,linkcolor=black]{hyperref} 
\usepackage[normalem]{ulem}
\usepackage{cancel, comment}
\usepackage{url, doi, cleveref}

\graphicspath{ {fig_GalSatpaper/} } 

\newcommand{\bvec}[1]{\boldsymbol{\textbf{#1}}}
\newcommand{\uvec}[1]{\boldsymbol{\hat{\textbf{#1}}}}

\makeatletter
\let\jnl@style=\rm
\def\ref@jnl#1{{\jnl@style#1}}
\def\aap{\ref@jnl{A\&A}}                
\def\aj{\ref@jnl{AJ}}                   
\def\grl{\ref@jnl{Geophys.~Res.~Lett.}} 
\def\icarus{\ref@jnl{Icarus}}           
\def\jgr{\ref@jnl{Journal of Geophysical Research (Planets)}}
\def\mnras{\ref@jnl{MNRAS}}             
\def\nat{\ref@jnl{Nature}}              
\def\planss{\ref@jnl{Planet.~Space~Sci.}} 
\def\psj{\ref@jnl{Planet.~Sci.~Journal}}  
\def\ssr{\ref@jnl{Space Science Reviews}}
\pgfplotsset{compat=1.18}

\makeatother
\begin{document}

\title{Transformation of orientation and rotation angles of synchronous satellites: Application to the Galilean moons}
\author{Marie Yseboodt and Rose-Marie Baland \\
Royal Observatory of Belgium, m.yseboodt@oma.be, \\
Avenue Circulaire 3, Brussels, Belgium}

\date{published in Icarus, 4/2/2026, https://doi.org/10.1016/j.icarus.2026.116977}

\maketitle 

\begin{abstract}

The orientation and rotation of a synchronous satellite can be referred to both its Laplace plane and the ICRF equatorial plane, in terms of Euler angles or spin axis Cartesian coordinates and Earth equatorial coordinates, respectively. 
We computed second-order analytical expressions to make the transformation between the two systems and applied them to the Galilean satellites (Io, Europa, Ganymede, and Callisto). 
If one term of the spin axis Cartesian coordinates series is dominant, trigonometric series can be generated for the inertial and orbital obliquities, node longitude and offset with respect to the Cassini plane.
Since the transformation avoids having to fit amplitudes and frequencies on the output numerical series, the physical meaning of the frequencies is preserved from the input series and the amplitudes can be directly related to the geophysical parameters of interest. 
We provide tables for the coordinates and angles' series assuming that the satellites are entirely solid, and considering two different orbital theories. The possible amplitude ranges for the main terms are also examined in the case where a liquid layer is assumed in the interior model.
We use our transformation method to propose an updated IAU WG solution which would result in an improvement with respect to zero obliquity models used so far. This method will also be useful for the interpretation of future Earth-based radar observations or Juice data.
\\


\end{abstract}

\section*{Highlights}

- We define the analytical transformation between Laplace plane and ICRF systems.

- We propose an updated orientation/rotation solution, based on a non-zero obliquity model.

- Second-order terms ensure transformation accuracy in the order of arcseconds.

- The transformation preserves the trigonometric form of the series.

\section{Introduction}

The synchronous rotation of the Jupiter’s Galilean satellites, Io, Europa, Ganymede, and Callisto, was only confirmed with certainty after decades of observations of their surface features from Earth telescopes, see \cite{Cam43, Dol74}.
The Galilean satellites are assumed to be in the so-called Cassini state \citep{Col66, Pea69}, which means that their rotation axis follows the long-term precession of their orbit normal while their rotation about the spin axis is periodically modified (physical librations) due to the torque exerted on their triaxial shape by Jupiter. The derivation of control networks for the Galilean satellites from images taken by Voyager 1 and 2 and Galileo did not reveal any significant deviation between their spin and orbit axes \citep{Dav81, Dav98}. The obliquity of the satellites, which is the angular separation of the rotation axis and the orbit normal, is therefore expected to be small (at most a few degrees). 

The obliquity and the libration amplitude of Ganymede will be measured with exquisite precision by Juice's 3GM experiment during the Ganymede orbital phase ($0.2$ and $0.4$ arcsec, respectively, see \citealt{Cap20}), slightly better than the expected accuracy with the GALA instrument (between $0.5$ and $1.7$ arcsec, see \citealt{Ste19}). The obliquity of Europa cannot be measured by 3GM due to the limited data provided by the close flyby, whereas the obliquity of Callisto can be determined to within $0.06^\circ-0.3^\circ$ \citep{Cap22}.
The Europa Clipper spacecraft's radio science and camera instruments will measure the orientation of Europa spin axis \citep{Maz23, Tur24,Ste26} with a precision of the order of $0.05^\circ$ and $0.001^\circ$, respectively. 
In parallel with space missions, the radar speckle tracking observations from Goldstone Solar System Radar and Green Bank Telescope between 2011 and 2023 provided estimates of the orientation of Europa and Ganymede's rotation axis with an accuracy of $0.01^\circ$, \cite{Mar25}.
Note that on Europa and Ganymede, at the equator, $1$ arcsec =~$8$~m and $13$~m, while $0.0001$ deg =~$3$~m and $5$~m, respectively. 
To make the most of these measurements, the rotation models for the satellites must be accurate, both in their description of the link between observables and internal parameters and in the orbital model used. This precise modeling has been an ongoing effort for about 20 years (e.g., \citealt{Hen05, Bil05, VanH08, Noy09, Ram11, Bal12, VanH13, Bal16, Coy26}) and has certainly not yet reached its conclusion.

The orientation and rotation of a body can be described using at least two different sets of angles.
The first set is the Euler angles with respect to a fixed plane with the inertial obliquity $\theta$, the node longitude $\psi$ and the prime meridian location $\phi$. This fixed plane can be the equatorial plane of the central planet at J2000 or, preferably, the Laplace plane. 
The second set of angles is the Earth equatorial coordinates with respect to the Earth equator at J2000 (= International Celestial Reference Frame ICRF equatorial plane): right ascension $\alpha$, declination $\delta$ and prime meridian location $W$ angle. 
These coordinates are adopted by the IAU Working Group on Cartographic Coordinates and Rotational Elements (WGCCRE) to define a rotation model for the main solar system bodies, see \cite{Arc18}.

The goal of this study is to find analytical expressions to transform the orientation angles of the Galilean satellites between the Laplace plane and the ICRF equatorial plane, correct up to the second order in small parameters like the spin axis obliquity.
This transformation can be applied if the orbital obliquity is assumed to be equal to $0$ (orbit normal = spin axis) or finite and given by a theoretical model or by observations. It is therefore an improvement over the zero obliquity model, see for example \cite{Sta18}.
This method is particularly well suited to the transformation of periodic series, as it preserves the periodic form of the input series by avoiding a fit of amplitudes and frequencies on the numerical output time series.
Our transformation model can be used to define the future IAU WG solution for the rotation and orientation using right ascension, declination, and prime meridian location, as predicted by a theoretical model for the spin orientation.
The link between the geophysical parameters of interest and the Earth equatorial coordinates will therefore be more direct than that obtained with a fit to a numerical series.
This study is also useful for the interpretation of future Earth based observation or Juice data, and can be applied to other satellites in synchronous rotation like Titan or the Uranian satellites.

Recently \cite{Yse23} described an analytical method that converts the Martian Euler angles into right ascension/declination and reciprocally, but this method cannot be applied without any change for different reasons. 
First, the node longitude of the Galilean satellites shows a large precession rate (periods of -7, -30, -130 and -560 years) while the Martian method works better for periodic motion with a small precession rate (period of about 26~ky for Mars). Therefore the precession term can be considered as a first-order quantity for Mars but not for the Galilean satellites.
Second, contrary to Mars obliquity $\varepsilon$ that is about $25^\circ$, the obliquity of the Galilean satellites is very small, probably much smaller than $1^\circ$. The denominators of the Mars transformation coefficients (see Eqs.~(21) of \citealt{Yse23}) are usually proportional to $\sin\varepsilon$, which may give approximate first-order coefficients with large value.
Lastly, the spin Cartesian coordinates $s_x$ and $s_y$ projected onto the Laplace plane are easily expressed as trigonometric series computed from the orbital forcing series and therefore are better suited as a starting point than the Euler angles $\theta$ and $\psi$ for a transformation to Earth equatorial coordinates. 

The paper is organized as follows. 
Section~\ref{sec_planes} defines the relevant planes and angles.
Sections~\ref{sec_transfo} and \ref{sec_rot} present the transformation equations for the spin orientation and rotation angles, respectively.
In Section~\ref{sec_dyn_mod}, we introduce a dynamical model for a solid planet, which serves as the basis for generating multiple trigonometric series in Section~\ref{sec_numericalvalues}. Section~\ref{sec_IAUsolution} provides recommendations for the IAU WGCCRE. 
Finally, Section~\ref{sec_discussion} discusses the implications of adopting an alternative ephemerides and the effects of a potential liquid layer.

\section{Planes and angles}
\label{sec_planes}

\begin{figure}[!htb]
      \begin{center}
\includegraphics[width=14cm]{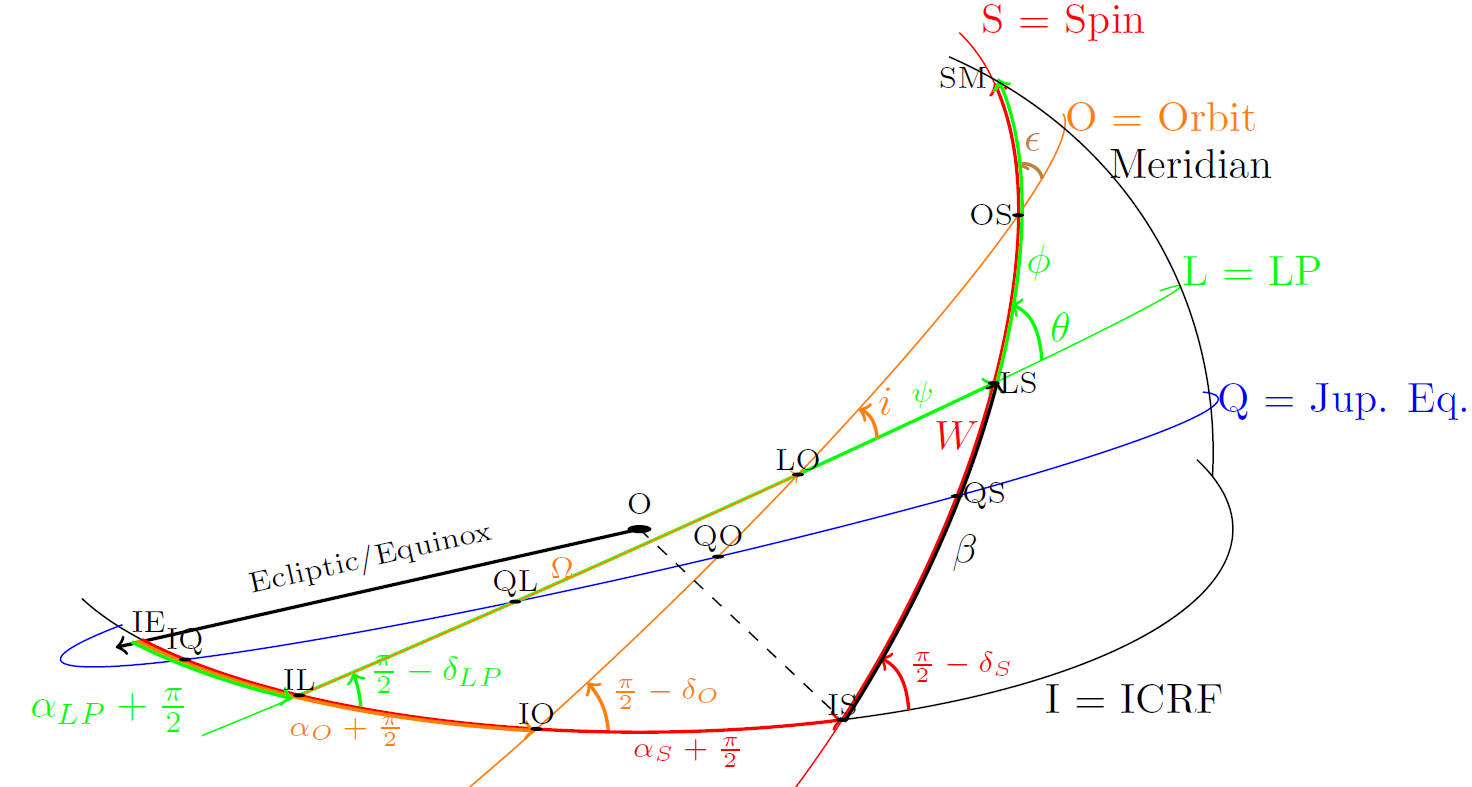}
\caption{Planes and angles.
Angles are not drawn to scale but are exaggerated for the purpose of illustration.
The ICRF equator is denoted by $I$, the Jupiter equator by $Q$, the Laplace plane by $L$, the orbital plane by $O$, and the plane perpendicular to the spin axis by $S$.
Nodes are the intersections between two planes and are labeled using the corresponding plane names. For example, $QO$ denotes the node of the orbital plane on the Jupiter equator. 
The angle $\Omega_{N_1 \, N_2}$ is measured from node $N_1$ to node $N_2$.
\label{fig_ang}}
\end{center}  
\end{figure}

The transformation of a position vector $\vec r$ expressed in the satellite body frame (BF) into its representation $\vec R$ in the Inertial Frame (IF) is:
\begin{equation}
\vec R = \mathbf{M} \; \vec r, 
\label{eq_landerpos}
\end{equation}
with $\mathbf{M}$ the rotation matrix from the BF to the IF. The BF frame is attached to the satellite and centered at its center of mass. Its $x$ and $y-$axes define the satellite's equator, and its $z-$axis defines its figure axis. The Galilean satellites should experience polar motion due to the gravitational torque exerted by their parent planet \citep{Coy26}. This is also the case for Titan, which in addition experiences polar motion due to the exchanges of angular momentum between a seasonally-varying atmosphere and the solid surface \citep{Coy16, Coy18}. However, for the sake of simplicity, we assume here that the figure axis coincides with the spin axis.
Including polar motion in the transformation would simply require additional rotations, see for example Eqs.~(1-3) of \cite{Yse23}. 
The IF is also centered at the center of mass, but its equatorial plane is the Earth mean equator at the J2000 reference epoch (or the equator of the International Celestial Reference Frame ICRF).
The IF $X-$axis points towards the ascending node of the ecliptic on the Earth equator (i.e.~towards the vernal point $IE$ in Fig.~\ref{fig_ang}). 
The transformation matrix $\mathbf{M}$ can be written in two different ways:
\begin{subequations}
\label{eq_M}
\begin{eqnarray}
\mathbf{M} &=& \mathbf{M}_{\theta\psi\phi} =
R_Z\left(-\frac{\pi}{2}-\alpha_{LP}\right) \cdot R_X\left(-\frac{\pi}{2}+\delta_{LP}\right) \cdot R_Z(-\psi) \cdot R_X(-\theta) \cdot R_Z(-\phi),
\label{eq_Mep} \\
&=& \mathbf{M}_{\alpha\delta W} = R_Z\left(-\frac{\pi}{2} - \alpha_S\right) \cdot R_X\left(-\frac{\pi}{2} + \delta_S\right) \cdot R_Z(-W),
\label{eq_Mad}
\end{eqnarray}
\end{subequations}
depending on whether Euler ($\theta, \psi, \phi$) or Earth equatorial ($\alpha_S, \delta_S,W$) angles are used for the spin position.
The rotation matrices are defined as
\begin{equation}
R_X[w] = \left( \begin{array}{ccc}
 1 & 0 & 0 \\
 0 & \cos w & \sin w \\
 0 & -\sin w & \cos w \\
\end{array} \right), \quad
R_Z[w]=\left(\begin{array}{ccc}
 \cos w & \sin w & 0 \\
 -\sin w & \cos w & 0 \\
 0 & 0 & 1 \\
\end{array} \right).
\end{equation}

The orientation of the $z-$axis is either defined by the inertial obliquity $\theta$ and the node longitude $\psi$ with respect to the Laplace plane, or by the right ascension $\alpha_S$ and declination $\delta_S$ with respect to the ICRF.
The Laplace plane (LP) is the mean plane around which the instantaneous orbital plane of the satellite precesses, see Section~\ref{sec_reference}. 
Note that here $\psi = \Omega_{IL\,LS}$ is the angle measured from the node of the Laplace plane on the ICRF ($IL$), a convention that differs from that of \cite{Bal12} and ending in $LS$, the node of the  satellite equator on the Laplace plane, where $\psi= \Omega_{QL\,LS}$ is measured from the node of the Laplace plane on the equator of Jupiter ($QL$) up to $LS$. These two angles both belong to the Laplace plane.
The inertial obliquity $\theta$ of the spin axis is measured from the Laplace plane pole, and should not be confused with the orbital obliquity $\varepsilon$, measured from the orbit normal. 

The longitude systems of Europa, Ganymede and Callisto have been defined by reference to a crater by the IAU WGCCRE, whereas that of Io has been defined by the mean sub-Jovian direction (see Table 2 of \citealt{Arc18}).
The prime meridian ($x$-axis of BF) is positioned by $\phi$ or $W$, depending on the angles’ set, see Fig.~\ref{fig_ang}. The difference $W-\phi$ is defined as the angle $\beta$, which varies with time, see Section~\ref{sec_rot}. The rotation angle $\phi$ of the satellite must include periodic variations, called librations, with respect to synchronous rotation. These variations are also included in $W$.

The Earth equatorial angles, as adopted by the Cartographic Coordinates and Rotational Elements Working Group from the IAU, can be used for data analysis. Euler angles are, however, preferred by the rotation modeling community. Dynamical models for spin precession caused by the gravitational torque exerted on the satellite by the parent planet generally resolve the orientation of the satellite with respect to a reference plane, which is the orbit or the Laplace plane, whereas libration models resolve for the rotation variations with respect to synchronicity with the orbital motion (see Section~\ref{sec_dyn_mod}). 

We position the Laplace plane for each satellite in the ICRF frame with two angles that are constant over time: the right ascension $\alpha_{LP}$ and the declination $\delta_{LP}$ (see Fig.~\ref{fig_ang}). 
$\Omega_{IE\,IL} = \alpha_{LP} + \pi/2$ is the node longitude along the ICRF equator from the ecliptic to the Laplace plane. 
$i_{IL} = \pi/2 - \delta_{LP}$ is the inclination of the Laplace plane on the ICRF equator. The orientation of the orbital plane is defined with $i$ and $\Omega$, the orbital inclination and longitude of the node with respect to the Laplace plane, or with $\alpha_O$ and $\delta_O$, the orbital right ascension and declination with respect to the ICRF. Note that the position of the parent planet in a frame centered on the satellite, measured from the orbital node on the Laplace plane, is defined by the sum $\omega-\pi+f$, with $\omega$ and $f$ the argument of the pericenter and the true anomaly, respectively, and by the semi-major axis $a$ and eccentricity~$e$.

\subsection{The orbital theory}
\label{sec_orbit}

The four Galilean satellites form a complex dynamical system. Their orbital variations, including precession, are mainly driven by the planet’s polar flattening on the one hand, and by mutual perturbations between the satellites and solar perturbations on the other hand, as well as by the Laplace resonance between Io, Europa and Ganymede.
Current solutions for the ephemerides of the Galilean moons are mainly based on ground-based astrometric observations and space-based data from Voyager, Galileo, and Juno.
For the numerical applications presented in this study (see Section~\ref{sec_numericalvalues}), we choose as orbital theory for the Galilean satellites the numerical ephemerides JUP387 computed by R.~Jacobson from JPL. They are available over the $1600-2400$ time period. 
In Section~\ref{sec_orbcomp}, we quantify the changes in spin position obtained using the last version of NOE \citep{Lai09} instead of JUP387.

\begin{figure}[htb!]
\centering
\includegraphics[width=13cm]{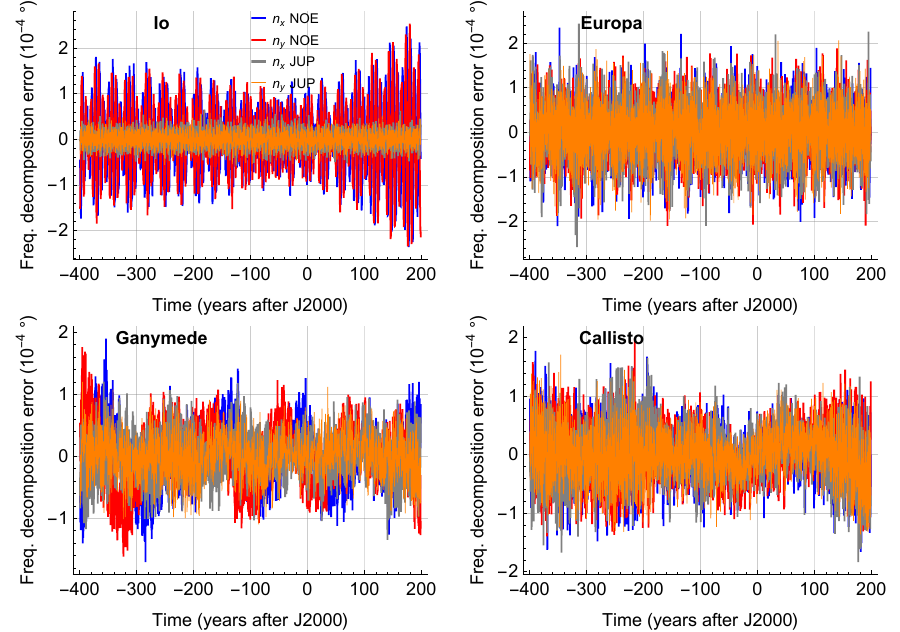}
\caption{Residuals of the orbit normal fit for the JUP and the NOE ephemerides for the four Galilean satellites, referred to their respective estimated Laplace plane.}
\label{fig_Residuals}
\end{figure}

Both JUP387 and the last version of NOE are represented by Chebyshev polynomials. However, a quasi-periodic decomposition of the orbital motion is particularly useful for computing the orientation/rotation of satellites. It is then possible to decompose the forcings into series and evaluate the rotational response term by term, retaining only the main terms of the solution. 
Using a Levenberg-Marquardt algorithm, we simultaneously fit the orientation of the Laplace plane (see Section~\ref{sec_reference}) over the $1600-2200$ time interval and a periodic series, with at least ten terms, describing the projected components of the orbit normal onto this plane ($n_x$, $n_y$, see Section~\ref{sec_proj}). The fit provides the amplitudes, frequencies, and phases of the periodic terms.
The frequencies correspond to integer combinations of the fundamental arguments of the dynamical system. 
The Laplace plane is given by the zero frequency term in our decomposition. 
We use the numerical values of \cite{Lai06} as a priori values for the fit. In order for each satellite to also follow the very slow (period of about $400 \, 000$ years, see \citealt{Lem16}) precession of Jupiter’s spin axis, a linear trend is also adjusted for each angle (see also Section~\ref{sec_IAUsolution}). 
The trend may be difficult to separate from very long-period terms, such as Callisto’s 560-year orbital precession. 
However, for Callisto this periodic orbital perturbation has the dominant amplitude and is relatively large (0.25°, see \citealt{Lai06}), which helps to distinguish it from the secular trend.
The fit residuals are usually less than $0.0002^\circ$, see Fig.~\ref{fig_Residuals}.

\subsection{The reference plane}
\label{sec_reference}

\begin{table}[!htb]
\small
\begin{center}
\begin{tabular}{cccccc}
\hline
& & Io & Europa & Ganymede & Callisto \\
\hline 
& & & & \\
Laplace pole: & & & & & \\
& & & & & \\
$\alpha_{LP}$ ($^\circ$) & & 268.0594 & 268.0850 & 268.2044 & 268.7322 \\
                         & & (268.05) & (268.08) & (268.20) & (268.72) \\ 
$\delta_{LP}$ ($^\circ$) & &  64.4968 &  64.5095 &  64.5683 &  64.8244 \\
                         & & (64.50)  & (64.51)  & (64.57)  & (64.83) \\ 
& & & & & \\
Transformation coefficients: & & & & & \\
 & & & & \\
$1/\cos\delta_{LP}$ & $\mathcal{O}1$ & 2.3226 & 2.3236 & 2.3286 & 2.3508 \\
$\tan \delta_{LP} \, /\cos\delta_{LP}$ & $\mathcal{O}2$ & 4.8686 & 4.8737 & 4.8971 & 5.0011 \\
$- 1/2 \, \tan \delta_{LP}$ & $\mathcal{O}2$ & -1.0481 & -1.0487 & -1.0515 & -1.0637 \\
 & & & & \\
$\cos\delta_{LP}$ & $\mathcal{O}1$ & 0.4306 & 0.4304 & 0.4294 & 0.4254 \\
$-\sin\delta_{LP}$ & $\mathcal{O}2$ & -0.9026 & -0.9027 & -0.9031 & -0.9050 \\
$ 1/4 \sin 2\delta_{LP}$ & $\mathcal{O}2$ & 0.1943 & 0.1942 & 0.1939 & 0.1925 \\
$-1/2 \, \cos^2\delta_{LP}$ & $\mathcal{O}2$ & -0.0927 & -0.0926 & -0.0922 & -0.0905 \\
\hline
\end{tabular}
\end{center}
\caption{Numerical values for the orientation angles of the Laplace plane for the four Galilean satellites and transformation coefficients, using the JUP387 ephemerides. The number of digits retained in this table is related to the order of magnitude of the residuals from the quasi-periodic decomposition of JUP387, see Fig.~\ref{fig_Residuals}. 
$\mathcal{O}1$ denotes a first-order coefficient, while $\mathcal{O}2$ denotes a second-order coefficient (see Eqs.~(\ref{eq_adapprox}-\ref{eq_sxapprox})). 
Values in parentheses are from \cite{Arc18}.
The Jupiter pole coordinates are $\alpha_{J}= 268.0564^\circ$, $\delta_{J} = 64.4953^\circ$ (see ephemerides JUP387 computed by R. Jacobson from JPL).}
\label{tab_refplanes}
\end{table}

As mentioned above, we need a fixed plane to serve as a reference plane for describing the orbital elements involved in the forcing torque and the resulting rotational response.
The equator of Jupiter at epoch J2000 may be proposed as a fixed reference plane, as is usually the case to describe the orbital motion of the Galilean satellites. However, due to the influence of the Sun, Jupiter's equator is not the best reference plane to describe the rotational response of the satellites (see e.g.~\cite{Noy09}). 
A relevant choice is the plane perpendicular to the center of the orbital cone for a regular motion, or for non-regular motion, that keeps the orbit inclination variations as small as possible. 
This reference plane is known as the Laplace plane. 
Since there are different ways to compute the Laplace plane (depending, for example, on the choice of the ephemerides, the time interval, or the quantity to minimize), different numerical estimations could be obtained. 
Our numerical values of the Laplace plane position for the four Galilean satellites are given in Table~\ref{tab_refplanes}, using the JUP387 ephemerides. The farther the satellite is from the planet, the farther the Laplace plane is from Jupiter's equator.

\subsection{Projection of the spin axis in the reference plane}
\label{sec_proj}

Following previous rotation studies (e.g.~\cite{Bil05}, \cite{Bal12}), we introduce the components ($n_x, n_y, n_z$) of the unit vector along the orbit normal expressed in the coordinates of the Laplace reference frame:
\begin{subequations}
\label{eq_nxnynz}
\begin{eqnarray}
n_x(t) &=& \quad \sin i(t) \; \sin\Omega(t) \approx \quad i(t) \, \sin\Omega(t), \\ 
n_y(t) &=&     - \sin i(t) \; \cos\Omega(t) \approx - \, i(t) \; \cos\Omega(t), \\
n_z(t) &=& \quad \cos i(t).
\end{eqnarray}
\end{subequations}
The angle $i$ corresponds to the angle $i_{LO}$ in Fig.~\ref{fig_ang} ($i \approx \sqrt{n_x^2+n_y^2}$) and the node longitude is $\Omega = \Omega_{IL\,LO}$. Similarly, the components $(s_x, s_y, s_z)$ of the unit vector along the spin axis expressed in the coordinates of the Laplace reference frame are given by:
\begin{subequations}
\label{eq_sxsythetapsi}
\begin{eqnarray}
s_x(t) &=& \quad \sin\theta(t) \; \sin\psi(t) \approx \; \theta(t) \; \sin\psi(t) , \label{eq_sx} \\ 
s_y(t) &=&     - \sin\theta(t) \; \cos\psi(t) \approx -\theta(t) \; \cos\psi(t) , \label{eq_sy} \\
s_z(t) &=& \quad \cos\theta(t).
\label{eq_sz}
\end{eqnarray}
\end{subequations}
The inertial obliquity $\theta(t)$ is the angle between the spin axis and the Laplace pole $i_{LS}$ ($\theta \approx \sqrt{s_x^2+s_y^2}$).
$\psi(t) = \Omega_{IL\,LS}$ is the node longitude of the equatorial plane along the Laplace plane, starting from the ICRF node. 

As we will show in the next section, the components $s_x$ and $s_y$ are more convenient to transform to the right ascension/declination angles than the angles $\theta$ and $\psi$ taken separately because $s_x/s_y$ mainly present small periodic variations, while $\psi(t)$ is dominated by a large precession trend.

\section{Transformation between the orientation angles \texorpdfstring{$\alpha_S$}{alphaS}, \texorpdfstring{$\delta_S$}{deltaS} and \texorpdfstring{$\theta$}{theta}, \texorpdfstring{$\psi$}{psi}}
\label{sec_transfo}

In this section, we describe the precise transformation between the Euler and Earth equator orientation angles of the spin axis, which is independent of the rotation angles $\phi$ and $W$. We use as intermediate variables the components of the unit vector along the spin axis in the coordinates of the Laplace frame, $\uvec{s} = (s_x, s_y, s_z)$ which are defined in terms of $\theta$ and $\psi$, see Eqs.~(\ref{eq_sxsythetapsi}).
The transformation described below is also valid for the orientation of the orbit normal, provided that the components $\uvec{n} = (n_x,n_y,n_z)$ of the unit vector along the orbit normal in the coordinates of the Laplace frame (Eqs.~\ref{eq_nxnynz}) are used as intermediate variables and subscript $O$ is used instead of subscript $S$ for the Earth equatorial angles. 

\subsection{Exact transformations}
\label{sec_exact}

Using the classical spherical trigonometry or equivalently the equality between the rotation matrices $\mathbf{M}_{\theta\psi\phi}$ and $\mathbf{M}_{\alpha\delta W}$ of Eqs.~(\ref{eq_M}), the exact transformations from the spin coordinates $(s_x, s_y, s_z)$ with respect to the Laplace frame to the Earth equatorial angles $(\alpha_S, \delta_S)$ are given by:
\begin{subequations}
\label{eq_fromsxtoad}
\begin{eqnarray}
\sin\delta_S &=& s_y \cos\delta_{LP} + \sin\delta_{LP} \, s_z,
\label{eq_sde} \\
\cos\delta_S \sin(\alpha_S-\alpha_{LP}) &=& s_x, 
\label{eq_sx2} \\
\cos\delta_S \cos(\alpha_S-\alpha_{LP}) &=& \cos\delta_{LP} \, s_z - s_y \sin\delta_{LP}.
\label{eq_sa} 
\end{eqnarray}
\end{subequations}
The opposite transformations from the Earth equatorial angles to the spin coordinates in the Laplace frame are given by:
\begin{subequations}
\label{eq_fromadtosx}
\begin{eqnarray}
s_x &=& \cos\delta_S \sin(\alpha_S-\alpha_{LP}),
\label{eq_sx3} \\
s_y &=& \sin\delta_S \cos\delta_{LP} - \cos\delta_S \sin\delta_{LP} \cos(\alpha_S-\alpha_{LP}), \label{eq_sy2} \\
s_z &=& \sin\delta_S \sin\delta_{LP} + \cos\delta_S \cos\delta_{LP} \cos(\alpha_S-\alpha_{LP}).
\label{eq_sz2} 
\end{eqnarray}
\end{subequations}
We aim to express each angle/unit vector component as a sum of different parts, including periodic series.
In order to maintain the representation in series for the periodic parts, approximate but accurate transformations are defined below.

\subsection{Approximate transformations}
\label{sec_app}

With the exception of $s_z$, each angle and unit vector component is expressed as the sum of the values of the Laplace pole coordinates, a slow linear term with a rate noted $tr$, and a series noted $\Delta$ of periodic terms with small amplitudes:
\begin{subequations}
\label{eq_sxform}
\begin{eqnarray}
s_x(t) &=& tr_x \, t + \Delta s_x(t), \label{sx} \\
s_y(t) &=& tr_y \, t + \Delta s_y(t), \label{sy} \\
s_z(t) &=& 1 + \Delta s_z(t),  
\\ 
\alpha_S(t) &=& \alpha_{LP} + tr_\alpha \, t + \Delta\alpha_S(t), \\
\delta_S(t) &=& \delta_{LP} + tr_\delta \, t + \Delta\delta_S(t),
\end{eqnarray}
\end{subequations}
where $t$ is the time. The Laplace pole location in the coordinates of the frame attached to the Laplace frame  is by definition given by $(0,0,1)$. 
The values of $\alpha_{LP}$ and $\delta_{LP}$ depend on the chosen orbital model, see Section~\ref{sec_planes}. 
The periodic terms $\Delta s_x$ and $\Delta s_y$ are the nutations in space of the spin axis (the dominant part of the nutational motion, if it exists, can be called precession by analogy with Earth and Mars). The periodic parts $\Delta\alpha_S(t)$ and $\Delta\delta_S(t)$ can also be called nutations. In the present context of approximate transformations, all these periodic variations are considered as first order terms. 
By construction, the spin models for $s_x$ and $s_y$ also contain secular terms. Since for the Galilean satellites, over a 100-year interval around J2000 (the time interval we target for the spin model), they are much smaller than the largest periodic terms, these secular terms are considered to be of second order. Therefore, since the spin vector is a unit vector, 
$\Delta s_z \simeq - \left(\Delta s_x^2 + \Delta s_y^2 \right)/2$ 
and is formally considered here as a second-order quantity. By construction, $s_z$ should also include quadratic and Poisson terms. Very small quadratic and Poisson terms (product between a linear term and a periodic series) are also neglected here in $\alpha_S/\delta_S$, because they are formally considered of order 3 or above and do not significantly improve accuracy in numerical applications. 

To obtain approximate expressions for $\Delta\alpha_S$ and $\Delta\delta_S$, assuming that we have expressions for $\Delta s_x$ and $\Delta s_y$, we express the exact trigonometric expressions of Eqs.~(\ref{eq_fromsxtoad}) correct up to the second-order in $\Delta$ quantities. Extracting $\Delta\alpha_S$ and $\Delta\delta_S$, we find, see also \cite{Lie93, Lie93b}:
\begin{subequations}
\label{eq_adapprox}
\begin{eqnarray}
\Delta\alpha_S &=& \frac{1}{\cos\delta_{LP}} \Delta s_x + \frac{\tan \delta_{LP}}{\cos\delta_{LP}} \, \Delta s_x \, \Delta s_y, 
\label{eq_alphatransfo} \\
\Delta\delta_S &=& \qquad \Delta s_y \quad - \frac{1}{2}\tan \delta_{LP} \, \Delta s_x^2. 
\label{eq_deltatransfo}
\end{eqnarray}
\end{subequations}
The inverse transformation is given by:
\begin{subequations}
\label{eq_sxapprox}
\begin{eqnarray}
\Delta s_x &=& \cos\delta_{LP} \, \Delta\alpha_S - \sin\delta_{LP} \, \Delta\alpha_S \, \Delta\delta_S , \\
\Delta s_y &=& \Delta\delta_S + \frac{1}{4} \sin 2\delta_{LP} \, \Delta\alpha_S^2, \\
\Delta s_z &=& \frac{-1}{\, 2} \left(\cos^2\delta_{LP} \, \Delta\alpha_S^2 + \Delta\delta_S^2 \right).
\end{eqnarray}
\end{subequations}
The transformation coefficients in Eqs.~(\ref{eq_adapprox}-\ref{eq_sxapprox}) depend on the Laplace plane orientation and slightly differ for each Galilean satellite. Numerical values of the first and second-order coefficients for the four Galilean satellites are listed in Table~\ref{tab_refplanes}. 
Similar expressions, but restricted to the first-order part of the transformation, apply to the small trends ($tr$ quantities):
\begin{subequations}
\label{eq_trends}
\begin{eqnarray}
\cos\delta_{LP} \, tr_\alpha &=& tr_x, \\
tr_\delta                    &=& tr_y.
\end{eqnarray}
\end{subequations}
Since quadratic and Poisson terms are neglected here, their transformation relations are not given. Such transformations have been provided by \cite{Yse23} in the case of Mars.

The first-order series in $\Delta\alpha_S$ and $\Delta\delta_S$, obtained from Eqs.~(\ref{eq_adapprox}) by neglecting second-order terms, have an analytical trigonometric form similar to that of $s_x, s_y$. In particular, they involve the same number of terms and the same set of frequencies.
The second-order terms result in additional periodic terms to the series, due to the recombination of sine and cosine factors. 
This recombination may introduce a small constant term in addition to the periodic series.
Note that the distinction between first and second order is purely formal, and not based on the actual values of the amplitudes. Some of these second-order terms may be of larger amplitude than some of the first-order terms. For this reason, the second-order terms allow to improve the accuracy of the transformation (from $0.01^\circ$ to less than $0.0001^\circ$ in $\alpha$, and from $0.006^\circ$ to $0.0001^\circ$ in $\delta$), see Table~\ref{tab_error} and Section~\ref{sec_admseries}.

\subsection{Back to the obliquity and node longitude}
\label{sec_obli}

The inertial obliquity and node longitude of the spin axis can be retrieved from the following exact relations:
\begin{subequations}
\begin{eqnarray}
\sin\theta &=& \sqrt{s_x^2 + s_y^2}, \label{eq_sintheta} \\
\psi &=& \arctan(-s_y, s_x).      
\label{eq_tanpsi} 
\end{eqnarray} 
\end{subequations}
Similarly, the inclination and node longitude of the orbit normal are obtained as
\begin{subequations}
\begin{eqnarray}
\sin i &=& \sqrt{n_x^2 +n_y^2},    \label{eq_i} \\
\Omega &=& \arctan(-n_y,n_x).  
\label{eq_Omega}
\end{eqnarray} 
\end{subequations}
The orbital obliquity (the angle between the orbit normal and the spin) is defined as:
\begin{equation}
\cos\varepsilon =  \uvec{n} \cdot \uvec{s}. 
\label{eq_eps}
\end{equation}

Following the notation of the built-in Mathematica ArcTan function, we use the notation $\arctan{(x,y)}$ in the definitions for $\psi$ and $\Omega$ to take into account the quadrant in which the point $(x,y)$ lies.
If we take into account the quadrant in which the projections of $\uvec{s}$ and $\uvec{n}$ onto the Laplace plane lie when calculating $\psi$ and $\Omega$, then $\theta$ and $i$ will be positive by default. This is convenient for satellites whose spin and orbit poles are on the same side with respect to the Laplace pole. For a satellite like the Moon, whose spin axis and orbit normal are located on opposite sides with respect to the Laplace pole, and thus their projections in opposite quadrants, it can be useful from a modeling point of view (where $\psi$ is generally assumed to be equal to or close to $\Omega$) to change the sign of $\theta$ and add $\pi$ to $\psi$. This is equivalent to redefining the descending node as the ascending node, but with a negative obliquity.

We can find valid approximations for the inertial obliquity $\theta$ and longitude $\psi$, either as a function of $\Delta s_x$ and $\Delta s_y$, or as a function of $\Delta\alpha_S$ and $\Delta\delta_S$, using the relationship between the two sets of periodic variations, up to order 2: 
\begin{subequations}
\label{eq_thetapsi}
\begin{eqnarray}
\theta &\approx& \sqrt{\Delta s_x^2 + \Delta s_y^2} + \frac{(tr_x \, \Delta s_x+tr_y \, \Delta s_y)}{\sqrt{\Delta s_x^2 + \Delta s_y^2}} \, t, 
\label{eq_thetasxsy} \\
 &\approx& \sqrt{\cos^2\delta_{LP} \, \Delta\alpha_S^2 + \Delta\delta_S^2} -\frac{\Delta\alpha_S^2 \, \Delta\delta_S \sin 2 \delta_{LP} }{4 \sqrt{\cos ^2\delta_{LP}\Delta\alpha_S^2+\Delta\delta_S^2}} + \frac{(\cos^2 \delta_{LP} \, tr_\alpha\, \Delta\alpha_S + tr_\delta \, \Delta\delta_s)}{\sqrt{\cos^2 \delta_{LP} \, \Delta\alpha_S^2 + \Delta\delta_S^2}} t,  
 \label{eq_thetaad} \\
\psi&\approx& \arctan{(-\Delta s_y , \Delta s_x)} + \frac{(tr_y \, \Delta s_x-tr_x\, \Delta s_y)}{\Delta s_x^2+\Delta s_y^2} \, t, 
\label{eq_psisxsy} \\
\nonumber &\approx& \arctan{(-\Delta\delta_S, \cos\delta_{LP} \, \Delta\alpha_S)} + \frac{1}{2} \Delta\alpha_S \sin\delta_{LP}\left(\frac{\Delta\alpha_S^2\cos^2\delta_{LP} + 2 \Delta\delta_S^2}{\Delta\delta_S^2 + \Delta\alpha_S^2 \cos^2\delta_{LP}}\right)\\
&&+ \frac{(tr_\delta \, \Delta\alpha_S - tr_\alpha \, \Delta\delta_S) \cos\delta_{LP}}{\Delta\delta_S^2 + \Delta\alpha_S^2 \cos^2\delta_{LP}}\, t. \nonumber \\
&& \label{eq_psiad}
\end{eqnarray}
\end{subequations}

Due to the linear terms in $s_x$ and $s_y$ (or in $\alpha_S$ and $\delta_S$), these relations contain some terms proportional to the linear rates $tr_{x/y/\alpha/\delta}$, or Poisson terms, which we retain since they ensure a higher degree of precision. 
For example, the resulting error on Europa’s inertial obliquity with respect to Eq.~(\ref{eq_sintheta}) is $\leq 0.0001^\circ$ when these terms are included, compared to $0.002^\circ$ when the Poisson terms are omitted. 
For Io, those terms contribute up to $6^\circ$ in $\psi$ after $100$ years, see section \ref{sec_obliquityseries}.

As mentioned in the introduction to this section, the transformations of Section~\ref{sec_app} also apply to $\uvec{n} = (n_x, n_y, n_z)$ and $(\alpha_O, \delta_O)$. Therefore similar relationships to Eqs.~(\ref{eq_thetapsi}) exist for the inclination $i$ and node longitude $\Omega$, with $\Delta n_x$ and $\Delta n_y$ instead of $\Delta s_x$ and $\Delta s_y$ and $\Delta\alpha_O$ and $\Delta\delta_O$ instead of $\Delta\alpha_S$ and $\Delta\delta_S$. For the same reason, this also allows us to find a workable expression for the orbital obliquity $\varepsilon$:
\begin{subequations}
\begin{eqnarray}
\varepsilon 
&\approx& \sqrt{(\Delta s_x- \Delta n_x)^2 + (\Delta s_y - \Delta n_x)^2},   
\label{eq_epsilon}  \\
\nonumber &\approx& \sqrt{\cos\delta_{LP}^2 (\Delta\alpha_S - \Delta\alpha_O)^2+(\Delta\delta_S - \Delta\delta_O)^2}\\
&&
    -\frac{(\Delta\alpha_O-\Delta\alpha_S)^2 (\Delta\delta_O+\Delta\delta_S) \sin2\delta_{LP}}{4 \sqrt{\cos\delta_{LP}^2 (\Delta\alpha_S - \Delta\alpha_O)^2+(\Delta\delta_S - \Delta\delta_O)^2}} .
\label{eq_epsilonadd}
\end{eqnarray}    
\end{subequations}
The orbital obliquity is unaffected by the long-term trends. 
The first-order part of Eq.~(\ref{eq_epsilonadd}) can be used to propagate errors from the IAU orientation angles to the orbital obliquity.

\subsection{Offset with respect to the Cassini plane}
\label{sec_offset}

If the satellite's orbit and spin precess regularly at a constant rate and if the satellite is unaffected by tidal dissipation, then the Laplace pole $\uvec{l}$ (the normal to the Laplace plane), the orbit normal $\uvec{n}$ and the spin axis $\uvec{s}$ belong to the same plane, known as the Cassini plane. 
The Cassini plane is defined as the plane containing both the Laplace pole and the orbit normal.
Due to quasi-periodic orbital precession and/or coupling with polar motion and/or tidal dissipation, the spin axis is in fact almost never in the Cassini plane.
The offset $\zeta$ of the spin axis with respect to the Cassini plane can be written as, see \cite{Yse06, Bal17}:
\begin{equation}
\sin\zeta = \frac{\uvec{l} \times \uvec{n}}{|| \, \uvec{l} \times \uvec{n} \, ||}\cdot \uvec{s}.
\label{eq_offset}
\end{equation}
With this sign convention, the offset is defined as a prograde angle. In the case of a dissipative body with a regular retrograde orbital precession, a positive offset indicates a spin axis lagging “behind” the Cassini plane. 

Using the expressions of the $\uvec{n}$ and $\uvec{s}$ vectors and since $\uvec{l}=(0,0,1)$ in the Laplace Frame, Eq.~(\ref{eq_offset}) can be reduced to
\begin{subequations}
\begin{eqnarray}
\nonumber \zeta &\simeq& \frac{ \Delta n_x \, \Delta s_y - \Delta n_y \, \Delta s_x }{\sqrt{ \Delta n_x^2 + \Delta n_y^2 }}\\
&&+ \frac{\left( \Delta n_x \, (\Delta n_x - \Delta s_x) + \Delta n_y \, (\Delta n_y - \Delta s_y) \right)
}{\left( \Delta n_x^2 + \Delta n_y^2 \right)^{3/2}} (tr_y \, \Delta n_x - tr_x \, \Delta n_y) \, t, 
\label{eq_zetasxsy} \\
&\simeq& \frac{
\left( \Delta\alpha_O \, \Delta\delta_S - \Delta\alpha_S \, \Delta\delta_O \right) \cos\delta_{LP}} {\sqrt{ \Delta\delta_O^2 + \Delta\alpha_O^2 \cos^2\delta_{LP} }} \nonumber \\
&& + t \, \frac{\left( \Delta\alpha_O \, tr_\delta - \Delta\delta_O \, tr_\alpha \right)  \left( \Delta\delta_O (\Delta\delta_O - \Delta\delta_S) + \Delta\alpha_O (\Delta\alpha_O - \Delta\alpha_S) \cos^2\delta_{LP} \right) \cos\delta_{LP} }{ \left( \Delta\delta_O^2 + \Delta\alpha_O^2 \cos^2\delta_{LP} \right)^{3/2}}  \nonumber \\
&& +\frac{\sin\delta_\text{LP} \, (\Delta\alpha_\text{S} - \Delta\alpha_\text{O})\left( 2\Delta\delta_\text{O}^3 \Delta\delta_\text{S} + \Delta\alpha_\text{O}^3 \Delta\alpha_\text{S} \cos^4\delta_\text{LP} \right)}{  2 \left( \Delta\delta_O^2 + \Delta\alpha_O^2 \cos^2\delta_{LP} \right)^{3/2}}
\nonumber \\
&& -\frac{\sin\delta_\text{LP} \, \Delta\alpha_\text{O}\, \Delta\delta_\text{O} \cos^2\delta_\text{LP}
\left( \Delta\alpha_\text{O}^2 \Delta\delta_\text{S} + 2 \Delta\alpha_\text{O} \Delta\alpha_\text{S} (\Delta\delta_\text{O} - \Delta\delta_\text{S}) - \Delta\alpha_\text{S}^2 \Delta\delta_\text{O}
\right)}{  2 \left( \Delta\delta_O^2 + \Delta\alpha_O^2 \cos^2\delta_{LP} \right)^{3/2}}.
\label{eq_offset2}
\end{eqnarray}    
\end{subequations}
The first term of Eq.~(\ref{eq_zetasxsy}) corresponds to Eq.~(9) of \cite{Bal11}.

\section{Transformation between the rotation angles \texorpdfstring{$\phi$}{phi} and \texorpdfstring{$W$}{W}}
\label{sec_rot}

The rotation angles $\phi$ and $W$ describe the location of the intersection between the prime meridian and the equator of the satellite, measured from the node of the equator over the Laplace plane and over the Earth mean equator of J2000, respectively. 
In this section, we now use the notation $\phi_{Euler}$ for the Euler rotation angle, to distinguish it from other rotation angles also noted with the letter $\phi$ and measured from a different origin, see Section~\ref{sec_defphi}. 

We assume for the rotation angles a generic form with an initial value at epoch, a linear term, and a series of small periodic variations ($\Delta\phi_{Euler}$ and $\Delta W$ are related to the librations in longitude, see Section~\ref{sec_defphi}):
\begin{subequations}
\begin{alignat}{4}
\phi_{Euler}(t) &= \phi_0 && + \dot\phi \, t && + \Delta\phi_{Euler}(t), \\
W(t) &= W_0 && + \dot W \, t && + \Delta W(t). 
\end{alignat}
\end{subequations} 

In this manuscript, $\dot\xi$ denotes the linear rate of the angle $\xi$, while $\xi'$ is used to denote time derivatives.

The difference between the rotation angles $W$ and $\phi_{Euler}$ is the angle $\beta$ corresponding to the angle $\Omega_{IS\,LS}$ ($IS$ is the intersection of the equator on the ICRF plane) in Fig.~\ref{fig_ang}:
\begin{equation}
W = \phi_{Euler} + \beta.   
\label{eq_W}
\end{equation}
It is measured along the equator of the satellite from the node with the ICRF to the node with the Laplace plane.

\subsection{Exact transformation} 
\label{sec_beta}

The $\beta$ angle is not constant with time due to the precessional motion of the equator but 
does not intervene in the transformations between orientation angles of Section~\ref{sec_transfo}. 
Its cosine and sine can be written as a function of the Euler orientation angles $\psi$ and $\theta$ and the declinations of the spin axis ($\delta_S$) and Laplace pole ($\delta_{LP}$) as follows: 
\begin{subequations}
\label{eq_csbeta} 
\begin{eqnarray}
\cos\beta 
&=& \frac{\cos\delta_{LP} \cos\psi \cos\theta + \sin\delta_{LP} \sin\theta}{\cos\delta_S}, \label{eq_cosbeta} \\
\sin\beta &=& \frac{\cos\delta_{LP} \sin\psi} {\cos\delta_S}, 
\label{eq_sinbeta}
\end{eqnarray}
\end{subequations}
which leads to (still using the notation $\arctan{(x,y)}$ for a point of coordinates $(x,y)$)
\begin{eqnarray}
\beta(t) &=& \arctan{(\cos\beta, \sin\beta)} = \arctan{( \cos\psi \, \cos\theta + \tan\delta_{LP} \sin\theta, \sin\psi)}, 
\label{eq_beta2}
\end{eqnarray}
where $\sin{\theta} = \sqrt{s_x^2+s_y^2}$ and $\cos{\theta} = \sqrt{1-s_x^2-s_y^2}$.

\subsection{Approximate transformation}
\label{sec_rotapp}

Using the approximate expressions for $s_x$ and $s_y$ of Eqs.~(\ref{sx}-\ref{sy}) and the transformation 
relationship between the different sets of orientation angles, we now find approximations for $\beta$, either as a 
function of $tr_x$, $\Delta s_x$ and $\Delta s_y$, or as a function of $tr_\alpha$, $\Delta\alpha_S$ and $\Delta\delta_S$ 
(we neglect quadratic and Poisson terms):
\begin{subequations}
\label{eq_betaapp} 
\begin{eqnarray}
\beta(t) &\approx& \psi(t) \, - \, \tan\delta_{LP} \, tr_x \, t- \, \tan\delta_{LP} \, \Delta s_x - \left(\tan^2\delta_{LP} + \frac{1}{2}\right) \, \Delta s_x \, \Delta s_y ,
\label{eq_beta3} \\
&\approx& \psi(t) \, - \, \sin\delta_{LP} \, tr_\alpha \, t - \, \sin\delta_{LP} \, \Delta\alpha_S - \frac{1}{2}\cos{\delta_{LP}}\, \Delta\alpha_S \, \Delta\delta_S.
\label{eq_beta4}
\end{eqnarray}
\end{subequations}
From these expressions, we can see that, although the two angles do not belong to the same plane, $\beta$ and $\psi$ have fairly close values at all times (maximum difference of around $0.7^\circ$ for Ganymede). They both exhibit a precession (linear trend) which causes them to deviate relatively rapidly from their epoch values. 
In the following, we define the difference between $\beta$ and $\psi$ as the angle $\mu$, that includes the difference between the trends and periodic variations of the two angles (see also \cite{Lie93, Lie93b} for $\Delta \mu$):
\begin{subequations}
\label{eq_mut}
\begin{eqnarray}
\mu &=& \beta - \psi \approx tr_\mu\, t + \Delta\mu (t), \label{eq_mu} \\
tr_\mu &\approx& - \tan\delta_{LP} \, tr_x 
\label{eq_mutra} \\
&\approx& - \sin\delta_{LP} \, tr_\alpha, 
\label{eq_mutr} \\
\Delta\mu (t) &\approx& - \, \tan\delta_{LP} \, \Delta s_x - \left(\tan^2\delta_{LP} + \frac{1}{2}\right) \, \Delta s_x \, \Delta s_y, 
\label{eq_musxsy} \\
&\approx& - \, \sin\delta_{LP} \, \Delta\alpha_S - \frac{1}{2}\cos{\delta_{LP}}\, \Delta\alpha_S \, \Delta\delta_S. 
\label{eq_muad}
\end{eqnarray}
\end{subequations}

\subsection{The periodic variations in $W$, including the physical librations}
\label{sec_defphi}

Librations in longitude are defined as the periodic variations of the satellite's rotation around its spin axis. We have just defined two different rotation angles ($\phi_{Euler}$ and $W$), whose periodic variations $\Delta\phi_{Euler}$ and $\Delta W$ are different from each other, since $\beta$ varies periodically with time. Clearly, if the definition of the rotation angle is not unique, neither is that of libration. 

To resolve for the rotation variations with respect to synchronicity with the orbital motion, the gravitational torque in libration models depends on the difference between a rotation angle and the orbital true longitude (see Section~\ref{sec_lib}), with both angles measured from the same origin. The choice of a fixed/inertial origin is necessary to correctly describe the torque and find a realistic libration motion.
The most elementary libration model considers that the satellite evolves on a non-precessing eccentric Keplerian orbit and that it undergoes diurnal libration such that the satellite's long axis points towards the central body at pericenter (e.g.~\citealt{VanH08}).
For modeling purposes only, this Keplerian orbit can therefore be used both as the reference plane (instead of the Laplace plane) and as a proxy for the satellite’s equatorial plane.
In this context, it makes sense to measure the angle of rotation on the plane of the orbit, starting from the pericenter. We note that rotation angle $\phi_{pericenter}$. It is also customary to refer the satellite's orbit not to its Laplace plane, but to Jupiter's equatorial plane, since in this modeling context, the Laplace plane is not of paramount importance. The geometric relationship between $\phi_{Euler}$ and $\phi_{pericenter}$, at first order in small angles $i$, $\theta$ and $\varepsilon$ is
\begin{eqnarray}
\Omega_{IQ\,IL} + \psi + \phi_{Euler} 
& = & \Omega_{JupEq} + \omega_{JupEq} + \pi + \phi_{pericenter}, 
\label{eq_phiInertial} 
\end{eqnarray}
where both the left and right hand sides are the broken angle starting at the point $IQ$ (the intersection of Jupiter equator on the ICRF plane) in Fig.~\ref{fig_ang} and ending at the meridian plane, whereas the $\phi$ angles start at different points. $\Omega_{JupEq}$ is the node longitude of the orbit starting from the intersection between the ICRF equator and Jupiter Equator at J2000 and $\omega_{JupEq}$ is the argument of the satellite pericenter with respect to Jupiter Equator.
$\omega_{JupEq} + \pi$ is the argument of the pericenter of the planet around the satellite.
The $\Omega_{IQ\,IL}$ angle is constant, since it is defined by three inertial planes. For a Keplerian non-precessing orbit, $\psi$, $\Omega_{JupEq}$ and $\omega_{JupEq}$ can be considered constant as well, so that $\Delta\phi_{Euler} = \Delta\phi_{pericenter}$ and is essentially diurnal.

However, the orbit pericenter is not a fixed point in space. The angle $\omega_{JupEq}$ is linearly and periodically perturbed. The modeler interested in a detailed rotation theory where librations also occur at periods other than the diurnal period needs to refer the rotation angle to another point. The node of the orbital plane is not fixed in space ($\Omega_{JupEq}$ is also linearly and periodically perturbed), and is not a more convincing starting point. The point $IQ$ is a suitable choice, so that the left-hand side in Eq.~(\ref{eq_phiInertial}) can be defined as the rotational angle $\phi_{Inertial}$:
\begin{equation}
\phi_{Inertial} = \Omega_{IQ\,IL} + \psi + \phi_{Euler}. 
\label{eq_phiInertial2}  
\end{equation}
The points $IL$ and $QL$ could also be suitable, since the triangle defined by the points $IQ$, $QL$, and $IL$ is fixed. Depending on the actual starting point ($IQ$, $IL$ or $QL$), $\phi_{Inertial}$ is defined to within one constant, just as $\Omega_{JupEq}$.

In the most elementary libration model (e.g.~\citealt{VanH08}), the rotation angle is expressed as the sum of the mean anomaly and the small libration angle:
\begin{equation}
\phi_{pericenter} = M(t) + \gamma = M_0 + n_M \, t + \gamma \qquad \textrm{[for a Keplerian orbit]},
\label{eq_phiper}
\end{equation}
where $M(t)$ only varies linearly at the rate $n_M$ and $\gamma$ is proportional to the eccentricity $e$ and $\sin M$, so that the satellite's major axis always points approximately towards the orbit's empty focus, and the angle of rotation is the sum of an epoch value, of a linear trend representative of synchronous rotation, and of a single periodic term. For more advanced modeling, a similar formulation should be used, where $M(t)$ is replaced by the mean mean longitude $L$ or by the mean longitude $\mathcal{L}$ (following here the notations in \citealt{Lai06}), both measured along Jupiter's equator:
\begin{subequations}
\label{eq_phiInertial34}
\begin{eqnarray}
\phi_{Inertial} &=& L(t) + \pi + \gamma_u, 
\label{eq_phiInertial3} \\
\phi_{Inertial} &=& \mathcal{L}(t) + \pi + \gamma_f,
\label{eq_phiInertial4} 
\end{eqnarray}
\end{subequations}
with
\begin{subequations}
\begin{eqnarray}
L(t) &=& L_0 + n_L \, t,  \label{eq_Lreg} \\
\mathcal{L}(t) &=& \Omega_{JupEq}(t) + \omega_{JupEq}(t) + M(t)= L (t) + \Delta\mathcal{L}, \label{eq_Lt} \\
\Delta\mathcal{L} &=& \Delta\Omega_{JupEq} + \Delta\omega_{JupEq} + \Delta M.
\label{eq_Lt2}
\end{eqnarray}
\end{subequations}
The origin of $L$ and $\mathcal{L}$ is
the intersection $IQ$ between the ICRF equator and Jupiter Equator at J2000.
By definition, $L$ varies only linearly, whereas $\Delta\mathcal{L}$ also includes the periodic variations in $\Omega_{JupEq}$, $\omega_{JupEq}$, and $M$. The mean motion $n_L$ of the mean longitude is the sum of the mean rates of the node longitude, of the pericenter argument, and of the mean anomaly, and differs from the mean motion of the mean anomaly $n_M$:
\begin{equation}
n_L = \dot\Omega_{JupEq} + \dot\omega_{JupEq} + n_M.
\label{eq_nL}
\end{equation}
If the orbit is nearly circular and has a low inclination (this is the case for the Galilean satellites), these trends may be difficult to fit separately.

Eqs.~(\ref{eq_phiInertial34}) define two different small libration angles ($\gamma_u$ and $\gamma_f$). $\gamma_u$ defines the rotation increment with respect to a uniform rotation at rate $n_L$, whereas $\gamma_f$ defines the increment with respect to a rotation that follows the forcing and is therefore in synchronicity with the mean longitude. $\gamma_f$ has an obvious dynamic meaning, but $\gamma_u$ might be more practical for describing observations. The connection between these two libration angles lies in the periodic variations of the mean longitude:
\begin{equation}
\gamma_f = \gamma_u - \Delta\mathcal{L}. 
\label{eq_gammaf}
\end{equation}
For a strictly Keplerian orbit, the two libration angles are equivalent: $\gamma_f = \gamma_u$.

Observations of a satellite's rotation can also be expressed not in terms of $\phi_{Euler}$ or $\phi_{Inertial}$, but in terms of the Earth equatorial angle $W$.
Making use of Eqs.~(\ref{eq_W}), (\ref{eq_mu}), (\ref{eq_phiInertial2}) and (\ref{eq_phiInertial34}), we have that 
\begin{subequations}
\label{eq_Wlong}
\begin{eqnarray}
W &=& \mu + \pi-\Omega_{IQ\,IL} + L + \gamma_u,   
\label{eq_W2} \\
  &=& \mu + \pi -\Omega_{IQ\,IL} + \mathcal{L} +\gamma_f. 
\label{eq_W3} 
\end{eqnarray}
\end{subequations}
The rate of $W$ is
\begin{equation}
\dot W = n_L + tr_\mu \approx n_L - \tan\delta_{LP} \, tr_x
\label{eq_Wrate}
\end{equation}
and the periodic part of $W$ is given by:
\begin{subequations}
\label{eq_DWgamma}
\begin{eqnarray}
\Delta W &=& \Delta\mu + \gamma_u,   \label{eq_DWgammau}\\
&=& \Delta\mu + \Delta\mathcal{L} + \gamma_f,   \label{eq_DWgammaf}
\end{eqnarray}
\end{subequations}
with $\Delta\mu$ the periodic part of $\mu = \beta-\psi$, see Eq.~(\ref{eq_mu}). 

With the latter expressions, the meaning of $\mu$ becomes clearer. If $\beta$ represents the transformation between $W$ and $\phi_{Euler}$, then $\mu$ represents, to within one additive constant, the transformation between $W$ and $L + \gamma_u$ (or equivalently between $W$ and $\mathcal{L} +\gamma_f$), or more simply, between $W$ and $\phi_{inertial}$:
\begin{equation}
W = \phi_{Inertial} + \mu - \Omega_{IQ\,IL}.
\label{eq_Wphiinert}
\end{equation}

\subsection{Explicit expressions for \texorpdfstring{$\phi_{Euler}$}{phiEuler}}
\label{sec_phiEuler}

We assume here that measuring the orientation and rotation of a satellite involves either the set of quantities ($\Delta s_x, \Delta s_y, \gamma_{f/u}$), or ($\Delta\alpha_S, \Delta\delta_S, \Delta W$), depending on whether the observer is using a representation close to Euler angles or terrestrial equatorial coordinates. The angle $\phi_{Euler}$, unlike $\phi_{Inertial}$ (and therefore $\gamma_{u/f}$), would be impractical as an observable, as it cannot be observed without a concomitant determination of $\psi$. However, $\phi_{Euler}$ is required to express the rotation matrix (Eq.~\ref{eq_Mep}) in terms of Euler angles.

From a measurement of $\gamma_{u/f}$ and the expression of $\phi_{Inertial}$, see Eq.~(\ref{eq_phiInertial34}), we can simply obtain the expression of $\phi_{Euler}$ from Eq.~(\ref{eq_phiInertial2}) as follows
\begin{subequations}
\begin{eqnarray}
    \phi_{Euler}&=& L(t) + \pi + \gamma_u-\Omega_{IQ\,IL} - \psi, \\
    \phi_{Euler}&=&\mathcal{L}(t) + \pi + \gamma_f-\Omega_{IQ\,IL} - \psi,   
\end{eqnarray}
\end{subequations}
where $\psi$ is obtained from measurements of $\Delta s_x$ and $\Delta s_y$ and Eq.~(\ref{eq_psisxsy}).

From a measurement of $\Delta W$ and Eq.~(\ref{eq_DWgamma}), it is possible to obtain the expression for $\gamma_{u/f}$ if the expression for $\Delta\mu$ is first derived from the measured ($\Delta\alpha_S, \Delta\delta_S$) with Eq.~(\ref{eq_muad}):
\begin{subequations}
\begin{eqnarray}
\gamma_{u} &=& \Delta W - \Delta\mu, \\
\gamma_{f} &=& \Delta W - \Delta\mu - \Delta\mathcal{L}.
\end{eqnarray}
\end{subequations}
$\phi_{Euler}$ is then obtained as described in the previous paragraph. In a more direct way, without going through $\gamma_{u/f}$ but calculating $\beta$ with Eq.~(\ref{eq_beta4}) from the measured ($\Delta\alpha_S, \Delta\delta_S$), we can use 
\begin{equation}
  \phi_{Euler} = W - \beta
\end{equation}
to obtain $\phi_{Euler}$ from the measured $W$.

The rate of $\phi_{Euler}$ is
\begin{equation}
\dot \phi_{Euler} = n_L - \dot\psi.
\label{eq_Phirate}
\end{equation}

\section{Dynamical models}
\label{sec_dyn_mod}

The Galilean satellites are assumed to be locked in the Cassini state, with their rotation variations described as a combination of variations in the rotation angle (librations, as defined in Section~\ref{sec_defphi}), variations in the orientation of the spin axis in space (nutations, as defined in Section~\ref{sec_app}), and variations in the orientation of the spin axis relative to the Body Frame (polar motion). These rotation variations are caused by the external gravitational torque exerted by Jupiter on the flattened shape of the satellites. 
In this section, we identify the relevant dynamical models used to construct Euler angle solutions, which will then be transformed into equatorial coordinates in Section~\ref{sec_admseries}. Since we provide a transformation method for series of rotation variations, we are only interested here in rotation models whose solution is expressed in series. 
Note that although the models described here apply essentially to all synchronous natural satellites, in the case of Titan, in addition to the external torque exerted by Saturn, angular momentum exchanges between the atmosphere and lakes and the solid surface also contribute to the rotation variations (the resulting variations in rotation around the spin axis are then called length-of-day variations). 

Series for rotational variations can be modeled within the angular momentum (AM) framework, by solving equations that relate the rate of change of the rotation AM to the external torques applied. The AM equations can be solved numerically, before a frequency analysis is applied to the solution (e.g.~\citealt{Ram11}). 
Fitted series for the rotational variations can also be obtained within an Hamiltonian framework (e.g.~\citealt{Noy09}). Another way of approaching the problem is to solve the AM equations analytically, using a semi-analytical orbital theory to express the forcing, so that the solution is directly expressed in series (e.g.~\citealt{Bil05, Bal12, Yse14}). The latter approach requires a number of working assumptions, particularly with regard to the quantities and terms that can be neglected or not. Such a series obtained from an analytical solution will generally contain fewer terms than a series fitted from a numerical solution. Nevertheless, by extending an analytical solution to a sufficient order, it should be possible to recover a sufficient number of terms and tend towards the numerical solution. For example, \cite{Coy26} recently extended an analytical AM theory to the second order in small quantities and recovered rotational variation terms that had previously only been identified from a frequency analysis. Here we will provide simplified dynamical models (analytical AM models developed to the first order in small quantities) identified in the literature. Providing a very detailed theory of the rotation of Galilean satellites is outside the scope of the present study and is not necessary to assess the accuracy of the transformations defined in Sections~\ref{sec_transfo} and \ref{sec_rot}.

Note that the transformation between $(s_x,s_y)$ and $(\alpha_S, \delta_S)$ requires only knowledge of the orientation of the Laplace plane, and is therefore independent of the details of the internal structure. On the other hand, the transformation between $W$ and $\phi_{Euler}$ (or $L + \gamma_u = \mathcal{L} + \gamma_f$), namely $\beta$ ($\mu$), requires knowledge of the orientation of both the Laplace plane and the spin axis, and therefore depends on the internal structure. In the absence of spin orientation measurements, this transformation is dependent on the theoretical model chosen to simulate the spin axis precession. 
Finally, note that neither the transformation between $(s_x, s_y)$ and $(\alpha_S, \delta_S)$ nor that between $W$ and $\phi_{Euler}$ depends on the librations $\gamma_{u/f}$.
We will illustrate the transformation and provide an example of numerical series for the spin orientation model ($\alpha_S$, $\delta_S$ and $W$) in the solid case, see Section~\ref{sec_numericalvalues}. 
The equations for the libration dynamical model are provided here for completeness and illustration purposes. Our goal in this study is not to solve them, but to link the different libration angles definitions that exist in the literature (see Section~\ref{sec_defphi}) and to show corresponding appropriate dynamical equations.

\subsection{Spin orientation model}
\label{sec_spin}

For a fully rigid triaxial satellite, we will use the model of \cite{Bal11}, where the variations of $\uvec{s}$ are described in an inertial reference frame linked to the Laplace plane:
\begin{subequations}
\label{eq_diff}
\begin{eqnarray}
n \, C \, \frac{d \uvec{s}}{d t} &=& n \, \kappa \, (\uvec{s} \wedge \uvec{n}), \\
\kappa &=& \frac{3}{2} M R^2\left(-C_{20} + 2 \, C_{22}\right) n = \frac{3}{2}(C-A) \, n.
\end{eqnarray}
\end{subequations}
This model is developed to first order in inclination and obliquity and assumes that spin precession in space is decoupled from librations and polar motion. As stated in Section~\ref{sec_proj}, $\uvec{s}$ is the unit vector along the spin axis and $\uvec{n}$ is the unit vector along the orbit normal. $n$ is the mean motion while the mass and the radius of the satellite are $M$ and $R$. 
$C_{20}$ and $C_{22}$ are the second-degree gravity field coefficients of the satellite, which can be expressed in terms of $A < B < C$, the principal moments of inertia. 
If the satellite's orbit is precessing uniformly at the rate $\dot\Omega$ and with inclination $i$, the orbital obliquity is obtained as
\begin{equation}
\varepsilon = -\frac{i\, \dot\Omega}{\dot \Omega+\kappa/C}, \label{eq_epsiloni}
\end{equation}
an expression that is consistent with the majority of the literature (e.g.~\citealt{Noy10, Hen04, Bou20, Coy26}). \cite{Bil05} and \cite{Bil11} obtain different solutions, due to their incorrect definitions of the coupling constant $\kappa$. The exact expression of Eq.~(\ref{eq_diff}) has been demonstrated in Appendix A of \cite{Bal12}.

Interestingly, Eq.~(\ref{eq_diff}) can be transformed into a couple of differential equations for the right ascension $\alpha_S$ and declination $\delta_S$ of the rotation axis with respect to the ICRF. With the ICRF coordinates ($\alpha_S, \delta_S$) and ($\alpha_O, \delta_O$), the spin and orbit unit vectors are expressed in the coordinates of the ICRF as
\begin{subequations}
\label{eq_snunit}
\begin{eqnarray}
\uvec{s} &=& (\cos\alpha_S \cos\delta_S, \cos\delta_S \sin\alpha_S, \sin\delta_S),\\
\uvec{n} &=& (\cos\alpha_O \cos\delta_O, \cos\delta_O \sin\alpha_O, \sin\delta_O),
\end{eqnarray}
\end{subequations}
so that
\begin{subequations}
\label{eq_dadd}
\begin{eqnarray}
\frac{d \alpha_S}{dt} & = & \frac{ \kappa }{C} \left(\cos\delta_O \tan\delta_S \cos(\alpha_O - \alpha_S)   - \sin\delta_O \right),  \\
\frac{d \delta_S}{dt} & = & \frac{ \kappa}{C} \cos\delta_O \sin(\alpha_O - \alpha_S).
\end{eqnarray}
\end{subequations}
In the limit of small orbital obliquity and neglecting tidal dissipation, Eqs.~(\ref{eq_dadd}) correct Eqs.~(E4-E5) of \cite{Bil22} as obtained from their Eq.~(4.8) (with $\uvec{s}.\uvec{n}\sim 1 $ and $p_3=0$).
Eqs.~(\ref{eq_dadd}) can be integrated numerically to provide a benchmark for assessing the accuracy of the transformation presented in section \ref{sec_transfo}.

For the case where the satellites harbor an internal global liquid layer, as will be considered in Section \ref{sec_liquid}, we will use the solution presented in Appendix 5 of \cite{Bal19}, which extends the model of \cite{Bal11, Bal12} by including the hydrodynamic pressure (Poincaré flow) at the interfaces between the ocean and the solid layers. That model assumes that the solid layers behave rigidly.

\subsection{Libration model}
\label{sec_lib}

The dynamical equation for solving librations can be obtained as the third component of the AM equation expressed in the rotating BF:
\begin{equation}
\frac{\partial \bvec{L}}{\partial t}+\bvec{$\Omega$} \wedge \bvec{L} = \bvec{N}, 
\label{eq_AMLib}
\end{equation}
where $\bvec{$\Omega$}=(\omega_x, \omega_y, \omega_z)$ is the rotation vector, $\bvec{N}=(N_x,N_y,N_z)$ is the total torque exerted on the satellite, and $\bvec{L}$ is the angular momentum. For a fully rigid triaxial satellite, $L_z = C \, \omega_z$ with $\omega_z\simeq \psi' + \phi'_{Euler} = \phi'_{Inertial}$ at first order in $\theta$ and the libration equation reads (see also Eqs.~(5.48) of \citealt{Mur00}):
\begin{equation}
 C \, \omega'_z - (A-B) \, \omega_x \, \omega_y = N_z .
 \label{eq_Nz}
\end{equation}
The $'$ notation is the time derivative.
In a simple first order theory, the product of the polar motion components $\omega_x$ and $\omega_y$ can be neglected and $N_z$ approximated at first order in small quantities, so that
\begin{equation}
C \, \phi''_{Inertial} = - \frac{3}{2} (B-A) \, \frac{G M_J}{d^3} \sin2\left(\phi_{Inertial}-l_{JupEq}\right), \label{eq_Lib}
\end{equation} 
with $d$ the distance between the satellite and the parent planet and $l_{JupEq} = \Omega_{JupEq} + \omega_{JupEq} + f$ the true longitude of the satellite. $M_J$ is the mass of Jupiter and $G$ is the universal gravitational constant. The difference $(\phi_{Inertial}-l_{JupEq})$ is the angle from the direction of the satellite long axis (corresponding to the principal moment of inertia $A$) to the direction to the central planet. Since the sine argument is a small quantity, it can replace the sine factor.

For a Keplerian orbit, with $GM_J=n^2a^3$, with $n$ and $a$ the mean motion and semi-major axis, respectively, Eq.~(\ref{eq_Lib}) simplifies to
\begin{equation}
 C \, \phi''_{pericenter} = - \frac{3}{2}n^2 (B-A) \, \left(\frac{a}{d}\right)^3 \sin2\left(\phi_{pericenter} - f \right) \qquad \textrm{[For a Keplerian orbit]},
\label{eq_LibKep}
\end{equation} 
with $f$ the true anomaly, see e.g.~Eq.~(1) of \cite{VanH08}. The right-hand side of the equation can first be expressed as series in eccentricity $e$ before solving the dynamical equation. The solution consists of a main diurnal libration and smaller terms at harmonics of the orbital period. 

Orbital perturbations lead to changes in the gravitational torque of Jupiter on the satellite compared to the torque in the Keplerian case, see for example \cite{Yse14}. Based on any orbital theory, the factor of $\sin2(\phi_{Inertial}-l_{JupEq})/d^3$ can first be expressed as series before solving the dynamical equation. Depending on the chosen definition for the libration angle (with respect to the forcing $\gamma_f$ or with respect to the uniform rotation $\gamma_u$), Eq.~(\ref{eq_Lib}) can take two different forms:
\begin{subequations}
\label{eq_LibOE}
\begin{eqnarray}
C \, (\gamma''_f + \mathcal{L}'') &=& - \frac{3}{2} (B-A) \, \frac{G M_J}{d^3} \sin (2 \, \gamma_f + 2 \, M - 2 \, f), \\
C \, \gamma''_u &=& - \frac{3}{2} (B-A) \, \frac{G M_J}{d^3} \sin (2 \, \gamma_u + 2 \, L - 2 \, l_{JupEq}). 
\end{eqnarray}
\end{subequations}
In addition to the diurnal (+ harmonics) libration, other libration terms arise due to the orbital perturbations.

\section{Spin orientation series and transformation accuracy}
\label{sec_numericalvalues} 

In this section, we assess the accuracy of the transformations defined in Sections~\ref{sec_transfo} and \ref{sec_rot} for the orientation and rotation angles of the Galilean satellites. As we pointed out at the start of Section 5, the coefficients for the transformation of the orientation angles depend only on the orientation of the Laplace plane (through $\alpha_{LP}$ and $\delta_{LP}$), whereas the $\beta$ and $\mu$ angles for the transformation of the rotation angles also depend on the orientation of the spin axis ($\alpha_S$, $\delta_S$), the latter being dependent on the internal structure of the satellites. We first build series for the spin axis coordinates ($s_x,s_y$) for satellites assumed to be entirely solid and rigid. This is an update of the corresponding series of \cite{Bal12} using more recent orbital ephemerides. Then we build the corresponding series in orientation and rotation angles, both for Euler ($\theta, \psi, \varepsilon, \zeta$) or Earth equatorial ($\alpha_S, \delta_S$) angles. The effect of an internal ocean on the orientation angles, and consequently on the rotation angles' transformation $\mu$, is discussed in Section~\ref{sec_liquid}.

\subsection{Spin orientation series for solid Galilean satellites, in \texorpdfstring{$(s_x,s_y)$}{sx, sy}}
\label{sec_sxsyseries}

Following the decomposition described in Section~\ref{sec_orbit}, the orbit normal coordinates ($n_x,n_y$) are written as the sum of a linear term (with the $tr_x$ and $tr_y$ rates, 
second order terms by construction) and of a trigonometric series (noted $\Delta$, and of first order by construction, see Section~\ref{sec_app}):
\begin{subequations}
\label{eq_nxnyt}
\begin{eqnarray}
n_x(t) &\approx& tr_x \, t + \Delta n_x = tr_x \, t + \; \sum_j i_j \, \sin(f_j \, t + \varphi_j), \\ 
n_y(t) &\approx& tr_y \, t + \Delta n_y = tr_y \, t \, - \sum_j i_j \, \cos(f_j \, t + \varphi_j),
\end{eqnarray}
\end{subequations}
where $i_j$ are the inclination amplitudes associated with the orbital node precession frequencies $f_j$ and the phases $\varphi_j$, and $t$ is the time.
For each inclination amplitude $i_j$ in the forcing, there is a corresponding inertial obliquity amplitude
\begin{equation}
\theta_j = i_j + \varepsilon_j  
\label{eq_thetaj}
\end{equation} 
in the satellite response, calculated using Eq.~(\ref{eq_epsiloni}) and the value of the parameters listed in Table~\ref{tab_MOI}, so that the spin axis coordinates ($s_x, s_y$) are written as follows (see also Eq.~\ref{eq_sxform}):
\begin{subequations}
\label{eq_sxsy}
\begin{eqnarray}
s_x(t) &=& tr_x \, t + \Delta s_x = tr_x \, t + \sum_j \theta_j \sin(f_j \, t + \varphi_j), \\
s_y(t) &=& tr_y \, t + \Delta s_y = tr_y \, t - \sum_j \theta_j \cos(f_j \, t + \varphi_j). 
\end{eqnarray}
\end{subequations}
The linear terms, due to their very long period, are left unchanged by the external gravitational torque, and are therefore identical for $n_x$ and $s_x$ and for $n_y$ and $s_y$.

\begin{table}[!htb]
\small
\begin{tabular}{cccccccc}
\hline
& $M$ ($10^{23}$ kg) & $R$ (km)& $n_L$ (rad/d) & $A/M R^2$ & $C/M R^2$ & $2\pi/\omega_f$ (y) & surface shift  (km/$^\circ$) \\
\hline
Io       & $0.892965$ & $1821.49$ & $3.551552313$ & $0.376425$ & $0.379538$ &  0.3937 & 31.791 \\ 
Europa   & $0.479857$ & $1560.8$  & $1.769322718$ & $0.354277$ & $0.354992$ &  3.2189 & 27.241 \\ 
Ganymede & $1.481478$ & $2631.2$  & $0.878207920$ & $0.311381$ & $0.311585$ & 19.9407 & 45.923 \\ 
Callisto & $1.075661$ & $2410.3$  & $0.376486233$ & $0.354869$ & $0.354922$ & 203.605 & 42.068 \\ 
\hline
\end{tabular}
\caption{Values for the mass ($M$), radius ($R$), mean motion ($n_L$), normalized principal moments of inertia $A/MR^2$ and $C/MR^2$, and free precession period ($2\pi/\omega_f$) of the four Galilean satellites. 
The moments of inertia $A$ and $C$ are derived from the observed $J_2$ and $C_{22}$ gravitational coefficients and from the mean moment of inertia $I$ (see \citet{And01} for Io, \citet{Gom20} for Europa and \citet{Sch04} for Ganymede and Callisto). 
The $n_L$ values are computed from JUP387 and differ from those in \cite{Lai06} by less than $3 \times 10^{-8}$ rad/d.
The last column evaluates the amplitude at the satellite surface of an arbitrary angular displacement of $1^\circ$.
\label{tab_MOI}}
\end{table}

\begin{table}[!htb]
\small
\centering
\begin{tabular}{ccccccc}
\hline
 & $tr_x$ ($^\circ$/y) & $tr_y = tr_\delta$ ($^\circ$/y) & $tr_\alpha$ ($^\circ$/y)  & $tr_\mu$ ($^\circ$/y) & $\dot W$ ($^\circ$/d) & $\delta_{S \, 0}$ ($^\circ$) \\
\hline
Io       & $-2.8 \times 10^{-5}$ & $2.4 \times 10^{-5}$ & $-6.5 \times 10^{-5}$ & $5.9 \times 10^{-5}$ & $203.488958424$ 
& $-1.4 \times 10^{-5}$ \\
Europa   & $-2.8 \times 10^{-5}$ & $2.4 \times 10^{-5}$ & $-6.4 \times 10^{-5}$ & $5.9 \times 10^{-5}$ & $101.374724491$ 
& $-2.49 \times 10^{-3}$ \\
Ganymede & $-2.8 \times 10^{-5}$ & $2.5 \times 10^{-5}$ & $-6.5 \times 10^{-5}$ & $5.9 \times 10^{-5}$ & $ 50.317607524$ 
& $-4.7 \times 10^{-4}$ \\
Callisto & $-2.8 \times 10^{-5}$ & $2.7 \times 10^{-5}$ & $-6.5 \times 10^{-5}$ & $5.9 \times 10^{-5}$ & $ 21.571072373$ 
& $-1.47 \times 10^{-3}$ \\
\hline
\end{tabular} 
\caption{Trends in the various quantities considered. The $tr_x$ and $tr_y$ trends, obtained from the JUP387 orbital theory, apply to both the orbit normal and spin axis coordinates $n/s_x$ and $n/s_y$, respectively. The $tr_\alpha$, $tr_\delta$ and $tr_\mu$ trends are obtained after the transformation described in Eqs.~(\ref{eq_trends}) and (\ref{eq_mutra}). The rotation rate in $W$ is obtained from Eq.~(\ref{eq_Wrate}).
The last column shows the small increment in $\delta_S$ due to the power reduction of a squared sine term in the second-order part of Eq.~(\ref{eq_deltat}). The number of digits retained in this table is related to the order of magnitude of the residuals from the quasi-periodic decomposition of JUP387, see Fig.~\ref{fig_Residuals}.}
\label{tab_trxtry}
\end{table}

\begin{figure}[htb!]
\centering
\includegraphics[width=10.5cm]{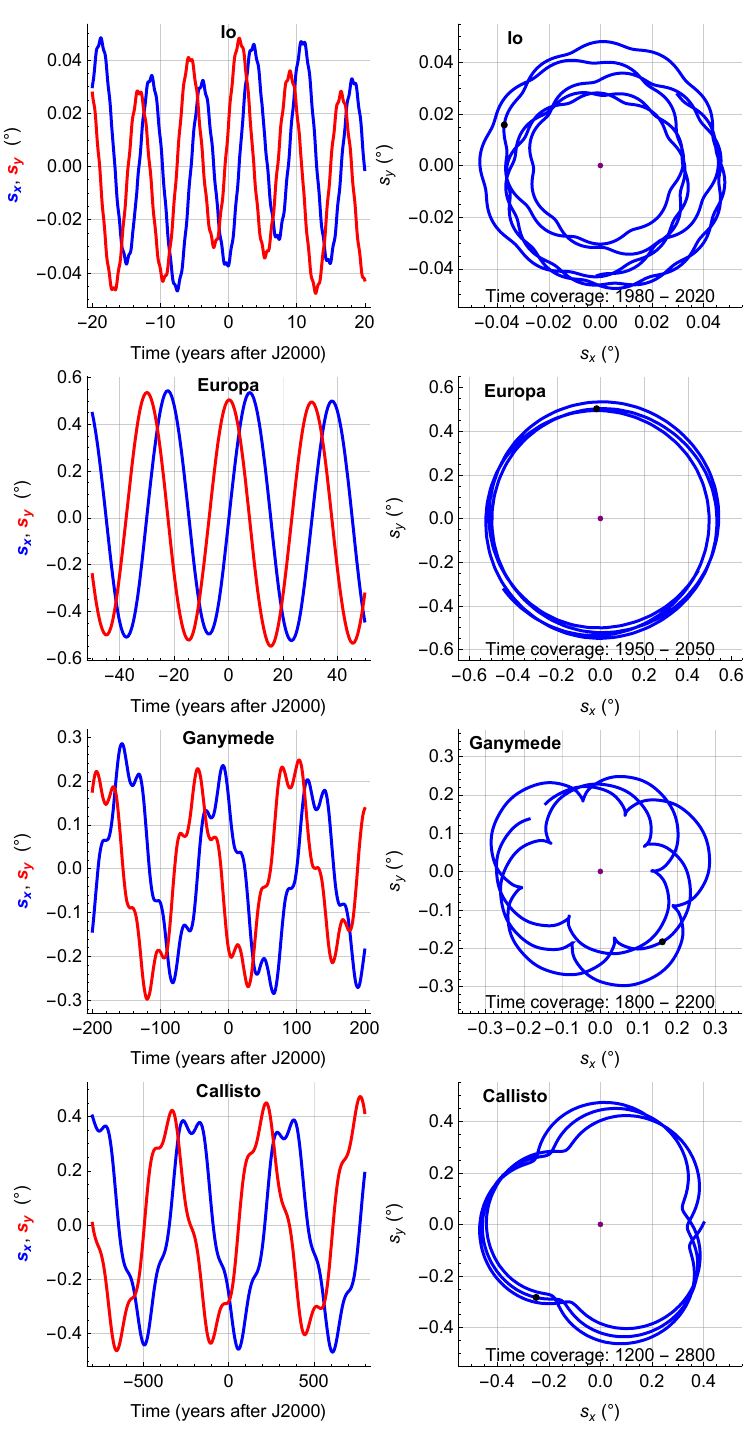}
\caption{Temporal evolution of the spin axis coordinates in the Laplace plane $s_x$ and $s_y$ for the four Galilean satellites for a solid satellite. The time coverage is different for each satellite.
The purple point is the Laplace pole, for which by definition $s_x=s_y=0$. The black dots in the right graphs are the J2000 spin positions. 
We use here reconstructed series, which explains why we can extend our plot beyond the time interval covered by the orbital theory. }
\label{fig_sxsy}
\end{figure}

The values of the trends $tr_x$ and $tr_y$ are given in Table~\ref{tab_trxtry}, whereas the values of the amplitudes $i_j$, $\varepsilon_j$ and $\theta_j$, frequencies $f_j$ and phases $\varphi_j$ are given in Table~\ref{tab_sxsy}, 
for the four Galilean satellites, considering a solid interior model and the JUP387 orbital theory. The corresponding temporal evolution of the spin axis coordinates ($s_x, s_y$) in the Laplace plane is shown in Fig.~\ref{fig_sxsy}. 

In Table~\ref{tab_sxsy}, we have considered a truncation threshold of the order of $0.0001^\circ$ on the $i_j$ terms, corresponding to the order of magnitude of the residuals in the orbit orientation angles between JUP387 and its frequency decomposition. We thus obtain $6$, $10$, $9$, and $8$ terms for Io, Europa, Ganymede, and Callisto, respectively. The arguments (frequencies and phases) of the series result from the orbital dynamics and are combinations of the fundamental arguments associated with the Jovian system (see last columns of Table~\ref{tab_sxsy}). However, arguments identified as being the same combination may have a slightly different value for each satellite, due to the fit to a numerical integration. 
The precision on the $i_j$, $\varepsilon_j$ and $\theta_j$ as given in Table~\ref{tab_sxsy} is $0.0001^\circ$ ($0.00001$ rad/y and $0.001^\circ$ in frequency and phase, respectively).
The typical precision of the time-series for $s_{x/y}(t)$ reconstructed from Table~\ref{tab_sxsy} over the $1900-2100$ interval is approximately $0.0003^\circ$ (see Table~\ref{tab_precision}). This corresponds to a relative truncation error below $1\%$. Neglecting terms related to the secular trends $tr_{x/y}$ would lead to typical errors of about $0.003^\circ$ after $100$ years in $s_{x/y}(t)$, greater than the truncation error.

\begin{table}[!htb]
\small
\centering
\begin{tabular}{ccccccccccc}
\hline
 & $i_j$ ($^\circ$) & $\varepsilon_j$ ($^\circ$) & $\theta_j$ ($^\circ$) & period (y) & $f_j$ (rad/y) & $\varphi_j$ ($^\circ$) & Argument & $\dot{J}_{IAU}$ \\
\hline
{\bf Io}   &&&&&&&& \\
 1 & 0.0358 & 0.0020 & 0.0378 & -7.42164 & -0.84660 & 260.604  & $\Omega_1$& $-\dot{J}_3$ \\
 2 & 0.0099 & 0.0001 & 0.0100 & -30.2008 & -0.20805 & 184.095 & $\Omega_2$& $-\dot{J}_4$ \\
 3 & 0.0013 & 0.0000 & 0.0013 &  -137.325 & -0.04575 & 59.788 & $\Omega_3$& $-\dot{J}_5$ \\
 4 & 0.0004 & 0.0006 & 0.0010 & -0.68144 & -9.22041 & 273.914 & $-2\nu-\Omega_2$ & \\
 5 & -0.0003 & 0.0000 & -0.0003 &  -564.38 & -0.01113 & 310.084 & $\Omega_4$ & $-\dot{J}_6$ \\
 6 & 0.0003 & 0.0000 & 0.0003 & 5.93122 & 1.05934 & 113.504  & $2L_S$ & $\dot{J}_8$ \\
\multicolumn{2}{l}{{\bf Europa}}  &&&&&&& \\
 1 & 0.4651 & 0.0555 & 0.5206 & -30.2008 & -0.20805 & 184.073 &$\Omega_2$ &$-\dot{J}_4$ \\
 2 & 0.0250 & 0.0006 & 0.0256 & -137.328 & -0.04575 & 59.767 & $\Omega_3$ &$-\dot{J}_5$ \\
 3 &  0.0056 &  0.0000 &  0.0056 & -560.607 & -0.01121 & 310.271 & $\Omega_4$ & $-\dot{J}_6$ \\
 4 & -0.0012 & -0.0009 & -0.0021 &   -7.42165 & -0.84660 & 260.589 &$\Omega_1$ &$-\dot{J}_3$ \\
 5 & -0.0012 &  0.0015 &  0.0003 &   -0.68144 & -9.22041 & 273.897 & $-2\nu-\Omega_2$ & \\
 6 &  0.0008 & -0.0003 &  0.0005 &    5.9312 & 1.05934 & 113.500 & $2L_S$ & $\dot{J}_8$ \\
 7 &  0.0002 &  0.0001 &  0.0003 &  -11.8638 & -0.52961 & 299.309 & $-L_S$ & $-\dot{J}_8/2$ \\
 8 &  0.0002 & -0.0002 &  0.0000 &   -0.669656 & -9.38270 & 38.1641 & $-2\nu-\Omega_3$ & \\
 9 &  0.0001 &  0.0000 &  0.0001 &   11.8639 & 0.52961 & 178.045 & $L_S$ & $\dot{J}_8/2$ \\
10 & -0.0001 &  0.0000 & -0.0001 &    4.95757 & 1.26739 & 68.1319 & $2L_S-\Omega_2+\Omega_0$& \\
\multicolumn{2}{l}{{\bf Ganymede}} &&&&&&& \\
 1 & 0.1860 & 0.0316 & 0.2176 & -137.328 & -0.04575 & 59.659 & $\Omega_3$&$-\dot{J}_5$ \\
 2 & 0.0382 & 0.0014 & 0.0396 & -560.839 & -0.01120 & 310.157 & $\Omega_4$&$-\dot{J}_6$ \\
 3 & -0.0165 & -0.0321 & -0.0486 & -30.2009 & -0.20805 & 183.966 & $\Omega_2$&$-\dot{J}_4$ \\
 4 &  0.0018 & -0.0014 &  0.0004 &   5.93122 & 1.05934 & 113.417 & $2L_S$ &$\dot{J}_8$ \\
 5 &  0.0003 & -0.0007 & -0.0004 & -11.8633 & -0.52963 & 298.979 & $-L_S$ & $-\dot{J}_8/2$ \\
 6 &  0.0002 & -0.0002 &  0.0000 &  -0.68144 & -9.22041 & 273.774 & $-2\nu-\Omega_2$ & \\
 7 &  0.0002 & -0.0002 &  0.0000 &   3.95426 & 1.58896 & 133.013 & $3L_S$ & $3\dot{J}_8/2$ \\
 8 &  0.0002 & -0.0001 &  0.0001 &  11.8641 & 0.52959 & 177.622 & $L_S$ & $\dot{J}_8/2$ \\
 9 & -0.0001 &  0.0001 &  0.0000 &   5.68547 & 1.10513 & 192.004 & $2L_S-2\Omega_3 +\Omega_0$ & \\
\multicolumn{2}{l}{{\bf Callisto}}   &&&&&&& \\
 1 & 0.2507 & 0.1429 & 0.3936 & -560.826 & -0.01120 & 309.669 &$\Omega_4$ & $-\dot{J}_6$ \\
 2 & -0.0302 & 0.0927 & 0.0626 & -137.327 & -0.04575 & 59.182 &$\Omega_3$ &$-\dot{J}_5$ \\
 3 &  0.0038 & -0.0037 & 0.0001 &   5.93112 &  1.05936 & 112.934 & $2L_S$ & $\dot{J}_8$ \\
 4 & -0.0006 &  0.0007 & 0.0001 & -30.2046  & -0.20802 & 183.619 & $\Omega_2$ & $-\dot{J}_4$ \\
 5 &  0.0006 & -0.0006 & 0.0000 & -11.8639  & -0.52960  & 298.653 & $-L_S$ & $-\dot{J}_8/2$ \\
 6 &  0.0005 & -0.0005 & 0.0000 &  11.865   &  0.52956 & 176.965 & $L_S$ & $\dot{J}_8/2$ \\
 7 &  0.0004 & -0.0004 & 0.0000 &   3.95429 &  1.58895 & 132.539 & $3L_S$ & $3\dot{J}_8/2$ \\
 8 & -0.0003 &  0.0003 & 0.0000 &   5.86872 &  1.07062 & 301.732 & $2L_S - \Omega_4 + \Omega_0$ & \\
\hline
\end{tabular} 
\caption{$(\Delta n_x, \Delta n_y)$ and $(\Delta s_x, \Delta s_y)$ series, according to the notations of Eqs.~(\ref{eq_nxnyt}-\ref{eq_sxsy}). Inclination, orbital and inertial obliquity amplitudes $i_j$, $\varepsilon_j$ and $\theta_j$ at the frequency and phase $f_j$ and $\varphi_j$ obtained from the JUP387 orbital theory and assuming that the satellites are entirely solid and rigid. 
These series can be used directly to reconstruct time series for $s_{x/y}(t)$, and indirectly to reconstruct the time series for the angles $\theta(t)$, $\varepsilon(t)$, $\psi(t)$, and $\zeta(t)$ (see Section~\ref{sec_obliquityseries}). 
The phases $\varphi_j$ are given with respect to J2000 epoch. The last two columns show the identification to the fundamental arguments of \cite{Lai06} and the fundamental IAU WG frequencies $\dot{J}_i$. The number of digits retained in this table is related to the order of magnitude of the residuals from the quasi-periodic decomposition of JUP387, see Fig.~\ref{fig_Residuals}.} 
\label{tab_sxsy}
\end{table}

\begin{figure}[htb!]
\centering
\includegraphics[width=12cm]{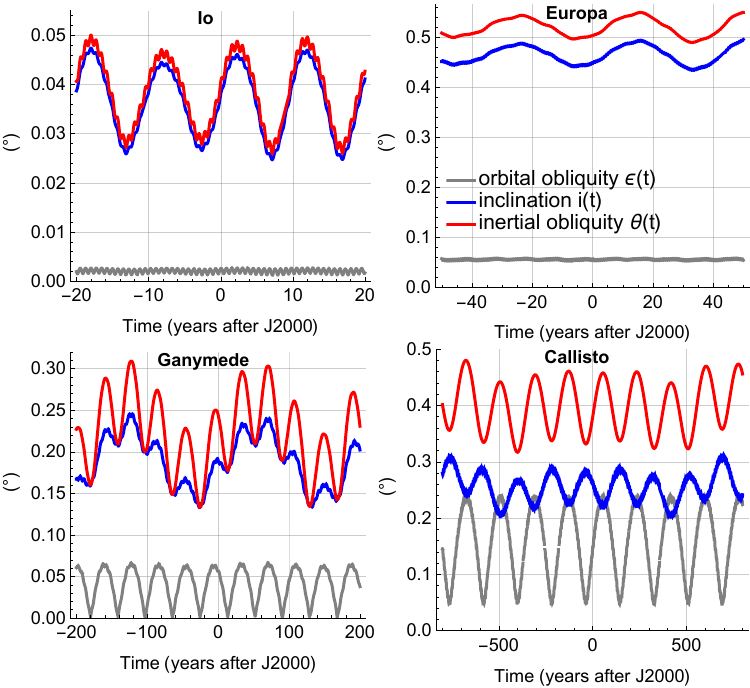}
\caption{Temporal evolution of the orbital ($\varepsilon$) and inertial ($\theta$) obliquities and of the orbital inclination ($i$) for the four Galilean satellites, assuming a solid interior model.}
\label{fig_obli}
\end{figure}

For each satellite, the orbital precession is dominated by the term with $j=1$ whose frequency corresponds to the node longitude proper frequency of that satellite. This may seem obvious, but note that this is not the case for all of the large satellites of Uranus, which strongly perturb each other \citep{Bal25}. The ratio $i_2/i_1$ is of $28\%$, $5\%$, $21\%$, and $12\%$, for Io, Europa, Ganymede, and Callisto by respectively. 
The dominant terms of the spin precession are here the same as the dominant terms of orbital precession, which will not necessarily be the case for satellites with an ocean, see Section \ref{sec_liquid}. The amplitude $\varepsilon_j$ can reach a value such that $\theta_j$ is large or very small, depending on the ratio of $\omega_f/\dot\Omega_j$ with $\omega_f=\kappa/C$ the natural frequency of the free retrograde precession. 
The free precession mode periods are given in Table~\ref{tab_MOI}.
In the limit case $\omega_f/\dot\Omega_j\rightarrow 0$ (relatively short forcing period), $\varepsilon_j$ tends towards $-i_j$ and $\theta_j$ tends towards $0$ (meaning that the spin axis tends to align with the Laplace pole, e.g.~term $j=3$ for Callisto). For $\omega_f/\dot\Omega_j\rightarrow \pm \infty$ (relatively long forcing period), $\varepsilon_j$ tends towards $0$ and $\theta_j$ tends towards $i_j$ (meaning that the spin axis tends to follow the precessing orbit normal, e.g.~term $j=5$ for Io). When the ratio $\omega_f/\dot\Omega_j$ is close to $-1$, $\varepsilon_j$ and $\theta_j$ can be resonantly amplified (e.g.~term $j=3$ for Ganymede). The ratio $\theta_2/\theta_1$ is of $26\%$, $5\%$, $18\%$, $16\%$ for Io, Europa, Ganymede and Callisto, respectively. Note that $\theta_3/\theta_1=22\%$ for Ganymede, while $i_3/i_1$ is only of $9\%$.
The largest orbital obliquity amplitudes is $\varepsilon_1$ for Io, Europa and Callisto, but $\varepsilon_3$ for Ganymede. The second largest orbital obliquity amplitudes are $\varepsilon_4$, $\varepsilon_5$, $\varepsilon_1$, and $\varepsilon_2$ for Io, Europa, Ganymede, and Callisto, respectively.

\begin{figure}[htb!]
\centering
\includegraphics[width=12cm]{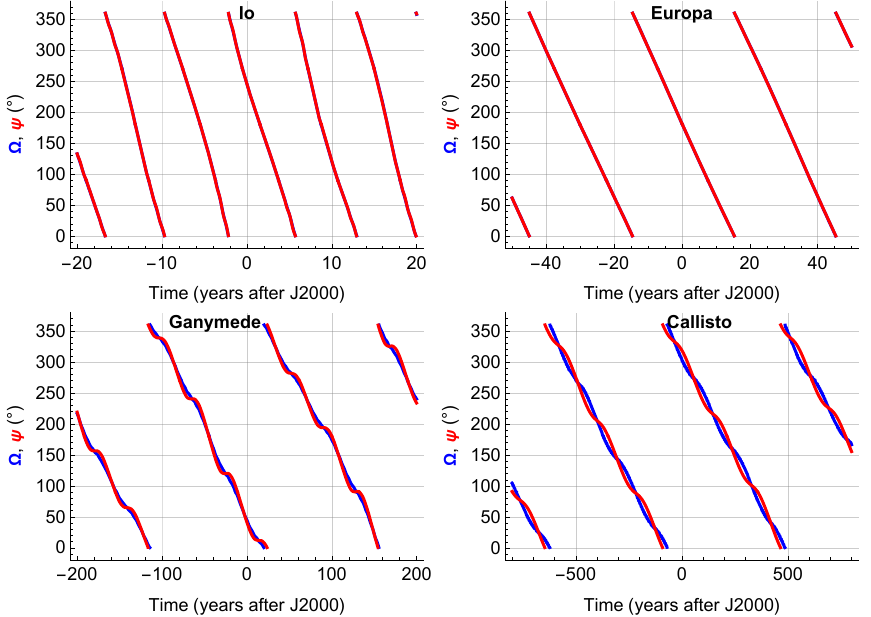}
\caption{Node longitude of the spin ($\psi$) and of the orbit ($\Omega$) as a function of time for the four Galilean satellites, assuming a solid interior model.}
\label{fig_psi}
\end{figure}

\begin{figure}[htb!]
\centering
\includegraphics[width=12cm]{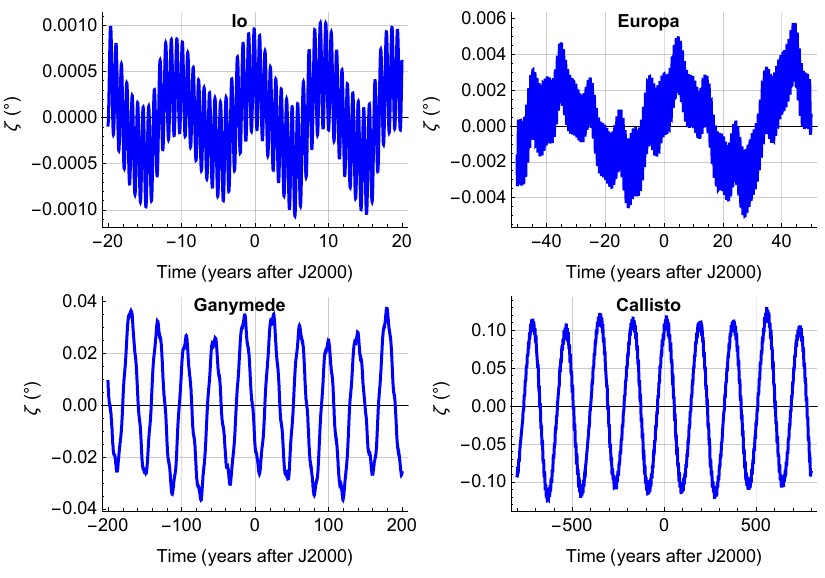}
\caption{Offset $\zeta$ of the spin axis with respect to the Cassini plane for the four Galilean satellites, assuming a solid interior model.}
\label{fig_offset}
\end{figure}

\begin{table}
\small\centering
\begin{tabular}{cccccc}
\hline
& $s_x, s_y$ & $\theta$ & $\psi$ & $\varepsilon$ & $\zeta$ \\
& ($10^{-4} \, {}^\circ$ / \%) & ($10^{-4} \, {}^\circ$ / \%) & (${}^\circ$ / \%) & ($10^{-4} \, {}^\circ$ / \%) & ($10^{-4} \, {}^\circ$ / \%) \\
\hline
\multicolumn{6}{l}{} \\
\multicolumn{6}{l}{\bf{Truncation error} of Table~\ref{tab_sxsy}} \\
{\bf Io}       & 3 / 0.6 \, & 3 / 0.6 \, & 0.6 \,/ 0.15 & 2 / 6 \,  & 2 / 10 \\
{\bf Europa}   & 4 / 0.06   & 3 / 0.04   & 0.04 / 0.01  & 2 / 0.4 & 3 / 5 \, \\
{\bf Ganymede} & 2 / 0.08   & 2 / 0.06   & 0.06 / 0.02  & 3 / 0.6 & 3 / 1 \, \\
{\bf Callisto} & 2 / 0.04   & 2 / 0.03   & 0.03 / 0.01  & 2 / 0.1 & 4 / 0.4 \\
\multicolumn{6}{l}{} \\
\multicolumn{6}{l}{\bf{Development error}} \\
{\bf Io} \, (order 3)     & - & $2$ / $0.3$   & $0.4$ / $0.1$   & $<1$ / $0.1$  & $<1$ / $2$  \, \\
{\bf Europa}\, (order 2)  & - & $<1$ / $<0.1$ & $0.01$ / $0.01$ & $<1$ / $<0.1$ & $<1$ / $0.2$ \\
{\bf Ganymede}  (order 5) & - & $1$ / $<0.1$  & $0.5$ / $0.2$   & $60$ / $9$ \,        & $<1$ / $0.1$ \\
{\bf Callisto}  (order 5) & - & $<1$ / $<0.1$ & $0.01$ / $0.01$ & $10$ / $0.4$      & $<1$ / $<0.1$ \\
\multicolumn{6}{l}{} \\
\multicolumn{6}{l}{\bf{Error after 100 years if the linear/Poisson terms are neglected}} \\
{\bf Io}       & \, 30 / 0.05 & 40 / 7   & 7 / 2      & - & 3 / 30 \\
{\bf Europa}   & 30 / 0.5 & \, 35 / 0.6 & 0.3 / 0.1  & - & 4 / 7  \, \\
{\bf Ganymede} & 30 / 0.3 & 30 / 1   & \, 2 / 0.6    & - & 8 / 2  \, \\
{\bf Callisto} & 30 / 0.4 & 40 / 1   & 0.4 / 0.1  & - & 20 / 2 \; \\
\hline
\end{tabular}
\caption{Comparison of different errors affecting the angles over the 1900–2100 period.
The first source of error arises from the truncation applied in Table~\ref{tab_sxsy}. Each term in the $s_{x/y}$ series is truncated at a level of $0.0001^\circ$ in amplitude, which is also the truncation threshold applied to the series. The cumulative errors on $s_{x/y}(t)$, and therefore on $\theta(t)$, $\varepsilon(t)$, and $\zeta(t)$, lead to a total truncation error of about $0.0003^\circ$. The cumulative truncation errors on the node longitude $\psi$ reach up to $0.6^\circ$ for Io.
The second source of error is the development error, arising when $\theta$, $\psi$, $\varepsilon$, and $\zeta$ are developed as series themselves.
A third possible source of error stems from the omission of the linear (for $s_x, s_y$, see Eq.~\ref{eq_sxsy}) and Poisson terms 
(for $\theta$, $\psi$, $\varepsilon$, and $\zeta$, see Eqs.~\ref{eq_trigo}).}
\label{tab_precision}
\end{table}

\subsection{Series for obliquity, node longitude, and offset}
\label{sec_obliquityseries}

The temporal evolution of the inertial ($\theta$, Eq.~\ref{eq_sintheta}) and orbital ($\varepsilon$, Eq.~\ref{eq_eps}) obliquities, node longitude ($\psi$, Eq.~\ref{eq_tanpsi}), and offset ($\zeta$, Eq.~\ref{eq_offset}) obtained from the series of Table~\ref{tab_trxtry} and Table~\ref{tab_sxsy} are shown in Figs.~\ref{fig_obli}, \ref{fig_psi}, and \ref{fig_offset}. We see that the deviation from the Cassini plane is small for Europa (usually less than $0.006^\circ$). For comparison with the spin Euler angles, the inclination ($i$, Eq.~\ref{eq_i}) and orbital node longitude ($\Omega$, Eq.~\ref{eq_Omega}) are also shown in Fig.~\ref{fig_obli} and \ref{fig_psi}, respectively.
The typical precision (or truncation error) of the time-series $\varepsilon(t)$ and $\theta(t)$ over the 1900-2100 interval, reconstructed from Table~\ref{tab_sxsy}, is by construction similar to that of the time-series $s_{x/y}(t)$, that is approximately $0.0003^\circ$ (see Table~\ref{tab_precision}). This corresponds to a relative error below $1\%$, except in the case of Io's orbital obliquity which is particularly small. 
In contrast, the longitude of the ascending node is a circulating angle, and the error associated with its reconstruction using Table~\ref{tab_sxsy} and Eq.~(\ref{eq_tanpsi}) is significantly larger but remains below $0.6^\circ$ and therefore below the percent level. The absolute precision on the reconstructed $\zeta(t)$ is similar to that of $\varepsilon(t)$, but its relative precision may be larger than that of $\varepsilon(t)$ when the offset is small (e.g.~Europa, $5\%$). As for $s_{x/y}(t)$, neglecting terms related to trends would lead after $100$ years to typical errors larger than the truncation error in $\theta$, $\psi$, and $\zeta$ ($\varepsilon$ is not affected by the trends). 

When we inject the series ($\Delta n_x, \Delta n_y$) and ($\Delta s_x, \Delta s_y$) from Section~\ref{sec_sxsyseries}, Table~\ref{tab_sxsy}, into the exact expressions for $\theta$ (Eq.~\ref{eq_sintheta}), $\psi$ (Eq.~\ref{eq_tanpsi}), $\varepsilon$ (Eq.~\ref{eq_epsilon}), and $\zeta$ (Eq.~\ref{eq_offset}) given in Sections~\ref{sec_obli} and \ref{sec_offset}, we obtain expressions which do not have the form of a periodic series.
Periodic series can be obtained instead by developing the angles' expressions of Eqs.~(\ref{eq_thetasxsy}, \ref{eq_psisxsy}, \ref{eq_epsilon}, \ref{eq_zetasxsy}) around the dominant terms of the series ($\Delta n_x, \Delta n_y$) and ($\Delta s_x, \Delta s_y$), when they exist.
If the first periodic term (with amplitudes $i_1$, $\theta_1$ and $\varepsilon_1$) dominates the other periodic terms ($i_j$, $\theta_j$ and $\varepsilon_j$ with $j>1$) and considering that the trends $tr_x$ and $tr_y$ are small, we can write the following developments for the inertial and orbital obliquities, the node longitude and the offset from the Cassini plane, up to order 2 in the small amplitudes of the other terms and in the trends: 
\begin{subequations}
\label{eq_dominant}
\begin{eqnarray}
\theta(t) &\approx& \theta_1 + (tr_x \sin\xi_1 - tr_y \cos\xi_1) \, t - (tr_x \cos\xi_1 + tr_y \sin\xi_1 ) \, t \sum_{j=2}^{N} \frac{\theta_j}{\theta_1} \sin(\xi_1-\xi_j) \nonumber \\
&& + \sum_{j=2}^{N} \theta_j \cos(\xi_1-\xi_j)
+ \sum_{k=2}^{N} \sum_{j=2}^{N} \frac{\theta_k \, \theta_j}{2 \, \theta_1} \sin(\xi_1-\xi_k) \, \sin(\xi_1-\xi_j), 
\label{eq_thetadominant} \\
\psi(t) &\approx& \xi_1 + \frac{1}{\theta_1}(\, tr_x \cos\xi_1 + tr_y \sin\xi_1 ) \, t
- t\,  \sum_{j=2}^{N} \frac{\theta_j}{\theta_1^2} \left(\, tr_x \cos(2\xi_1-\xi_j)+tr_y  \sin(2\xi_1-\xi_j) \right) \nonumber \\
&& - \sum_{j=2}^{N} \frac{\theta_j}{\theta_1} \sin(\xi_1-\xi_j) + \sum_{k=2}^{N} \sum_{j=2}^{N} \frac{\theta_k \, \theta_j}{\theta_1^2} \cos(\xi_1-\xi_k) \sin(\xi_1-\xi_j), 
\label{eq_psidominant} \\
\varepsilon(t) &\approx& \varepsilon_1 + \sum_{j=2}^{N} \varepsilon_j \cos(\xi_1-\xi_j )+ \sum_{k=2}^{N} \sum_{j=2}^{N} \frac{\varepsilon_k \, \varepsilon_j}{2 \, \varepsilon_1} \sin(\xi_1-\xi_k) \, \sin(\xi_1-\xi_j),
\label{eq_epsilondominant} \\
\zeta(t) &\approx& - (tr_x \cos\xi_1 + tr_y \sin \xi_1) \frac{\varepsilon_1}{i_1} \, t \nonumber \\
&& + \, t\, \sum_{j=2}^{N} \left(\frac{2\varepsilon_1 i_j-i_1\varepsilon_j}{2 i_1^2}\left(tr_x \cos(2\xi_1-\xi_j)+tr_y\sin(2\xi_1-\xi_j)\right)-\frac{i_1\varepsilon_j}{2i_1^2} (tr_x \cos \xi_j+tr_y\sin \xi_j)\right) \nonumber \\
&& + \sum_{j=2}^{N} \frac{\varepsilon_1\, i_j-\varepsilon_j\, i_1}{i_1}\sin(\xi_1-\xi_j) + \sum_{k=2}^{N} \sum_{j=2}^{N} \frac{i_j \, (\varepsilon_k\, i_1-\varepsilon_1\, i_k)}{i^2_1}\cos(\xi_1-\xi_k) \, \sin(\xi_1-\xi_j), 
\label{eq_zetadominant}
\end{eqnarray}
\end{subequations}
where we use the notation $\xi_j = f_j \, t + \varphi_j$. $i_j$, $\varepsilon_j$, $\theta_j$, $f_j$, and $\varphi_j$ are given in columns 2, 3, 4, 6, and 7 in Table~\ref{tab_sxsy}. As already noted after Eq.~(\ref{eq_epsilon}), $\varepsilon$ is unaffected by the long-term trends.
Developments of a higher order are usually necessary to reach a sufficient accuracy in the time domain. 

Note that the presence, absence, or even the position in the series of a single dominant term, when it exists, may change significantly when considering a different interior model than the solid one used here. Some amplitudes may be amplified or diminished, resulting in a markedly different configuration.

\begin{table}[!htb]
\small
\centering
\begin{tabular}{ccccccc}
\hline
& $\mathcal{E}_j$ ($^\circ$)& $\mathcal{Z}_j$ ($^\circ$) & period (y) & $f_j$ (rad/y) & $\varphi_j$ ($^\circ$) & Argument \\
\hline
\multicolumn{6}{l}{} \\
{\bf Io}  & $\mathcal{E}_{cst}$ = 0.0020 &&&&& \\
1 & 0.0006 & 0.0006 & -0.75034 & -8.37381 & 13.310 & $-2\nu-\Omega_1-\Omega_2$ \\
2 & 0.0001 & 0.0005 & -9.83977 & -0.63855 & 76.509 & $\Omega_1-\Omega_2$ \\
\multicolumn{6}{l}{} \\
{\bf Europa} & $\mathcal{E}_{cst}$ = 0.0555 &&&&&  \\ 
1 &  0.0015 &  0.0016 &  -0.69717  & -9.01236 &  89.824 & $-2\nu-2\Omega_2$ \\ 
2 & -0.0009 & -0.0008 &  -9.83977  & -0.63855 &  76.516 & $\Omega_1-\Omega_2$ \\ 
3 &  0.0006 &  0.0024 & -38.7134   & -0.1623  & 124.306 & $\Omega_2-\Omega_3$ \\ 
4 &  0.0000 & -0.0007 &  31.9203   &  0.19684 & 126.198 & $-\Omega_2+\Omega_4$ \\
\multicolumn{6}{l}{} \\
{\bf Ganymede} & $\mathcal{E}_{cst}$ = 0.0403 &&&&& \\
 1 & -0.0279 & -0.0293 & -38.7134 & -0.16230 & 124.307 & $\Omega_2-\Omega_3$ \\
 2 & -0.0060 & -0.0013 & -19.3567 & -0.32460 & 248.614 & $2\Omega_2-2\Omega_3$ \\
 3 & -0.0041 &  0.0000 & -12.9045 & -0.48690 &  12.921 & $3\Omega_2-3\Omega_3$ \\
 4 & -0.0026 &  0.0000 & -9.67835 & -0.64920 & 137.228 & $4\Omega_2-4\Omega_3$ \\
 5 & -0.0010 & -0.0017 &  5.68568 &  1.10509 &  53.758 & $2L_S-\Omega_3$ \\
 6 &  0.0010 & -0.0051 &  181.858 &  0.03455 & 250.498 & $-\Omega_3+\Omega_4$ \\
 7 & -0.0008 &  0.0032 &  31.9186 &  0.19685 & 126.191 & $-\Omega_2+\Omega_4$ \\
 8 &  0.0008 & -0.0002 & -4.95758 & -1.26739 &  70.549 & $\Omega_2-2L_S$ \\
 9 &  0.0004 &  0.0028 & -49.1834 & -0.12775 &  14.805 & $\Omega_2-2\Omega_3+\Omega_4$ \\
10 & -0.0005 & -0.0008 & -12.985  & -0.48388 & 239.320 & $-\Omega_3-L_S$ \\
\multicolumn{6}{l}{} \\
{\bf Callisto}  & $\mathcal{E}_{cst}$ = 0.1584 &&&&& \\
 1 &  0.0875 & -0.1100 & 181.858  &  0.03455 & 250.487 & $-\Omega_3+\Omega_4$ \\
 2 & -0.0133 & -0.0066 & 90.9289  &  0.06910 & 140.974 & $-2\Omega_3+2\Omega_4$ \\
 3 &  0.0042 &  0.0000 & 60.6193  &  0.10365 &  31.461 & $3\Omega_4-3\Omega_3$ \\
 4 & -0.0033 &  0.0059 & -5.86906 & -1.07056 & 196.735 & $\Omega_4-2L_S$ \\
 5 & -0.0017 &  0.0000 & 45.4644  &  0.13820 & 281.948 & $4\Omega_4-4\Omega_3$ \\
 6 & -0.0013 & -0.0005 &  5.68557 &  1.10511 &  53.752 & $2L_S-\Omega_3$ \\
 7 &  0.0010 &  0.0012 & -6.06479 & -1.03601 &  87.222 & $-\Omega_3+2\Omega_4-2L_S$ \\
 8 &  0.0009 &  0.0000 & 36.3716  &  0.17275 & 172.435 & $5\Omega_4-5\Omega_3$ \\
 9 &  0.0006 & -0.0010 & 31.9235  &  0.19682 & 126.050 & $-\Omega_2+\Omega_4$ \\
10 & -0.0005 &  0.0009 & 12.1203  &  0.51840 &  11.016 & $\Omega_4+L_S$ \\
11 & -0.0005 &  0.0000 & -6.27403 & -1.00146 & 337.709 & $-2\Omega_3+3\Omega_4-2L_S$ \\
12 & -0.0004 &  0.0008 & -11.6192 & -0.54076 & 132.704 & $\Omega_4-L_S$ \\
13 & -0.0004 &  0.0006 & -3.92662 & -1.60015 & 177.130 & $\Omega_4-3L_S$ \\
14 &  0.0003 & -0.0005 & -5.80798 & -1.08182 &   7.937 & $2 \Omega_4-\Omega_0-2L_S$ \\
\hline
\end{tabular}
\caption{Periodic development of the orbital obliquity $\varepsilon$ and the Cassini plane offset $\zeta$ for the four Galilean satellites, considered as solid. 
We consider developments up to orders 3, 2, 5, and 5, for Io, Europa, Ganymede, and Callisto, respectively. The series presented in this table are provided for illustrative purposes, with the smallest terms omitted for the sake of brevity.
For each satellite, the first term of the obliquity is the constant mean value $\mathcal{E}_{cst}$. The phases $\varphi_j$ are given with respect to J2000 epoch.}
\label{tab_obliquityzeta}
\end{table}

Rewriting products of trigonometric functions in terms of trigonometric functions with combined arguments, each development can then be written as a trigonometric series:
\begin{subequations}
\label{eq_trigo}
\begin{eqnarray}
\theta(t) &\approx& \mathcal{T}_{cst} + \sum_{j}\mathcal{T}_j \, \cos\xi_j + t \, \sum_{k} \left(\mathcal{T}^c_k \, \cos\xi_k + \mathcal{T}^s_k \, \sin\xi_k\right),
\label{eq_thetatrigo} \\
\psi(t) &\approx& \xi_1 \;\; + \sum_{j}\mathcal{P}_j \, \sin\xi_j + t\, \sum_{k} \left(\mathcal{P}^c_k \, \cos\xi_k+ \mathcal{P}^s_k \, \sin\xi_k\right),
\label{eq_psitrigo} \\
\varepsilon(t) &\approx& \mathcal{E}_{cst} + \sum_{j} \mathcal{E}_j \, \cos\xi_j, 
\label{eq_epsilontrigo} \\
\zeta(t) &\approx & \qquad \; \, \sum_{j}\mathcal{Z}_j \, \sin\xi_j + t\, \sum_{k} \left(\mathcal{Z}^c_k \, \cos\xi_k+ \mathcal{Z}^s_k \, \sin\xi_k\right).
\label{eq_zetatrigo} 
\end{eqnarray}
\end{subequations}
The general form of these expansions remains valid for developments at any order. The series expansions for $\theta$ and $\varepsilon$ contain a mean term, $\mathcal{T}_{cst}$ and $\mathcal{E}_{cst}$, respectively, which differ from $\theta_1$ and $\varepsilon_1$ for high order developments, while the mean Cassini plane offset value is zero. The first term of $\psi$ is $\xi_1$, the first argument of the $n/s$ series. Each angle has a periodic cosine or sine series. Except for $\varepsilon$, all angles have a Poisson series with both cosine and sine amplitudes multiplied by the time $t$. 
The arguments of the periodic and Poisson series are always a linear combination of $\xi_j$ arguments of Table~\ref{tab_sxsy}. 

These trigonometric developments can be very accurate. If we go up to orders 3, 2, 5, and 5 respectively, for each satellite, the "development error" is usually less than $1\%$, except for the orbital obliquity of Ganymede ($9\%$, see Table~\ref{tab_precision}). Considering a higher order for Ganymede’s $\varepsilon$ does not solve the accuracy issue, as the development cannot converge due to $|\varepsilon_1/\varepsilon_3|\sim1$. For Europa, it is possible to neglect the second-order terms of the development while maintaining a reasonable accuracy (e.g.~$10^{-4} \,^\circ$ or about $0.2\%$ in $\varepsilon$). Compared with a purely numerical solution obtained from ephemerides, these development errors should be added to the truncation errors in the original series in $n/s$, discussed in Section~\ref{sec_sxsyseries} and summarized in the first four lines of Table~\ref{tab_precision}, to obtain the total error budget.

\begin{figure}[htb!]
\hspace*{-1cm}
\includegraphics[width=18.5cm]{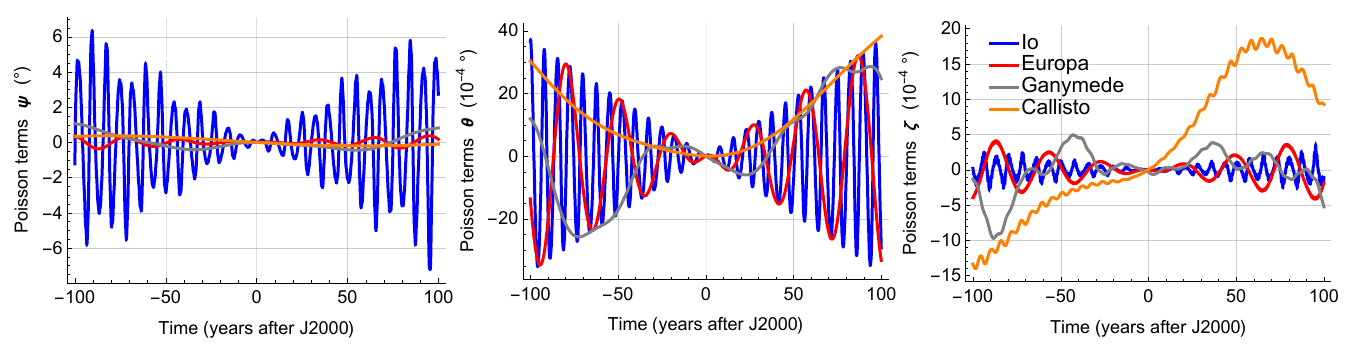}
\caption{Poisson terms for the angles $\psi$, $\theta$ and $\zeta$ for the four Galilean satellites over the calendar period 1900-2100. }
\label{fig_Poiss}
\end{figure}

We do not show the trigonometric expansions here in their entirety, as they can include dozens and dozens of terms. For illustration purposes only, we provide truncated, and therefore imprecise, expansions in
Tables~\ref{tab_obliquityzeta} to \ref{tab_Poisson}, for the periodic and Poisson terms. 
At least as far as the most important terms are concerned, the $\theta$ and $\psi$ series, as well as the $\varepsilon$ and $\zeta$ series, tend to contain terms with the same arguments. The series for $\psi$ are multiplied by $\theta_1$ to obtain amplitudes of the same order as those of $\theta$.
The Poisson series for the angles $\psi$, $\theta$ and $\zeta$ are plotted on Fig.~\ref{fig_Poiss}. Between 2030 and 2035, during the Juice mission timeframe, Ganymede's Poisson terms can reach approximately up to $0.001^\circ$, $1^\circ$, and $0.0005^\circ$ for $\theta$, $\psi$, and $\zeta$, respectively. 
The periodic series of the orbital obliquity for Callisto closely matches the solution of \cite{Noy09} (see his Table~10), which was derived through frequency decomposition of a numerical integration of the Hamiltonian equations. Differences in amplitudes and frequencies are generally within a few percent, and are probably partly due to the use of a different ephemeris. This comparison of our analytical results and this numerical series provides support for our approach. A key advantage of our method is its ability to clearly identify the various contributing terms, including the Poisson terms in $\theta$, $\psi$ and $\zeta$, see Eq.~(\ref{eq_trigo}). 

For Io and Europa, $\mathcal{T}_{cst}\sim \theta_1$ and $\mathcal{E}_{cst}\sim \varepsilon_1$, while most of their periodic terms are such that $|\mathcal{T}_j|\sim |\theta_1 \mathcal{P}_j|\sim|\theta_j|$ as can be understood from the first-order term in Eq.~(\ref{eq_thetadominant}). Note, however, that the argument of each term is now a combination of two distinct arguments, rather than the original single argument of Table~\ref{tab_sxsy}, so the periodicity will be different from the $s_x/s_y$ series. Most of their amplitudes $\mathcal{E}_j$ are also obtained essentially as first-order terms in Eq.~(\ref{eq_epsilondominant}) and therefore tend to be similar to the amplitudes $\varepsilon_j$, but are not organized in the same order. For those first order terms, $\mathcal{Z}_j$ are not necessarily expected to have an amplitude similar to that of $\varepsilon_j$, however, see Eq.~(\ref{eq_zetadominant}). For Ganymede (Callisto), because of the relatively large amplitudes of the second and third terms (second term) in Table~\ref{tab_sxsy}, the amplitudes in Tables~\ref{tab_inertialobliquity} and \ref{tab_obliquityzeta} do not necessarily correspond to those in Table~\ref{tab_sxsy} (especially for $\varepsilon$ in the case of Ganymede) and there are many high-order terms involved.

With the exception of the orbital obliquity of Ganymede, the series of $\theta$, $\psi$, $\varepsilon$, and $\zeta$ help us to understand the temporal behavior of these angles (Figs.~\ref{fig_sxsy}, \ref{fig_obli}, \ref{fig_psi}, and \ref{fig_offset}). 
The second terms of the series perturb the inertial obliquity of Io, Europa, Ganymede and Callisto by $26\%$, $5\%$, $22\%$, and $16\%$, respectively. The third term of Ganymede's series perturbs its inertial obliquity by $18\%$, see third panel of Fig.~\ref{fig_obli}.
The main perturbation of $\varepsilon$ comes from $\mathcal{E}_1$ which is similar (or at least mainly originates from) to $\varepsilon_4$, $\varepsilon_5$, and $\varepsilon_2$ for Io, Europa, and Callisto, respectively. Europa's orbital obliquity varies by $3\%$ over time, while its spin axis describes an almost circular trajectory in space, see Fig.~\ref{fig_sxsy}. For Io and Callisto, $\varepsilon$ varies by $30\%$ and $55\%$, respectively. For Ganymede, $\varepsilon$ varies by almost $100\%$ around its mean value, between $0$ and $0.6^\circ$,  but this is poorly described by the $\varepsilon$ series.
Callisto has the largest offset of the four satellites. Its periodic variations are dominated by the first term of the offset series with an amplitude $\mathcal{Z}_1=0.11^\circ$.

\begin{table}[!htb]
\small
\centering
\begin{tabular}{ccccccc}
\hline
& $\mathcal{T}_j$ ($^\circ$)& $\theta_1\mathcal{P}_j$ ($^\circ$) & period (y) & $f_j$ (rad/y) & $\varphi_j$ ($^\circ$) & Argument \\
\hline
\multicolumn{7}{l}{} \\
{\bf Io}  & $\mathcal{T}_{cst}$ = 0.0385 & & & & & \\
1 &  0.0099 & -0.0100 & -9.83977 & -0.63855 &  76.509 & $\Omega_1-\Omega_2$ \\
2 &  0.0013 & -0.0013 & -7.84565 & -0.80085 & 200.816 & $\Omega_1-\Omega_3$\\
3 &  0.0010 &  0.0010 & -0.75034 & -8.37381 &  13.310 & $-2\nu-\Omega_1-\Omega_2$ \\
4 & -0.0007 &  0.0014 & -4.92655 & -1.27537 & 146.715 & $2 \Omega_1-2\Omega_2$ \\
\multicolumn{7}{l}{} \\
{\bf Europa}  & $\mathcal{T}_{cst}$ = 0.5209 & &  & & &  \\
1 &  0.0256 & -0.0256 & -38.7134 & -0.16230 & 124.306 & $\Omega_2-\Omega_3$ \\
2 &  0.0056 & 0.0056 &  31.9203 &  0.19684 & 126.198 & $-\Omega_2+\Omega_4$ \\
3 & -0.0021 &  -0.0021 & -9.83977 & -0.63855 &  76.516 & $\Omega_1-\Omega_2$ \\
4 &  0.0005 & -0.0005 & -4.95758 & -1.26739 &  70.573 & $\Omega_2-2L_S$ \\
\multicolumn{7}{l}{} \\
{\bf Ganymede} & $\mathcal{T}_{cst}$ = 0.2222 & &  & & &  \\
1 & -0.0479 &  -0.0486 & -38.7134 & -0.1623  & 124.307 & $\Omega_2-\Omega_3$ \\
2 &  0.0389 & 0.0396 & 181.858  &  0.03455 & 250.498 & $-\Omega_3+\Omega_4$ \\
3 & -0.0045 &  0.0000 &  31.9186 &  0.19685 & 126.191 & $-\Omega_2+\Omega_4$ \\
4 &  0.0043 & 0.0088 & -49.1834 & -0.12775 &  14.805 & $\Omega_2-2 \Omega_3+\Omega_4$ \\
5 & -0.0026 &  -0.0054 & -19.3567 & -0.3246  & 248.614 & $2 \Omega_2-2 \Omega_3$ \\
6 & -0.0017 &  -0.0036 &  90.9289 &  0.0691  & 140.996 & $-2 \Omega_3+2 \Omega_4$ \\
7 &  0.0007 &  0.002 & -21.6624 & -0.29005 & 139.112 & $2 \Omega_2-3 \Omega_3+\Omega_4$ \\
8 & -0.0006 &  -0.0016 & -67.4162 & -0.0932  & 265.303 & $\Omega_2-3 \Omega_3+2 \Omega_4$ \\
\multicolumn{7}{l}{} \\
{\bf Callisto} & $\mathcal{T}_{cst}$ = 0.3961 & &  & & &  \\
1 &  0.0624 & -0.0626 & 181.858 & 0.03455 & 250.487 & $-\Omega_3+\Omega_4$ \\
2 & -0.0025 &  0.005 & 90.9289 & 0.0691  & 140.974 & $-2 \Omega_3+2\Omega_4$ \\
 \hline
\end{tabular}
\caption{Periodic development of the inertial obliquity $\theta$ and the node longitude $\psi$ for the four Galilean satellite, considered as solid. We consider developments up to orders 3, 2, 5, and 5, for Io, Europa, Ganymede, and Callisto, respectively. The series presented in this table are provided for illustrative purposes, with the smallest terms omitted for the sake of brevity.
For each satellite, the first term of the inertial obliquity is the constant mean value $\mathcal{T}_{cst}$. The phases $\varphi_j$ are given with respect to J2000 epoch.}
\label{tab_inertialobliquity}
\end{table}

\begin{table}[!htb]
\small
\centering
\begin{tabular}{ccccccccccc}
\hline
& $ \mathcal{T}^c_j$& $ \mathcal{T}^s_j$ & $ \theta_1\mathcal{P}^c_j$& $\theta_1 \mathcal{P}^s_j$ & $ \mathcal{Z}^c_j$& $ \mathcal{Z}^s_j$ & period (y) & $f_j$ (rad/y) & $\varphi_j$ ($^\circ$) & Argument \\
\hline
\multicolumn{11}{l}{} \\
{\bf Io}  &&&&&&&&&& \\
1 & -24 & -27 & -28 & 24 & 2 & -1 & -7.42167 & -0.8466 & 260.604 & $\Omega_1$ \\
2 &  4 & 4 & 7 & -6 & 0 &0 & -4.23067 & -1.48515 & 337.113 & $2 \Omega_1-\Omega_2$ \\
\multicolumn{11}{l}{} \\
{\bf Europa}  &&&&&&&&&& \\
1 & -24 & -28 & -28 & 24 & 3 & -3 & -30.2004 & -0.20805 & 184.073 & $\Omega_2$ \\
\multicolumn{11}{l}{} \\
{\bf Ganymede}  &&&&&&&&&& \\
1 & -23. & -27  & -28 & 24 & 5 & -4 &-137.337 & -0.04575 & 59.659 & $\Omega_3$ \\
2 & -3. & -3 & -6 & 5 & -2 & 2 &53.8619 & 0.11665 & 296.538 & $2 \Omega_3-\Omega_2$ \\
\multicolumn{11}{l}{} \\
{\bf Callisto}  &&&&&&&&&& \\
1 & -25. & -28 & -28 & 25 & 16& -15& -560.999 & -0.0112 & 309.669 & $\Omega_4$ \\
2 & 2. & 2 & 5 & -4 & 0 & 0 & 269.67 & 0.0233 & 199.92 & $2\Omega_4-\Omega_3$ \\
\hline
\end{tabular}
\caption{Poisson series for the angles $\theta$, $\psi$ and $\zeta$, including cosine and sine terms. Amplitude units are $10^{-6} \,^\circ$/y. The phases $\varphi_j$ are given with respect to J2000 epoch.}
\label{tab_Poisson}
\end{table}

\subsection{Series for \texorpdfstring{$(\alpha_S, \delta_S)$}{alpha, delta} and \texorpdfstring{$W$}{W}}
\label{sec_admseries}

We assume for $\alpha_S$ and $\delta_S$ a form similar to that of $s_x$ and $s_y$, but with non-zero mean terms: 
\begin{subequations}
\label{eq_adbt}
\begin{eqnarray}
\alpha_S(t) &=& \alpha_{LP} + tr_\alpha \, t + \sum_j \alpha^S_j \sin(f_j \, t + \varphi_j), 
\label{eq_alphat} \\
\delta_S(t) &=& \delta_{LP} + \delta_{S\,0} + tr_\delta \, t  + \sum_j \delta^S_j \cos(f_j \, t + \varphi_j). 
\label{eq_deltat}
\end{eqnarray}
\end{subequations}
The mean value of $\alpha_S$ is that of the Laplace plane ($\alpha_{LP}$), as already noticed in Eq.~(\ref{eq_sxform}), see Section~\ref{sec_transfo}. For $\delta_S$, the mean value is the sum of $\delta_{LP}$ and of $\delta_{S\,0} \sim \theta_1^2 \tan\delta_{LP} /4$, the latter being a small increment due to the constant term induced by the power reduction of a squared sine term in the second-order part of Eq.~(\ref{eq_deltatransfo}). As part of the transformation, this incremental quantity depends on the assumed interior model for the satellites (for instance, entirely solid or not).
For $W$, we reexpress Eq.~(\ref{eq_Wlong}) assuming the following form:
\begin{equation}
W = W_0 + \dot W \, t + \gamma_u + \sum_j \mu_j \, \sin(f_j \, t + \varphi_j),  
\label{eq_Wmu}
\end{equation}
where $\dot W$ is defined as $n_L + tr_\mu$, see Eq.~(\ref{eq_Wrate}), with $tr_\mu$ the rate in angle $\mu$. $\gamma_u$ is the libration angle defined with respect to a uniform rotation, see Eq.~(\ref{eq_phiInertial3}). The last term is the series $\Delta\mu$ expressed in a form similar to that of $\Delta s_{x/y}$. 

\begin{figure}[htb!]
\centering
\includegraphics[width=10cm]{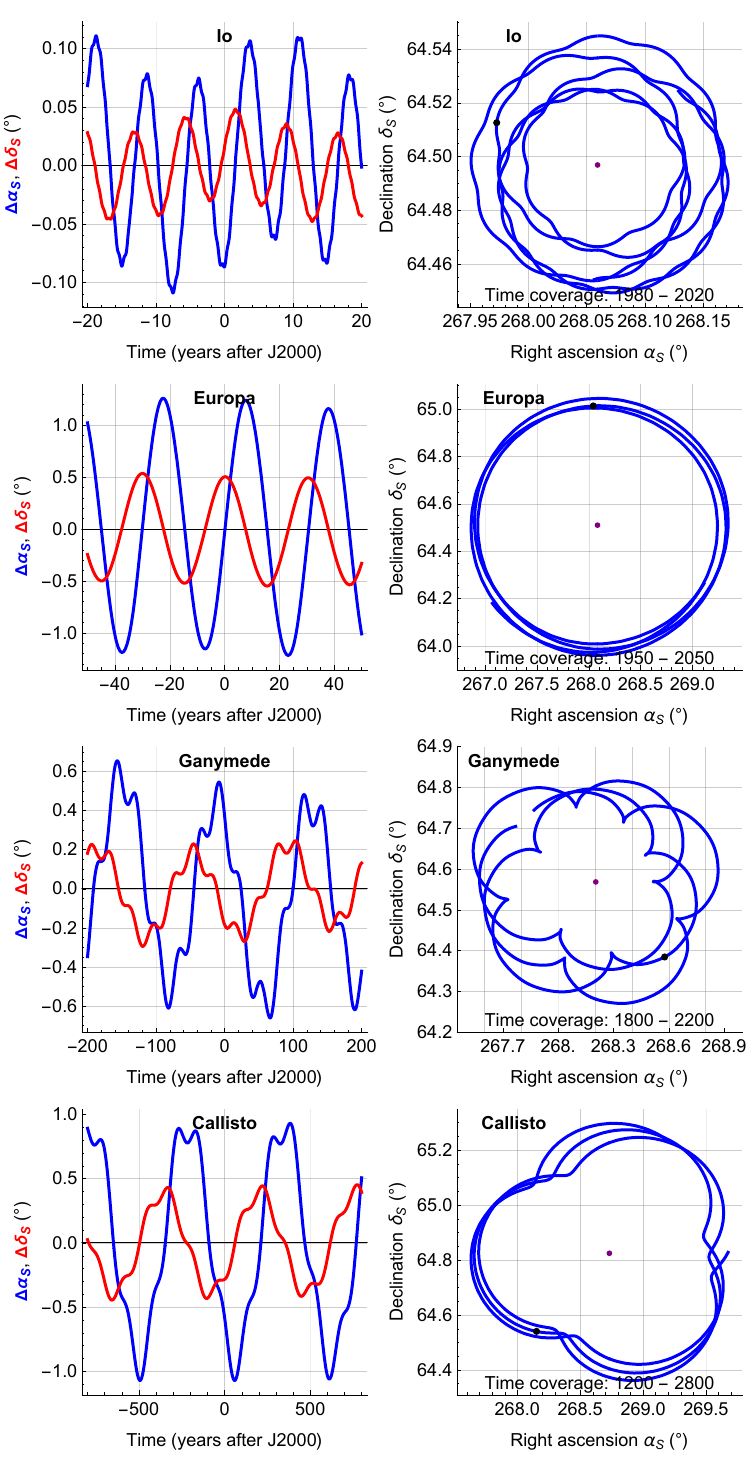}
\caption{Temporal evolution of the spin axis angle (right ascension $\alpha_S$ and declination $\delta_S$) for the four Galilean satellites.
The left plot show the periodic variations while the right plot shows the projection of the motion in the ICRF plane for a solid satellite. 
The purple point is the Laplace plane ($\alpha_{LP}, \delta_{LP}$) while the black dot is the J2000 spin position. }
\label{fig_ad}
\end{figure}

\begin{figure}[htb!]
\centering
\includegraphics[width=13cm]{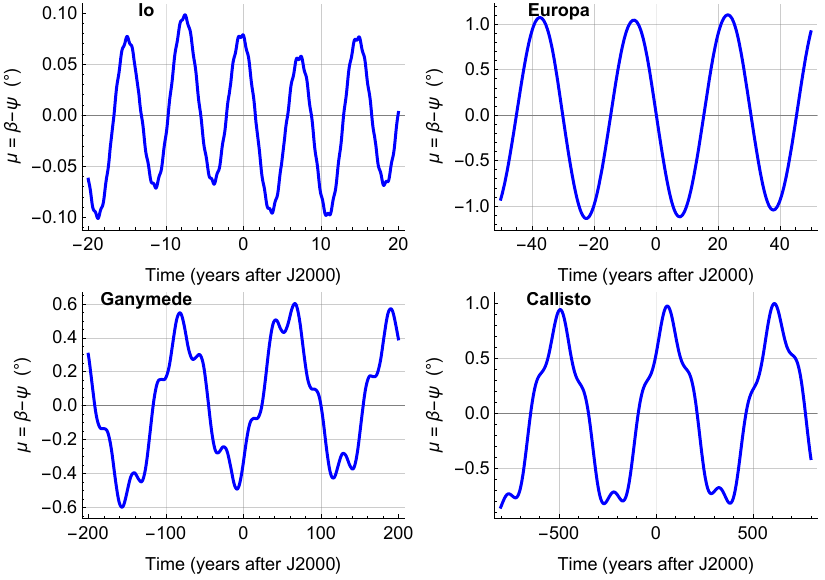}
\caption{Temporal evolution of the $\mu$ angle ($\mu = \beta - \psi$) for the four Galilean satellites, assuming a solid interior model.}
\label{fig_mu}
\end{figure}

Using the transformations of Eqs.~(\ref{eq_adapprox}), (\ref{eq_trends}), and (\ref{eq_mut}) on the spin model developed in Section~\ref{sec_sxsyseries}, we find values for the rates and for the amplitudes of the periodic series in $\alpha$, $\delta$ and $\mu$, as well as for the increment $\delta_{S \, 0}$, see Tables~\ref{tab_trxtry} and \ref{tab_spinad}. $\delta_{S\, 0}$ ranges from $-10^{-5}$ deg for Io to $-2.5 \times 10^{-3}$ deg for Europa, see Table~\ref{tab_trxtry}, which have respectively the smallest and largest dominant inertial obliquity amplitude $\theta_1$. Figure~\ref{fig_ad} shows the temporal evolution of the right ascension $\alpha_S$ and declination $\delta_S$ angles, whereas the evolution of $\mu$ is presented in Fig.~\ref{fig_mu}.

\begin{table}[!htb]
\small
\centering
\begin{tabular}{lcccccccc}
\hline
 & $\alpha^S_j$ ($^\circ$) & $\delta^S_j$ ($^\circ$) & $\mu_j$ ($^\circ$) & period (y) & $f_j$ (rad/y) & $\varphi_j$ ($^\circ$) & Argument & $\dot{J}_{IAU}$ \\
\hline
\textbf{Io} &&&&&&&& \\
 1 & 0.0878 & -0.0378 & -0.0793 & -7.42164 & -0.84660 & 260.604 & $\Omega_1$ & ${\bf -\dot{J}_3}$ \\
 2 & 0.0233 & -0.0100 & -0.0210 & -30.2008 & -0.20805 & 184.095 & $\Omega_2$ & ${\bf -\dot{J}_4}$ \\
 3 & 0.0030  & -0.0013 & -0.0027 & -137.325 & -0.04575 & 59.788 & $\Omega_3$ & $-\dot{J}_5$ \\
 4 & 0.0023  & -0.0010 & -0.0021 & -0.68144 & -9.22041 & 273.914 & $-2\nu-\Omega_2$ & \\
 5 & -0.0008 &  0.0003 & 0.0007 & -564.38 & -0.01113 & 310.084 & $\Omega_4$ & $-\dot{J}_6$ \\
 6 & 0.0006  & -0.0003 & -0.0006 & 5.93122 & 1.05934 & 113.504 & $2L_S$ & $\dot{J}_8$ \\
\textbf{Europa} &&&&&&&& \\
 1 & 1.2097 & -0.5206 & -1.0920 & -30.2008 & -0.20805 & 184.073 & $\Omega_2$ & ${\bf -\dot{J}_4}$ \\
 2 & 0.0595 & -0.0256 & -0.0537 & -137.328 & -0.04575 & 59.767 & $\Omega_3$ & ${\bf -\dot{J}_5}$ \\
 3 & 0.0131 & -0.0056 & -0.0118 & -560.607 & -0.01121 & 310.271 & $\Omega_4$ & ${\bf -\dot{J}_6}$ \\
 4 ($\mathcal{O}2$) & -0.0115 & 0.0025 & 0.0116 & -15.1004 & -0.41609 & 8.146 & $2\Omega_2$ & ${\bf -\dot{J}_7} = -2 \dot{J}_4$ \\
 5 & -0.0049 & 0.0021 & 0.0044 & -7.42165 & -0.84660 & 260.589 & $\Omega_1$ & $-\dot{J}_3$ \\
 6 & 0.0013 & -0.0005 & -0.0011 & 5.9312 & 1.05934 & 113.500 & $2L_S$ & $\dot{J}_8$ \\
 7 & 0.0007 & -0.0003 & -0.0007 & -0.68144 & -9.22041 & 273.897 & $-2\nu-\Omega_2$ & \\
 8 & 0.0007 & -0.0003 & -0.0006 & -11.8638 & -0.52961 & 299.309 & $-L_S$ & $-\dot{J}_8/2$ \\
9 ($\mathcal{O}2$) & 0.0000 & -0.0002 & 0.0000 & -38.7149 & -0.16229 & 124.306 & $\Omega_2 - \Omega_3$ & $-\dot{J}_4 + \dot{J}_5$ \\
10 ($\mathcal{O}2$) & -0.0011 & 0.0002 & 0.0011 & -24.7564 & -0.25380 & 243.840 & $\Omega_2 + \Omega_3$ & $-\dot{J}_4 - \dot{J}_5$ \\
\textbf{Ganymede} &&&&&&&& \\
 1 & 0.5066 & -0.2176 & -0.4575 & -137.328 & -0.04575 & 59.659 & $\Omega_3$ & ${\bf -\dot{J}_5}$ \\
 2 & -0.1132 & 0.0486 & 0.1022 & -30.2009 & -0.20805 & 183.966  & $\Omega_2$ & ${\bf -\dot{J}_4}$ \\
 3 & 0.0922 & -0.0396 & -0.0833 & -560.839 & -0.01120 & 310.157 & $\Omega_4$ & ${\bf -\dot{J}_6}$ \\
 4 ($\mathcal{O}2$) & -0.0020 & 0.0004 & 0.0020 & -68.664 & -0.09151 & 119.318  & $2 \Omega_3$ & $-2 \dot{J}_5$ \\
 5 & -0.0010 & 0.0004 & 0.0009 & -11.8633 & -0.52963 & 298.979 & $-L_S$ & $-\dot{J}_8$/2 \\
 6 & 0.0010 & -0.0004 & -0.0009 & 5.93122 & 1.05934 & 113.417 & $2L_S$ & $\dot{J}_8$ \\
 7 ($\mathcal{O}2$) & -0.0000 &  0.0002 &  0.0000 &  -38.715 & -0.16229 & 124.307 & $\Omega_2 - \Omega_3$ & $-\dot{J}_4 + \dot{J}_5$ \\
 8 ($\mathcal{O}2$) &  0.0009 & -0.0002 & -0.0009 &  -24.7565 & -0.25380 & 243.626 & $\Omega_2 + \Omega_3$ & $-\dot{J}_4 - \dot{J}_5$ \\
 9 ($\mathcal{O}2$) &  0.0000 & -0.0002 &  0.0000 &  181.858 & 0.03455 & 250.498 & $-\Omega_3 + \Omega_4$ & $\dot{J}_5 - \dot{J}_6$ \\
10 ($\mathcal{O}2$) & -0.0007 &  0.0002 &  0.0007 & -110.316 & -0.05696 & 9.816 & $\Omega_3 + \Omega_4$ & $-\dot{J}_5 - \dot{J}_6$ \\
\textbf{Callisto} &&&&&&&& \\
1 & 0.9253 & -0.3936 & -0.8374 & -560.826 & -0.01120 & 309.668  & $\Omega_4$ & ${\bf -\dot{J}_6}$ \\
2 & 0.1470 & -0.0626 & -0.1331 & -137.327 & -0.04575 & 59.182  & $\Omega_3$ & ${\bf -\dot{J}_5}$ \\
3 ($\mathcal{O}2$) & -0.0068 & 0.0014 & 0.0068 & -280.413 & -0.02241 & 259.337 & $2 \Omega_4$ & $-2\dot{J}_6$ \\
4 ($\mathcal{O}2$) & 0.0000 & -0.0005 & 0.0000 &  181.858 & 0.03455 & 250.487 & $-\Omega_3 + \Omega_4$ & $\dot{J}_5 - \dot{J}_6$ \\
5 ($\mathcal{O}2$) & -0.0022 & 0.0005 & 0.0022 & -110.315 & -0.05696 & 8.850 & $\Omega_3 + \Omega_4$ & $-\dot{J}_5 - \dot{J}_6$ \\
6 & 0.0003 & -0.0001 & -0.0002 & 5.93112 & 1.05936 & 112.934 & $2 L_S$ & ${\bf \dot{J}_8}$ \\
\hline
\end{tabular}
\caption{Spin variations series $\Delta\alpha_S$, $\Delta\delta_S$ and $\Delta\mu$ of the Galilean satellites for the solid interior model and the JUP387 orbital theory.
The truncation level is $0.0001^\circ$ on the $\delta^S_j$ amplitudes.
The last columns show the identification to the fundamental arguments of \cite{Lai06} and to the IAU WG frequencies $\dot{J}_i$. The $\dot{J}_i$ frequencies appearing in \cite{Arc18} are indicated in bold. Second order terms, not present in the $\Delta s_{x/y}$ series, are labeled with the symbol $\mathcal{O}2$. These series can be used to reconstruct time series for $\alpha/\delta_S(t)$ and $\mu(t)$, and indirectly to reconstruct the time series for the angles $\theta(t)$, $\varepsilon(t)$, $\psi(t)$, and $\zeta(t)$. The phases $\varphi_j$ are given with respect to J2000 epoch.}
\label{tab_spinad}
\end{table}

The chosen truncation level of the Table~\ref{tab_spinad} is $0.0001^\circ$ on the $\delta^S_j$ individual amplitudes. The last columns show the identification to the fundamental arguments of \cite{Lai06} and to the IAU WG frequencies. The $\dot{J}_i$ frequencies appearing in \cite{Arc18} are indicated in bold: 2 lines for Io, 4 for Europa and 3 for Ganymede and Callisto. Even though the frequencies and phases of the fundamental arguments are not affected by the transformation, the second order terms of the transformation give rise to new argument combinations in Table~\ref{tab_spinad}, labeled with the symbol $\mathcal{O}2$, compared to Table~\ref{tab_sxsy}, see for instance the term with argument $2 \, \Omega_2$ for Europa, already present in \cite{Arc18} after \cite{Lie79} (see his Eq.~18). 
Because of the truncation levels and of second-order terms, the number of terms for Europa, Ganymede and Callisto differ from the number of terms retained in Table~\ref{tab_sxsy} in the first place. For Io, all the second order terms are below the truncation level and each term in Table~\ref{tab_spinad} corresponds to the original term in Table~\ref{tab_sxsy}. 
For the $k^{th}$ satellite, the $\Delta\alpha_S$, $\Delta\delta_S$, and $\Delta\mu$ series are dominated by the $\Omega_k$ term.

We now compare in the time domain the approximate series of Table~\ref{tab_spinad} for $\alpha_S$ and $\delta_S$ to their exact transformation from Table~\ref{tab_sxsy} and Eqs.~(\ref{eq_sde})-(\ref{eq_sx2}). Depending on whether or not the second-order transformation terms of Eqs.~(\ref{eq_adapprox}) are used or not, the error is generally less than $0.0002^\circ$ or $0.02^\circ$, respectively, see Table~\ref{tab_error} for the Galilean satellites assumed to be entirely solid.
For $\mu = \beta - \psi$, the comparison between the approximate series of Table~\ref{tab_spinad} and the exact transformation of Eqs.~(\ref{eq_tanpsi}, \ref{eq_beta2}) indicates an accuracy generally better than $0.0002^\circ$ or $0.013^\circ$ for the second or first-order transformations, respectively, see Table~\ref{tab_error}. Compared with a purely numerical solution obtained from ephemerides, these transformation errors should be added to the truncation errors in the original series in $n/s$, see Table~\ref{tab_precision}, to obtain the total error budget. Neglecting secular terms would lead to typical errors of about $0.007^\circ$, $0.003^\circ$, and $0.006^\circ$ after $100$ years in $\alpha_S$, $\delta_S$, and $\mu$, respectively, larger than the combined truncation and second-order transformation errors.

Starting from the periodic series $\Delta\alpha_S$ and $\Delta\delta_S$ and Eqs.~(\ref{eq_thetaad}, \ref{eq_psiad}, \ref{eq_epsilonadd}, \ref{eq_offset2}), it is possible to obtain periodic series for the angles $\theta$, $\varepsilon$, $\psi$ and $\zeta$. We do not carry out this exercise here, as it would be very similar to the one based on the series $\Delta n_{x/y}$ and $\Delta s_{s/y}$, see Section~\ref{sec_obliquityseries}.

\begin{table}[!htb]
\small
\centering
\begin{tabular}{ccccccc}
\hline
Errors ($^\circ$) &  \multicolumn{2}{c}{$\alpha_S$} &
\multicolumn{2}{c}{$\delta_S$} & \multicolumn{2}{c}{$\mu$} \\
& 1st order & 2nd order & 1st order & 2nd order & 1st order & 2nd order \\
\hline
Io       & 0.0001 & 0.0001 & 0.0000 & 0.0000 & 0.0001 & 0.0000 \\
Europa   & 0.0131 & 0.0005 & 0.0056 & 0.0001 & 0.0132 & 0.0002 \\
Ganymede & 0.0038 & 0.0001 & 0.0015 & 0.0001 & 0.0037 & 0.00003 \\
Callisto & 0.0090 & 0.0003 & 0.0039 & 0.0001 &  0.0094 & 0.0001 \\
\hline
\end{tabular}
\caption{Maximal value of the errors of the first and second order estimations (in degree) for the four Galilean satellites for the spin right ascension $\alpha_S$, the declination $\delta_S$ and the angle $\mu$ for a solid satellite.}
\label{tab_error}
\end{table}

\section{Recommendations for updating the IAU WGCCRE solution}
\label{sec_IAUsolution}

The IAU Working Group on Cartographic Coordinates and Rotational Elements (WGCCRE) provides rotational elements of the planets and satellites, describing the direction of the North pole of rotation and the position of the prime meridian, see \cite{Arc18} for the last report. The Galilean satellites’ IAU WG solutions for $\alpha_S$ and $\delta_S$, here denoted $\alpha^{IAU}$ and $\delta^{IAU}$, have the same form as Eq.~(\ref{eq_adbt}), while the rotation angle $W$, here denoted $W^{IAU}$, is expressed as Eqs.~(\ref{eq_Wmu}), see Table~\ref{tab_IAU}.
The arguments of the periodic terms are listed as $J_j$ with $j=3,\dots,8$.
The mean values for $\alpha^{IAU}$ and $\delta^{IAU}$, which correspond to the Laplace plane coordinates, are similar for the four satellites, whereas their linear terms are identical. 

\begin{table}[!htb]
\small
\centering
\begin{tabular}{cccccccccc}
\hline
 & Cst ($^\circ$) & Linear term ($^\circ$) & $J_3$ & $J_4$ & $J_5$ & $J_6$ & $J_7$ & $J_8$ \\
\multicolumn{9}{l}{} \\
\hline
Period (y)    &&&  7.4216 & 30.2191 & 137.3522 & 559.8756 & 15.1095 & 5.9308 \\
$f_j$ (rad/y) &&& 0.84661 & 0.20792 & 0.045745 & 0.011222 & 0.41584 & 1.0594 \\
$f_j$ ($^\circ$/cent.) &&&  $4850.7$ & $1191.3$ & $262.1$ & $64.3$ & $2382.6$ & $6070.0$ \\
$\varphi_j$ ($^\circ$) &&&  283.90 & 355.80 & 119.90 & 229.80 & 352.25 & 113.35 \\
\multicolumn{9}{l}{} \\
\textbf{Io} &&&&&&&& \\
$\alpha^{IAU}$ & 268.05 & $-0.009 \, T$      &  0.094 &  0.024 & -- & -- & -- & -- \\
$\delta^{IAU}$ & 64.50  & $0.003 \, T$       &  0.040 &  0.011 & -- & -- & -- & -- \\
$W^{IAU}$      & 200.39 & $203.4889538 \, d$ & -0.085 & -0.022 & -- & -- & -- & -- \\
\textbf{Europa}  &&&&&&&& \\
$\alpha^{IAU}$ & 268.08 & $-0.009 \, T$      & -- &  1.086 &  0.060 &  0.015 &  0.009 & -- \\
$\delta^{IAU}$ & 64.51  & $0.003 \, T$       & -- &  0.468 &  0.026 &  0.007 &  0.002 & -- \\
$W^{IAU}$      & 36.022 & $101.3747235 \, d$ & -- & -0.980 & -0.054 & -0.014 & -0.008 & -- \\
\textbf{Ganymede}  &&&&&&&& \\
$\alpha^{IAU}$  & 268.20 & $-0.009 \, T$     & -- & -0.037 &  0.431 &  0.091 & -- & -- \\
$\delta^{IAU}$  & 64.57  & $0.003 \, T$      & -- & -0.016 &  0.186 &  0.039 & -- & -- \\
$W^{IAU}$       & 44.064 & $50.3176081 \, d$ & -- &  0.033 & -0.389 & -0.082 & -- & -- \\
\textbf{Callisto}  &&&&&&&& \\ 
$\alpha^{IAU}$  & 268.72 & $-0.009 \, T$     & -- & -- & -0.068 &  0.590 & -- &  0.010 \\
$\delta^{IAU}$  & 64.83  & $0.003 \, T$      & -- & -- & -0.029 &  0.254 & -- & -0.004 \\
$W^{IAU}$       & 259.51 & $21.5710715 \, d$ & -- & -- &  0.061 & -0.533 & -- & -0.009 \\
\hline
\end{tabular}
\caption{Rotation solution of the IAU WGCCRE for the four Galilean satellites \citep{Arc18}. The IAU WG arguments $J_j$ are expressed in the form $f_j \, T + \varphi_j$ in Eqs.~(\ref{eq_adbt}-\ref{eq_Wmu}), with $f_j$ and $\varphi_j$ the frequency and phase, respectively. $\alpha^{IAU}$ and $W^{IAU}$ are expressed as sine series whereas $\delta^{IAU}$ is expressed as a cosine series. 
The amplitudes are given in degrees and $T$ ($d$) is the time in Julian centuries (days) from J2000.}
\label{tab_IAU}
\end{table}

Since its very first report \citep{Dav80}, the WGCCRE has openly assumed that the North pole of rotation of the Galilean satellites lies along the normal to their orbit, neglecting their small orbital obliquities in the expression for $\alpha^{IAU}$ and $\delta^{IAU}$. They also neglect the mean longitude variations and small librations in the periodic series of $W^{IAU}$. 
It can be indeed verified numerically that in the IAU WG solution, $\Delta W$ consists only in $\Delta\mu$, a projection of the $\Delta\alpha$ series, and assumes $\gamma_u = \Delta\mathcal{L} + \gamma_f = 0$, see Eqs.~(\ref{eq_muad} and \ref{eq_DWgamma}).
In addition to neglecting orbital obliquity and librations, the IAU WG model is also outdated. In particular, the periodic series based on Lieske ephemerides \citep{Lie79} have remained unchanged ever since. 

Pending an actual measurement of orientation and rotation angles, and with some simplifications as detailed below, the solution obtained in Section~\ref{sec_admseries} above assuming that the Galilean satellites are entirely solid and are locked in the multi-frequency Cassini state with non-zero obliquities and considering a recent orbital theory would be a perfectly acceptable replacement of the current IAU WG solution. The only thing we do not calculate here is the value of the libration $\gamma_u$. Libration computations can be found in \cite{Ram11} or \cite{VanH13}. 
The periodic variations $\Delta \mu$ that are included in the rotation angle $W$ are the periodic terms of the transformation from $\phi_{Inertial}$, see Eq.~(\ref{eq_Wphiinert}), and therefore does not include $\gamma_u$.

\begin{table}[!htb]
\footnotesize
\centering
\vspace*{-1.cm}
\begin{tabular}{l@{\,}c@{\,}@{\,}c@{\,}@{\,}c@{\,}@{\,}c@{\,}@{\,}c@{\,}@{\,}c@{\,}@{\,}c@{\,}@{\,}c@{\,}@{\,}c@{\,}@{\,}c@{\,}@{\,}c@{\,}@{\,}c@{\,}@{\,}c@{\,}}
\hline
& Cst ($^\circ$) & Linear term ($^\circ$) & $\mathcal{J}_3$ & $\mathcal{J}_4$ & $\mathcal{J}_5$ & $\mathcal{J}_6$ & $\mathcal{J}_7$ & $\mathcal{J}_8$ & $\mathcal{J}_9$ \\
\hline
 & & & $\Omega_1$ & $\Omega_2$ & $\Omega_3$ & $\Omega_{3}$ & $\Omega_4$ & $\Omega_{4}$ & $-2 \nu -\Omega_2$ \\ 
Period (y) & & & -7.4217 & -30.200 & -137.34 & -137.34 & -561.00 & -561.00 & -0.68144  \\
$f_j$ (rad/y) & & & -0.84660 & -0.20805 & -0.04575 & -0.04575 & -0.01120 & -0.01120 & -9.22041 \\
$f_j$ ($^\circ$/cent.) & & & -4850.7 & -1192.0 & -262.13 & -262.13 & -64.171 & -64.171 & -52829.1 \\
$\varphi_j$ ($^\circ$) & & & 260.604 & 184.073 & 59.659 & 59.182 & 309.668 & 310.157 & 273.914 \\
\multicolumn{10}{l}{} \\
\textbf{Io} &&&&&&&&& \\
$\alpha_S$   & 268.0594 & -0.0065 $T$          &  0.0878 &  0.0233 &  0.0030 & - & -0.0008 & - &  0.0023 \\
$\delta_S$   &  64.4968 &  0.0024 $T$          & -0.0378 & -0.0100 & -0.0013 & - &  0.0003 & - & -0.0010 \\
$W-\gamma_u$ & 200.3980 & $203.488958424 \, d$ & -0.0793 & -0.0210 & -0.0027 & - &  0.0007 & - & -0.0021 \\
\textbf{Europa} &&&&&&&&& \\
$\alpha_S$   & 268.0850 & -0.0065 $T$          & -0.0049 &  1.2097 &  0.0595 & - &  0.0131 & - &  0.0007 \\
$\delta_S$   &  64.5070 &  0.0024 $T$          &  0.0021 & -0.5206 & -0.0256 & - & -0.0056 & - & -0.0003 \\
$W-\gamma_u$ & 35.9172 & $101.374724491 \, d$ &  0.0044 & -1.0920 & -0.0537 & - & -0.0118 & - & -0.0007 \\
\textbf{Ganymede} &&&&&&&&& \\
$\alpha_S$   & 268.2044 & -0.0065 $T$         & - & -0.1132 &  0.5066 & - & - &  0.0922 & - \\
$\delta_S$   &  64.5678 &  0.0024 $T$         & - &  0.0486 & -0.2176 & - & - & -0.0396 & - \\
$W-\gamma_u$ & 44.1652 & $50.317607524 \, d$ & - &  0.1022 & -0.4575 & - & - & -0.0833 & - \\
\textbf{Callisto} &&&&&&&&& \\
$\alpha_S$   & 268.7322 & -0.0065 $T$         & - & - & - &  0.1470 &  0.9253 & - & - \\
$\delta_S$   &  64.8229 &  0.0024 $T$         & - & - & - & -0.0626 & -0.3936 & - & - \\
$W-\gamma_u$ & 259.7839 & $21.571072373 \, d$ & - & - & - & -0.1331 & -0.8374 & - & - \\
\multicolumn{10}{l}{} \\
\hline
& & & $\mathcal{J}_{10}$ & $\mathcal{J}_{11}$ & $\mathcal{J}_{12}$ & $\mathcal{J}_{13}$ & $\mathcal{J}_{14}$ & $\mathcal{J}_{15}$ & $\mathcal{J}_{16}$ \\
\hline
 & & & $2 L_S$ & $2 \Omega_2$ & $-L_S$ & $\Omega_2 + \Omega_3$ & $2 \Omega_3$ & $\Omega_3 + \Omega_4$ & $2 \Omega_4$ \\ 
Period (y) & & & 5.9312 & -15.101 & -11.863 & -24.756 & -68.661 & -110.31 & -280.37 \\
$f_j$ (rad/y) & & & 1.05934 & -0.41609 & -0.52963 & -0.25380 & -0.09151 & -0.05696 & -0.02241 \\
$f_j$ ($^\circ$/cent.) & & & 6069.6 & -2384.0 & -3034.6 & -1454.2 & -524.31 & -326.36 & -128.40 \\
$\varphi_j$ ($^\circ$) & & & 113.500 & 8.146 & 298.879 & 243.626 & 119.318 & 8.850 & 259.337 \\
\multicolumn{10}{l}{} \\
\textbf{Io} &&&&&&&&& \\
$\alpha_S$   & & & - & - & - & - & - & - & - \\
$\delta_S$   & & & - & - & - & - & - & - & - \\
$W-\gamma_u$ & & & - & - & - & - & - & - & - \\
\textbf{Europa} &&&&&&&&& \\
$\alpha_S$   & & &  0.0013 & -0.0115 &  0.0007 & -0.0011 & - & - & - \\
$\delta_S$   & & & -0.0005 &  0.0025 & -0.0003 &  0.0002 & - & - & - \\
$W-\gamma_u$ & & & -0.0011 &  0.0116 & -0.0006 &  0.0011 & - & - & - \\
\textbf{Ganymede} &&&&&&&&& \\
$\alpha_S$   & & &  0.0010 & - & -0.0010 &  0.0009 & -0.0020 & - & - \\
$\delta_S$   & & & -0.0004 & - &  0.0004 & -0.0002 &  0.0004 & - & - \\
$W-\gamma_u$ & & & -0.0009 & - &  0.0009 & -0.0009 &  0.0020 & - & - \\
\textbf{Callisto} &&&&&&&&& \\
$\alpha_S$   & & &  - & - & - & - & - & -0.0022 & -0.0068 \\
$\delta_S$   & & &  - & - & - & - & - &  0.0005 &  0.0014 \\
$W-\gamma_u$ & & &  - & - & - & - & - &  0.0022 &  0.0068 \\
\hline
\end{tabular}
\tiny
\caption{Proposed update for the solution ($\alpha_S$, $\delta_S$ and $W$) of the Galilean satellites considered as entirely rigid and using the JUP387 orbital theory, with a limited number of terms ($5$, $9$, $7$, and $4$, respectively) and a limited set of arguments ($14$). The arguments $\mathcal{J}_j$ are expressed in the form $f_j \, T + \varphi_j$ in Eqs.~(\ref{eq_adbt}-\ref{eq_Wmu}), with $f_j$ and $\varphi_j$ the frequency and phase, respectively. $\alpha_{S}$ and $W_{S}$ are expressed as sine series whereas $\delta_{S}$ is expressed as a cosine series. The amplitudes are given in degrees and $T$ ($d$) is the time in Julian centuries (days) from J2000.
The column header shows the identification to the fundamental arguments, using the reduced set of arguments $\mathcal{J_j}$, whose physical origin is recalled on the first line. Compared to the full solution of Table~\ref{tab_spinad} for the periodic terms, the precision is degraded to about $0.001^\circ$. 
Our model does not include any physical libration ($\gamma_u=0$).}
\label{tab_spinad_IAU}
\end{table}

The Earth equatorial coordinates $\alpha_{LP}$ and $\delta_{LP}$ of the Laplace planes of our solution are close to those of the IAU WG solutions (they differ at most by $0.01^\circ$ and $0.005^\circ$, respectively), but more precise, see Table~\ref{tab_refplanes}. We therefore recommend updating the mean values of $\alpha^{IAU}$ and $\delta^{IAU}$ with $\alpha_{LP}$ and $\delta_{LP}+\delta_{S\,0}
$ as given in Tables~\ref{tab_refplanes} and \ref{tab_trxtry}, using at least three digits after the decimal point to take advantage of the improved precision achieved by recent ephemerides. The mean value for $\delta$ should also include the small increment $\delta_{S\,0}$, see Eq.~(\ref{eq_deltat}), because of the power reduction of a squared sine term in the second-order part of Eq.~(\ref{eq_deltatransfo}).
Computing a new independent $W_0$ requires control network computations which are beyond the scope of the present study. Here, the numerical value of $W_0$ is chosen to ensure that our defined prime meridian location ($ W + \alpha_S \sin\delta_S$, where the $\alpha_S \sin\delta_S$ term accounts for the equator precession) matches that of the IAU WG model at the J2000 epoch. 
The corresponding corrections are $-0.25$, $2.8$, $-4.6$, and $-11.5$ km, respectively.

Originally, the IAU WG adopted the Jovian $\alpha$ and $\delta$ rates ($-9 \times 10^{-5}$ deg/y and $3 \times 10^{-5}$ deg/y in $\alpha$ and $\delta$, respectively) from \cite{Lie79} and used them as the rates for the Galilean satellites as well. It was not until the 2005 report (see \citealt{Sei07}) that the Jovian rates were updated, but the IAU WG did not update those of the satellites at the same time.
Here, with the JUP387 ephemerides, we find rates in $\alpha_S$ and $\delta_S$ (e.g.~$-6.5 \times 10^{-5}$ deg/y  and $2.4 \times 10^{-5}$ deg/y for Io, see Table~\ref{tab_trxtry}) that are in good agreement with the rates in Jupiter's equatorial coordinates of the current IAU WG solution ($-6.499 \times 10^{-5}$ deg/y and $2.413 \times 10^{-5}$ deg/y). We therefore recommend updating the $\alpha$ and $\delta$ IAU WG rates for the Galilean satellites to the currently adopted rate for Jupiter.
For the rates in $W$, those of Table~\ref{tab_trxtry}, based on recent ephemerides, could replace the IAU WG rates. 

Each term in the periodic series of the IAU WG $(\alpha, \delta, W)$ solution can be replaced by the equivalent term of our solution for a solid satellite, see Table~\ref{tab_spinad}, which has (almost) the same period as the IAU WG term, but a different amplitude, which, as we will see later, ultimately leads to the neglect of one of the replaced terms (the term for Callisto at period 5.93 years).
Thanks to the use of more recent and precise ephemeris, the second-order part of the transformation to $\alpha$ and $\delta$, and the non-zero obliquity, we find several additional terms that can be considered for an update of the IAU WG solution, depending on the precision required. For Callisto, three of these additional terms have amplitudes larger than the sixth term of the series, which corresponds to the third term of the IAU WG series (the very one we will be neglecting below). 
Note that when considering different interior structures for a satellite with a global ocean (see for example section~\ref{sec_liquid}), some amplitudes may be significantly different from their solid counterparts if the forcing frequency is close enough to one of the free mode frequencies, so that the order of the different terms may differ.

The IAU WG arguments are characterized by positive rates while in our solution (Table~\ref{tab_spinad}), the rates of the arguments are mostly negative, which makes more sense from a dynamic point of view (precessions are usually retrograde motions). Throughout our successive developments, from ephemerides to $\Delta\alpha_S, \Delta\delta_S$ and $\Delta\mu$ series, we have kept, at least for the first-order terms of the transformation, signs of rates that correspond to the convention used for the periodic series of Eq.~(\ref{eq_nxnyt}) for $n_{x/y}$, where a retrograde precession of the orbital node is characterized by a negative rate. This convention allows for the use of the dynamical Eq.~(\ref{eq_eps}) with $\dot\Omega=f_j$ to get the amplitudes (Eq.~\ref{eq_thetaj}) of the periodic series of Eq.~(\ref{eq_sxsy}) for $s_{x/y}$. In this way, we can easily compare the forced frequencies $f_j$ with the $\kappa/C$ free frequency to identify the potential for resonant amplification.
With this convention, the phases remain unchanged by the transformation from $\Delta s_x/\Delta s_y$ to $\Delta\alpha_S/\Delta\delta_S/\Delta\mu$ and, as a result, differ from those of the IAU WG arguments, while
the different terms of $\Delta\alpha_S$ and $\Delta\delta_S$ have amplitude of opposite signs for the same argument, when the signs are the same in the IAU WG series.

\begin{figure}[htb!]
\centering
\includegraphics[width=13cm]{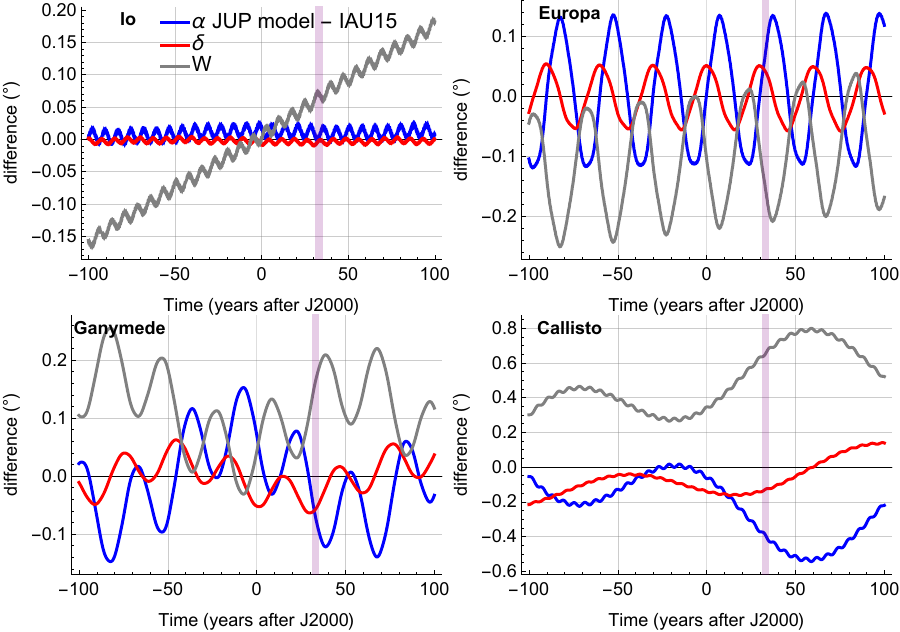}
\caption{Differences in right ascension, declination, and rotation angle between our model and the IAU WG solution \citep{Arc18} for each of the four Galilean satellites. The purple box represents the timing of the JUICE mission.}
\label{fig_diff}
\end{figure}

The current IAU WG solution includes a limited number of terms and is provided with a presumed accuracy of $0.01^\circ$ on each nutation amplitudes (we assumed the precision is one order higher than the given number of digits). Our solution is by construction more accurate (to within about $0.001^\circ$ degrees for the time-series in $\alpha$, due to the truncation error of Table~\ref{tab_spinad}), but is based on the assumption that the satellites are entirely rigid. Therefore, it is not certain that, for the time being, an update of the IAU WG solution should include as many terms as ours. If a simplified model with an accuracy of the order of $0.001^\circ$ on each nutation term is considered sufficient for the four satellites, some simplifications can be made to the periodic series of Table~\ref{tab_spinad}.
Firstly, we propose to retain only the $5$, $9$, $7$, and $4$ largest terms, respectively, for the four satellites. 
Thus, the $5.93-$year periodicity reported for Callisto in \cite{Arc18} and also observed in Table~\ref{tab_sxsy} (third line) is neglected here (its period is much shorter than the free precession period of approximately $204$ years, so that the associated inertial obliquity amplitude is significantly smaller than the inclination amplitude).
Next, we suggest reducing the number of arguments used to express the solution.
Remember that some arguments listed in Table~\ref{tab_spinad} correspond to identical combinations of fundamental frequencies but may differ slightly in value between satellites due to fitting to numerical integrations (e.g., $\dot\Omega_4=-0.01113$ rad/year in the Io series, versus $-0.01120$ rad/year in the Callisto series). When such differences do not degrade the model's accuracy beyond the $0.0005^\circ$ threshold, we use a common value for the argument across satellites.
For example, for $\Omega_1$, which dominates the periodic series of Io, but is only fifth in the series for Europa, we choose the value of the argument as found Io's series, see Table~\ref{tab_spinad_IAU}.
The differences from the original arguments are usually smaller than $10^{-4}$~rad/year and $0.5^{\circ}$ in frequency and phase, respectively.
Only differences in phase value can cause differences in the time domain that exceed the threshold.
The precision of the time series obtained using $14$ different arguments rather than the $25$ arguments of Table~\ref{tab_spinad}
is smaller than $0.0002^\circ$. 
The proposed update for the IAU WG solution, with the simplified periodic series, is given in Table~\ref{tab_spinad_IAU}. There is no physical libration in the expression for $W$, only periodic variations due to the projection of the spin precession/nutations.
The differences between our model based on the JUP orbital theory and the IAU WG solution of \cite{Arc18} can reach up to $0.4^\circ$ after $100$ years, see Fig.~\ref{fig_diff}.

\section{Discussion}
\label{sec_discussion}

\subsection{Sensitivity to the choice of ephemeris}
\label{sec_orbcomp}

\begin{figure}[htb!]
\centering
\includegraphics[width=13cm]{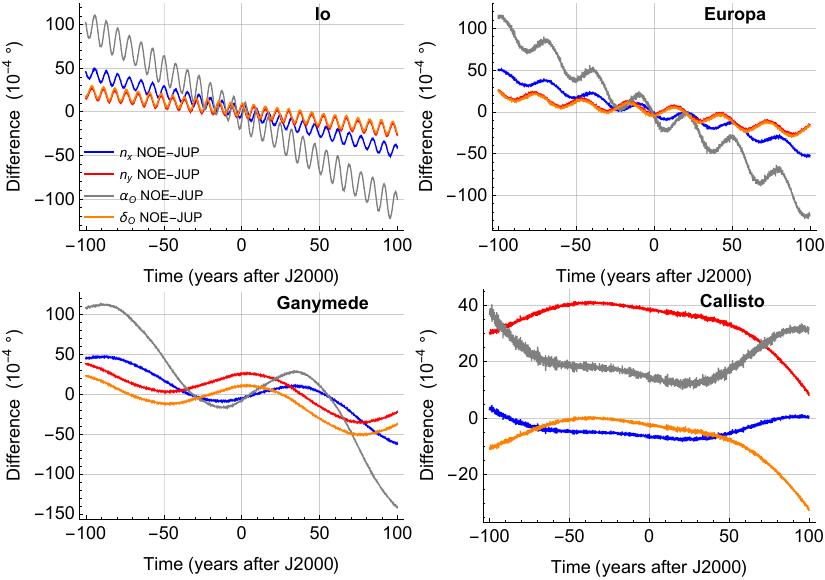}
\caption{Differences in orbit coordinates $n_{x,y}$ and orbital right ascension $\alpha_O$ and declination $\delta_O$ between the JUP387 and NOE ephemerides as a function of time, for the four Galilean satellites.}
\label{fig_NOEJUP}
\end{figure}

\begin{table}[!htb]
\small
\centering
\begin{tabular}{cccccc}
\hline\noalign{\smallskip}
 & & Io & Europa & Ganymede & Callisto \\
\hline
\multicolumn{5}{l}{} \\
Laplace pole: & & & & \\
$\alpha_{LP}$ ($^\circ$) & 268.0590 & 268.0850 & 268.2049 & 268.7352 \\
& (+0.0004) & (+0.0001) & (-0.0005) & (-0.0030) \\
$\delta_{LP}$ ($^\circ$) &  64.4970 &  64.5093 &  64.5668 &  64.8203 \\ 
& (-0.0002) & (+0.0002) & (0.0015) & (+0.0041) \\
\multicolumn{5}{l}{} \\
Transformation coefficients: & & & & & \\
$1/\cos\delta_{LP}$                    & $\mathcal{O}1$ &  2.3226 &  2.3236 &  2.3285 &  2.3504 \\
$\tan \delta_{LP} \, /\cos\delta_{LP}$ & $\mathcal{O}2$ &  4.8687 &  4.8736 &  4.8965 &  4.9995 \\
$- 1/2 \, \tan \delta_{LP}$            & $\mathcal{O}2$ & -1.0481 & -1.0487 & -1.0514 & -1.0635 \\
\multicolumn{5}{l}{} \\
$\cos\delta_{LP}$                      & $\mathcal{O}1$ &  0.4306 &  0.4304 &  0.4295 &  0.4255 \\
$-\sin\delta_{LP}$                     & $\mathcal{O}2$ & -0.9026 & -0.9027 & -0.9031 & -0.9050 \\
$ 1/4 \sin 2\delta_{LP}$               & $\mathcal{O}2$ &  0.1943 &  0.1942 &  0.1939 &  0.1925 \\
\hline
\end{tabular} \\
\caption{Numerical values for the orientation angles of the Laplace plane for the four Galilean satellites and transformation coefficients, using the NOE ephemerides. The values in parentheses for $\alpha_{LP}$ and $\delta_{LP}$ indicate the increments needed to obtain the JUP387 values, see Table~\ref{tab_refplanes}.}
\label{tab_refplanesNOE}
\end{table}

\begin{table}[!htb]
  \small
  \centering
  \begin{tabular}{ccccccccc}
  \hline
  & $tr_x$ ($^\circ$/y) & $tr_y = tr_\delta$ ($^\circ$/y) & $tr_\alpha$ ($^\circ$/y) & $tr_\mu$ ($^\circ$/y) & $\dot W$ ($^\circ$/d) & $\delta_{S 0}$ ($^\circ$) \\
  \hline
  Io       & $-7.3 \times 10^{-5}$ & $2.8 \times 10^{-6}$ & $-1.7 \times 10^{-4}$ & $1.5 \times 10^{-4}$ & 203.4889584317 & $-1.4 \times 10^{-4}$ \\
  Europa   & $-7.3 \times 10^{-5}$ & $3.0 \times 10^{-6}$ & $-1.7 \times 10^{-4}$ & $1.5 \times 10^{-4}$ & 101.3747246256 & $-2.5 \times 10^{-3}$ \\
  Ganymede & $-7.2 \times 10^{-5}$ & $4.1 \times 10^{-6}$ & $-1.7 \times 10^{-4}$ & $1.5 \times 10^{-4}$ &  50.3176077039 & $-4.6 \times 10^{-4}$ \\
  Callisto & $-6.7 \times 10^{-5}$ & $8.6 \times 10^{-6}$ & $-1.6 \times 10^{-4}$ & $1.4 \times 10^{-4}$ &  21.5710726895 & $-1.4 \times 10^{-3}$ \\
  \hline
  \end{tabular} 
  \caption{Trends in the various quantities considered, obtained for NOE orbital theory. See Table~\ref{tab_trxtry} for more details on the definition of the quantities.}
  \label{tab_trxtryNOE}
  \end{table}

A comparison between the orbit orientations of the Galilean satellites obtained with JUP387 or NOE reveals differences, on both short and long time scales, see Fig.~\ref{fig_NOEJUP}. 
As with JUP387, we numerically fit the Laplace plane coordinates ($\alpha_{LP},\delta_{LP}$), a periodic series and one secular term for the projected coordinates ($n_x,n_y$) of the orbit normal onto the Laplace plane. The Laplace plane coordinates differ by as much as $0.004^\circ$ from those associated to JUP387, which also affects some of the transformation coefficients, see Table~\ref{tab_refplanesNOE}.
The small secular rates ($tr_{x/y}$) associated to the two ephemerides differ significantly, see Tables~\ref{tab_trxtry} and \ref{tab_trxtryNOE}, with discrepancies in $n_{x/y}$ reaching up to $0.005^\circ$ after $100$ years, corresponding to a difference up to $150$ km in the position of the Galilean satellite around Jupiter. 
The differences in the amplitudes of the $n_{x/y}$  
series between JUP387 and NOE range from approximately $3 \times 10^{-4}$ deg (for Io) to $2 \times 10^{-3}$ deg (for Callisto), see Tables~\ref{tab_sxsy} and \ref{tab_sxsyNOE}.
The differences in the periods are the largest for term $j=5$ of Io’s series ($7$ years of difference). This term, with argument $\Omega_4$, has a small amplitude and a long period and is therefore more difficult to estimate uniquely than a term with a large amplitude and short period. For the other satellites, the $\Omega_4$ terms are relatively more important, and their period is generally better constrained to around $-560$ years.
Note that for each satellite, since the Laplace planes of JUP387 and NOE are different, the $n_{x/y}$ coordinates corresponding to each ephemerides are expressed in a different reference frame. The transformation into equatorial coordinates $\alpha_O$ and $\delta_O$ allows for a comparison of the orbit orientation in an identical reference frame (see gray and orange curves in Fig.~\ref{fig_NOEJUP}).

\begin{table}[!htb]
\small
\centering
\begin{tabular}{ccccccccccc}
\hline
 & $i_j$ ($^\circ$) & $\varepsilon_j$ ($^\circ$) & $\theta_j$ ($^\circ$) & period (y) & $f_j$ (rad/y) & $\varphi_j$ ($^\circ$) & Argument & $\dot{J}_{IAU}$ \\
\hline
{\bf Io}   &&&&&&&& \\
1 &  0.0361 & 0.0020 &  0.0382 & -7.42169 & -0.84660 & 259.686 & $\Omega_1$  & $-\dot{J}_3$ \\
2 &  0.0099 & 0.0001 &  0.0100 & -30.2013 & -0.20804 & 184.127 & $\Omega_2$ & $-\dot{J}_4$ \\
3 &  0.0013 & 0.0000 &  0.0013 & -137.748 & -0.04561 &  60.350 & $\Omega_3$ & $-\dot{J}_5$ \\
4 &  0.0004 & 0.0006 &  0.0010 & -0.68144 & -9.22041 & 273.900 & $-2\nu-\Omega_2$ & \\
5 & -0.0003 & 0.0000 & -0.0003 & -557.341 & -0.01127 & 312.374 & $\Omega_4$ & $-\dot{J}_6$ \\
6 &  0.0003 & 0.0000 &  0.0003 &  5.93112 &  1.05936 & 113.576 & $2L_S$ & $\dot{J}_8$ \\
7 & -0.0001 & 0.0000 & -0.0001 & -11.8632 & -0.52963 & 298.256 & $-L_S$ & $-\dot{J}_8/2$ \\
\multicolumn{2}{l}{{\bf Europa}}  &&&&&&& \\
1 &  0.4645 &  0.0554 &  0.5200 & -30.2013 & -0.208084 & 184.097 & $\Omega_2$ & $-\dot{J}_4$ \\
2 &  0.0249 &  0.0006 &  0.0254 & -137.543 & -0.04568 &  60.120 & $\Omega_3$ & $-\dot{J}_5$ \\
3 &  0.0055 &  0.0000 &  0.0055 & -560.525 & -0.01121 & 309.300 & $\Omega_4$ & $-\dot{J}_6$ \\
4 & -0.0012 & -0.0009 & -0.0021 & -7.42169 & -0.84660 & 259.662 & $\Omega_1$ & $-\dot{J}_3$ \\
5 & -0.0012 &  0.0015 &  0.0003 & -0.68144 & -9.22041 & 273.890 & $-2\nu-\Omega_2$ & \\
6 &  0.0008 &  0.0003 &  0.0005 &  5.93117 &  1.05935 & 113.503 & $2L_S$ & $\dot{J}_8$ \\
7 &  0.0002 & -0.0002 &  0.0000 & -0.66965 & -9.38278 &  37.839 & $-2\nu-\Omega_3$ & \\
8 &  0.0002 &  0.0001 &  0.0003 & -11.8636 & -0.52962 & 299.287 & $-L_S$ & $-\dot{J}_8/2$ \\
9 & -0.0001 &  0.0000 & -0.0001 &   4.9576 &  1.26738 &  68.087 & $2L_S-\Omega_2+\Omega_0$ & \\
\multicolumn{2}{l}{{\bf Ganymede}} &&&&&&& \\
1 &  0.1845 &  0.0313 &  0.2158 & -137.532 & -0.04569 &  59.999 & $\Omega_3$ & $-\dot{J}_5$ \\
2 &  0.0376 &  0.0014 &  0.0390 & -560.304 & -0.01121 & 309.329 & $\Omega_4$ & $-\dot{J}_6$ \\
3 & -0.0165 & -0.0320 & -0.0485 & -30.2014 & -0.20804 & 183.990 & $\Omega_2$ & $-\dot{J}_4$ \\
4 &  0.0018 & -0.0014 &  0.0004 &  5.93126 &  1.05933 & 113.428 & $2L_S$ & $\dot{J}_8$ \\
5 &  0.0003 & -0.0007 & -0.0004 & -11.8629 & -0.52965 & 298.928 & $-L_S$ & $-\dot{J}_8/2$ \\
6 &  0.0002 & -0.0002 &  0.0000 & -0.68144 & -9.22041 & 273.804 & $-2\nu-\Omega_2$ & \\
7 &  0.0002 & -0.0002 &  0.0000 &  3.95426 &  1.58897 & 132.963 & $3 L_S$ & $3\dot{J}_8/2$ \\
8 &  0.0002 & -0.0001 &  0.0001 &  11.8637 &  0.52961 & 178.089 & $L_S$ & $\dot{J}_8/2$ \\
9 & -0.0001 &  0.0001 &  0.0000 &   5.6858 &  1.10507 & 191.565 & $2L_S-2\Omega_3 +\Omega_0$ & \\
\multicolumn{2}{l}{{\bf Callisto}}   &&&&&&& \\
1 &  0.2486 &  0.1421 & 0.3907 & -559.98  & -0.01122 & 308.856 & $\Omega_4$ & $-\dot{J}_6$ \\
2 & -0.0301 &  0.0927 & 0.0626 & -137.535 & -0.04568 &  59.519 & $\Omega_3$ & $-\dot{J}_5$ \\
3 &  0.0038 & -0.0037 & 0.0001 & 5.93109  &  1.05936 & 112.748 & $2L_S$ & $\dot{J}_8$ \\
4 & -0.0006 &  0.0007 & 0.0001 & -30.2024 & -0.20804 & 183.563 & $\Omega_2$ & $-\dot{J}_4$ \\
5 &  0.0006 & -0.0006 & 0.0000 & -11.8638 & -0.52961 & 298.699 & $-L_S$ & $-\dot{J}_8/2$ \\
6 &  0.0005 & -0.0005 & 0.0000 &  11.8645 &  0.52958 & 177.598 & $L_S$ & $\dot{J}_8/2$ \\	
7 &  0.0004 & -0.0004 & 0.0000 &  3.9543  &  1.58895 & 132.474 & $3 L_S$ & $3\dot{J}_8/2$ \\
8 & -0.0003 &  0.0003 & 0.0000 &  5.86882 &  1.07060 & 302.232 & $2L_S - \Omega_4 + \Omega_0$ & \\
\hline
\end{tabular} 
\caption{Inclination, orbital and inertial obliquity amplitudes $i_j$, $\varepsilon_j$ and $\theta_j$ at the frequency and phase $f_j$ and $\varphi_j$ obtained from the NOE orbital theory and assuming that the satellites are entirely solid and rigid. The phases $\varphi_j$ are given with respect to J2000 epoch. 
The last columns show the identification to the fundamental arguments of \cite{Lai06} and the fundamental IAU WG frequencies $\dot{J}_i$.} 
\label{tab_sxsyNOE}
\end{table}

\begin{figure}[htb!]
\centering
\includegraphics[height=8cm]{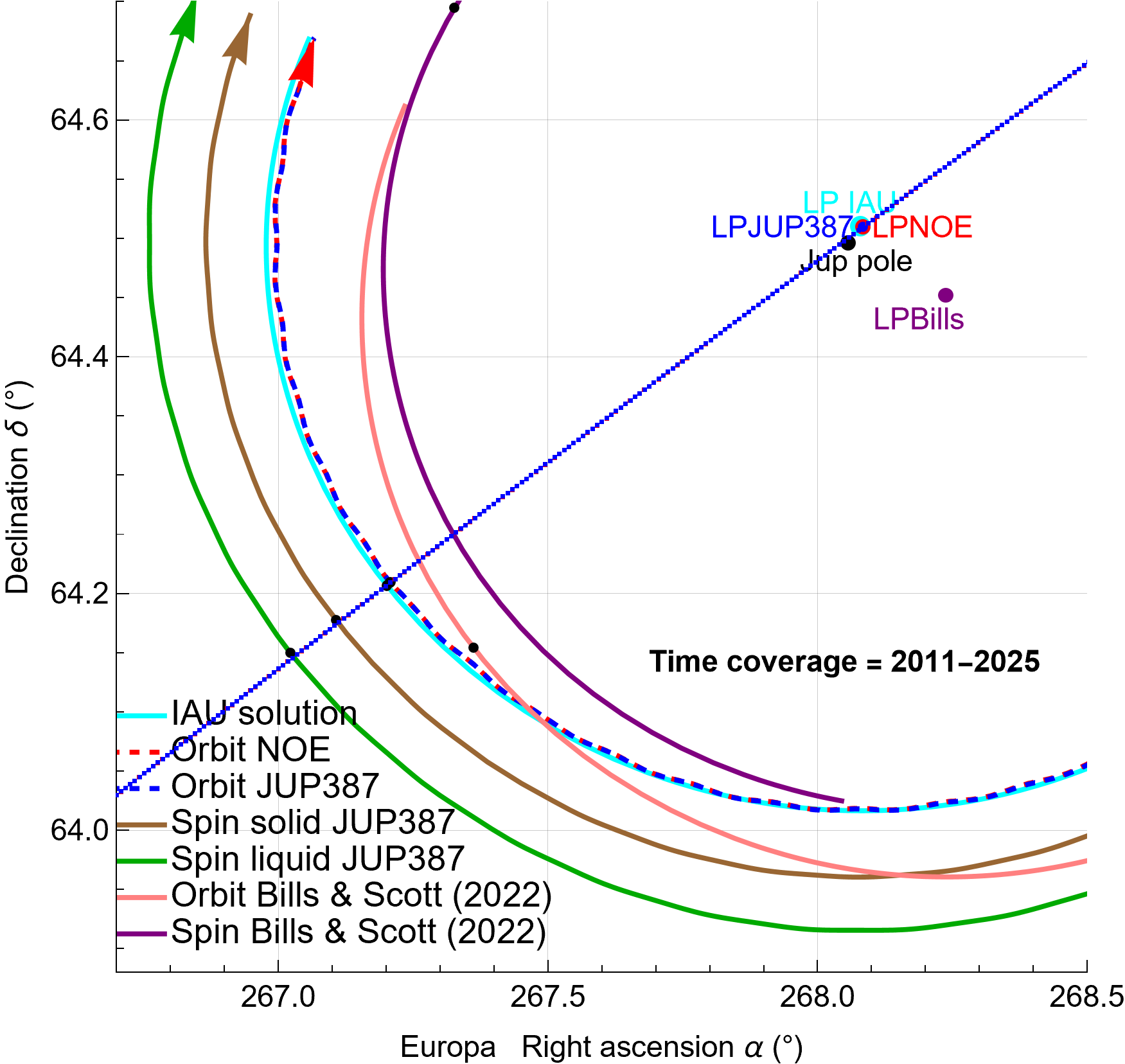}
\caption{Comparison of the orbit and the spin position (for a solid satellite and one with a liquid layer) in the right ascension/declination plane for Europa.
Three orbital theories are compared: JUP387 (blue), NOE (red, below the JUP387 curve), IAU WG solution (cyan). The spin position is plotted in a solid case (brown) and a liquid case (green, obliquity of 0.1 deg).
The time range covered by the plot is 2011-2025.
The black dots are evaluated in 2020.0.
The instantaneous Cassini plane is also plotted with dotted lines. 
The Laplace plane associated with each orbit is given in the same color as the orbit.} 
\label{fig_changeOrb}
\end{figure}

\begin{table}[!htb]
\small
\centering
\begin{tabular}{lcccccccc}
\hline
 & $\alpha^S_j$ ($^\circ$) & $\delta^S_j$ ($^\circ$) & $\mu_j$ ($^\circ$) & period (y) & $f_j$ (rad/y) & $\varphi_j$ ($^\circ$) & Argument & $\dot{J}_{IAU}$ \\
\hline
\textbf{Io} &&&&&&&& \\
1 &  0.0886 & -0.0382 & -0.0800 & -7.42169 & -0.8466 & 259.686 & $\Omega_1$ & ${\bf -\dot{J}_3}$ \\
2 &  0.0233 & -0.0100 & -0.0210 & -30.2013 & -0.2080 & 184.127 & $\Omega_2$ & ${\bf -\dot{J}_4}$ \\
3 &  0.0029 & -0.0013 & -0.0027 & -137.748 & -0.0456 &  60.350 & $\Omega_3$ & $-\dot{J}_5$ \\
4 &  0.0023 & -0.0010 & -0.0020 & -0.68144 & -9.2204 & 273.900 & $-2\nu-\Omega_2$ & \\
5 & -0.0008 &  0.0003 &  0.0007 & -557.341 & -0.0113 & 312.374 & $\Omega_4$ & $-\dot{J}_6$ \\
6 &  0.0006 & -0.0003 & -0.0005 &   5.9311 &  1.0594 & 113.576 & $2L_S$ & $\dot{J}_8$ \\
7 & -0.0002 &  0.0001 &  0.0002 & -11.8632 & -0.5296 & 298.235 & $-L_S$ & $-\dot{J}_8/2$ \\
\textbf{Europa} &&&&&&&& \\
1 &  1.2082 & -0.5200 & -1.0906 & -30.2013 & -0.2080 & 184.097 & $\Omega_2$ & ${\bf -\dot{J}_4}$ \\
2 &  0.0591 & -0.0254 & -0.0534 & -137.543 & -0.0457 & 60.1204 & $\Omega_3$ & ${\bf -\dot{J}_5}$ \\
3 &  0.0129 & -0.0055 & -0.0116 & -560.525 & -0.0112 & 309.299 & $\Omega_4$ & ${\bf -\dot{J}_6}$ \\
4 ($\mathcal{O}2$) & -0.0115 & 0.0025 & 0.0116 & -15.1006 & -0.4161 & 8.194 & $2\Omega_2$ & ${\bf -\dot{J}_7} = -2 \dot{J}_4$ \\
5 & -0.0049 &  0.0021 &  0.0044 & -7.4269 & -0.8466 & 259.658 & $\Omega_1$ & $-\dot{J}_3$ \\
6 &  0.0013 & -0.0005 & -0.0011 &  5.93117 & 1.0594 & 113.496 & $2L_S$ & $\dot{J}_8$ \\
7 &  0.0007 & -0.0003 & -0.0007 & -0.68144 & -9.2204 & 273.892 & $-2\nu-\Omega_2$ & \\
8 &  0.0007 & -0.0003 & -0.0006 & -11.8636 & -0.5296 & 299.274 & $-L_S$ & $-\dot{J}_8/2$ \\
9 ($\mathcal{O}2$) & 0.0000 & -0.0002 & 0.0000 & -38.6985 & -0.1624 & 123.977 & $\Omega_2 - \Omega_3$ & $-\dot{J}_4 + \dot{J}_5$ \\
10 ($\mathcal{O}2$) & -0.0011 & 0.0002 & 0.0011 & -24.7637 & -0.2537 & 244.217 & $\Omega_2 + \Omega_3$ & $-\dot{J}_4 - \dot{J}_5$ \\
\textbf{Ganymede} &&&&&&&& \\
1 &  0.5025 & -0.2158 & -0.4538 & -137.532  & -0.04569 &  59.999 & $\Omega_3$ & ${\bf -\dot{J}_5}$ \\
2 & -0.1129 &  0.0485 &  0.1019 &  -30.2014 & -0.20804 & 183.990 & $\Omega_2$ & ${\bf -\dot{J}_4}$ \\
3 &  0.0907 & -0.0390 & -0.0819 & -560.304 & -0.01121 & 309.328 & $\Omega_4$ & ${\bf -\dot{J}_6}$ \\
4 ($\mathcal{O}2$) & -0.0020 &  0.0004 &  0.0020 &  -68.766   & -0.09137 & 119.999 & $2 \Omega_3$ & $-2 \dot{J}_5$ \\
5 & -0.0010 &  0.0004 &  0.0009 &  -11.8629  & -0.52965 & 298.928 & $-L_S$ & $-\dot{J}_8$/2 \\
6 &  0.0010 & -0.0004 & -0.0009 &    5.93126 &  1.05933 & 113.428 & $2L_S$ & $\dot{J}_8$ \\
7 ($\mathcal{O}2$)  &  0.0000 &  0.0002 &  0.0000 &  -38.6997 & -0.16236 & 123.991 & $ \Omega_2 - \Omega_3$ & $-\dot{J}_4 + \dot{J}_5$ \\
8 ($\mathcal{O}2$)  &  0.0009 & -0.0002 & -0.0009 &  -24.7635 & -0.25373 & 243.990 & $ \Omega_2 + \Omega_3$ & $-\dot{J}_4 - \dot{J}_5$ \\
9 ($\mathcal{O}2$)  &  0.0000 & -0.0002 &  0.0000 &  182.273  &  0.03447 & 249.329 & $-\Omega_3 + \Omega_4$ & $ \dot{J}_5 - \dot{J}_6$ \\
10 ($\mathcal{O}2$) & -0.0007 &  0.0002 &  0.0007 & -110.427  & -0.05690 &   9.328 & $ \Omega_3 + \Omega_4$ & $-\dot{J}_5 - \dot{J}_6$ \\
\textbf{Callisto} &&&&&&&& \\
1 &  0.9183 & -0.3907 & -0.8310 & -559.68  & -0.01122 & 308.856 & $\Omega_4$ & ${\bf -\dot{J}_6}$ \\
2 &  0.1473 & -0.0626 & -0.1333 & -137.535 & -0.04568 &  59.519 & $\Omega_3$ & ${\bf -\dot{J}_5}$ \\
3 ($\mathcal{O}2$) & -0.0067 &  0.0014 &  0.0067 &  -279.99 &  0.02244 & 257.712 & $2 \Omega_4$ & $-2\dot{J}_6$ \\
4 ($\mathcal{O}2$) &  0.0000 & -0.0005 &  0.0000 &  182.312 &  0.03446 & 249.337 & $-\Omega_3 + \Omega_4$ & $ \dot{J}_5 - \dot{J}_6$ \\
5 ($\mathcal{O}2$) & -0.0021 &  0.0005 &  0.0021 & -110.416 &  0.05690 & 8.375   & $ \Omega_3 + \Omega_4$ & $-\dot{J}_5 - \dot{J}_6$ \\
6 &  0.0003 & -0.0001 & -0.0002 &    5.93109 &  1.05936 & 112.748 & $2L_S$ & ${\bf \dot{J}_8}$ \\
\hline
\end{tabular} \\
\caption{Spin position variations $\Delta\alpha_S$, $\Delta\delta_S$ and $\Delta\mu$ for the Galilean satellites for the solid interior model and the NOE orbital theory.
The truncation level is $0.0001^\circ$.
The last columns show the identification to the fundamental frequencies in \cite{Lai06} and to the IAU WG frequencies $\dot{J}_i$. The $\dot{J}_i$ frequencies appearing in \cite{Arc18} are indicated in bold. Order 2 terms, not present in the $\Delta s_{x/y}$ series, are labeled with the symbol $\mathcal{O}2$.}
\label{tab_spinadNOE}
\end{table}

As we did with the JUP387 orbital theory, we compute the values of the amplitudes $\varepsilon_j$ and $\theta_j$ considering a solid interior model, see Table~\ref{tab_sxsyNOE}, to obtain the complete $(s_x,s_y)$ solution. Then we transform the latter into solutions for the equatorial coordinates $\alpha_S, \delta_S$ and $\mu$, including rates, see Table~\ref{tab_trxtryNOE}, and periodic terms, see Table~\ref{tab_spinadNOE}.
We applied the same level of truncation as in Table~\ref{tab_sxsy}, and obtained the same number of terms for Europa, Ganymede and Callisto as with the JUP387 theory, but one more term for Io.
By applying similar simplifications as in Section \ref{sec_IAUsolution} for the solution derived from the JUP387 theory, we propose an alternative update for the IAU WG solution, but based on the NOE orbital theory, see Table~\ref{tab_spinad_IAUNOE}. For the $\alpha$ and $\delta$ rates, we have retained the values derived from the fit of the orbital precession of the NOE orbital theory, which are quite different from the values we recommend for a replacement of the IAU WG solution based on the JUP387 orbital theory.

Figure~\ref{fig_changeOrb} shows the orbit normal and spin motion of Europa in the ICRF plane over a $9$-year period. The angular separation between the spin axis and the orbit normal (also referred to as the orbital obliquity) is clearly noticeable. The spin motion is shown for both a solid interior model and an interior model with an internal global liquid ocean (see discussion in Section~\ref{sec_liquid}). The difference between the JUP and NOE models is not noticeable at this scale. As expected, the IAU WG model remains close to the orbit normal path, but deviates significantly from the spin axis trajectory.
For reference, the solutions of \cite{Bil22} for the orbit normal and spin motion, which are inconsistent with both the IAU WG solution and our solutions, are also shown in the figure.

\begin{table}[!htb]
\footnotesize
\centering
\begin{tabular}{l@{\,}c@{\,}@{\,}c@{\,}@{\,}c@{\,}@{\,}c@{\,}@{\,}c@{\,}@{\,}c@{\,}@{\,}c@{\,}@{\,}c@{\,}@{\,}c@{\,}@{\,}c@{\,}}
\hline
& Cst ($^\circ$) & Linear term ($^\circ$) & $\mathcal{J}_3$ & $\mathcal{J}_4$ & $\mathcal{J}_5$ & $\mathcal{J}_6$ & $\mathcal{J}_7$ & $\mathcal{J}_8$ & $\mathcal{J}_9$ \\
\hline
& & & $\Omega_1$ & $\Omega_2$ & $\Omega_3$ & $\Omega_{3}$ & $\Omega_4$ & $\Omega_{4}$ & $-2 \nu -\Omega_2$ \\
Period (y) & & & -7.4217 & -30.202 & -137.49 & -137.55 & -560.00 & -560.50 & -0.68144  \\
$f_j$ (rad/y) & & & -0.84660 & -0.20804 & -0.0457 & -0.04568 & -0.01122 & -0.01121 & -9.2204 \\
$f_j$ ($^\circ$/cent.) & & & -4850.7 & -1191.8 & -261.79 & -261.73 & -64.286 & -64.229 & -52829.0 \\
$\varphi_j$ ($^\circ$) & & & 259.686 & 184.097 & 59.999 & 59.519 & 308.856 & 309.328 & 273.900 \\
\multicolumn{10}{l}{} \\
\textbf{Io} &&&&&&&&& \\
$\alpha_S$   & 268.0590 & -0.017 $T$          &  0.0886 &  0.0233 &  0.0029 & - & -0.0008 & - &  0.0023 \\
$\delta_S$   &  64.4969 &  0.00028 $T$        & -0.0382 & -0.0100 & -0.0013 & - &  0.0003 & - & -0.0010 \\
$W-\gamma_u$ & 200.3970 & $203.48895843 \, d$ & -0.0800 & -0.0210 & -0.0027 & - &  0.0007 & - & -0.0020 \\
\textbf{Europa} &&&&&&&&& \\
$\alpha_S$   & 268.0850 & -0.0169 $T$         & -0.0049 &  1.2082 &  0.0591 & - & - &  0.0129 &  0.0007 \\
$\delta_S$   &  64.5068 &  0.0003 $T$         &  0.0021 & -0.5200 & -0.0254 & - & - & -0.0055 & -0.0003 \\
$W-\gamma_u$ & 35.9187 & $101.37472463 \, d$ &  0.0044 & -1.0906 & -0.0534 & - & - & -0.0116 & -0.0007 \\
\textbf{Ganymede} &&&&&&&&& \\
$\alpha_S$   & 268.2049 & -0.0167 $T$         & - & -0.1129 &  0.5025 & - & - & 0.0907 & - \\
$\delta_S$   &  64.5663 &  0.00041 $T$        & - & 0.0485 & -0.2158 & - & - & -0.0390 & - \\
$W-\gamma_u$ & 44.1615 & $50.317607704 \, d$ & - & 0.1019 & -0.4538 & - & - & -0.0819 & - \\
\textbf{Callisto} &&&&&&&&& \\
$\alpha_S$   & 268.7352 & -0.0157 $T$         & - & - & - &  0.1473 &  0.9183 & - & - \\
$\delta_S$   &  64.8188 &  0.00086 $T$        & - & - & - & -0.0626 & -0.3907 & - & - \\
$W-\gamma_u$ & 259.7765 & $21.571072689 \, d$ & - & - & - & -0.1333 & -0.8310 & - & - \\
\multicolumn{10}{l}{} \\
\hline
& & & $\mathcal{J}_{10}$ & $\mathcal{J}_{11}$ & $\mathcal{J}_{12}$ & $\mathcal{J}_{13}$ & $\mathcal{J}_{14}$ & $\mathcal{J}_{15}$ & $\mathcal{J}_{16}$ \\
\hline
& & & $2 L_S$ & $2 \Omega_2$ & $-L_S$ & $\Omega_2 + \Omega_3$ & $2 \Omega_3$ & $\Omega_3 + \Omega_4$ &  $2 \Omega_4$ \\
Period (y) & & & 5.9309 & -15.100 & -11.863 & -24.766 & -68.766 & -110.42 & -280.00 \\
$f_j$ (rad/y) & & & 1.0594 & -0.4161 & -0.52965 & -0.2537 & -0.09137 & -0.05690 & -0.02244 \\
$f_j$ ($^\circ$/cent.) & & & 6069.9 & -2384.1 & -3034.7 & -1453.6 & -523.51 & -326.01 & -128.57 \\
$\varphi_j$ ($^\circ$) & & & 113.496 & 8.194 & 298.928 & 244.217 & 119.999 & 8.375 & 257.712 \\
\multicolumn{10}{l}{} \\
\textbf{Io} &&&&&&&&& \\
$\alpha_S$   & & & - & - & - & - & - & - & - \\
$\delta_S$   & & & - & - & - & - & - & - & - \\
$W-\gamma_u$ & & & - & - & - & - & - & - & - \\
\textbf{Europa} &&&&&&&&& \\
$\alpha_S$   & & &  0.0013 & -0.0115 &  0.0007 & -0.0011 & - & - & - \\
$\delta_S$   & & & -0.0005 &  0.0025 & -0.0003 &  0.0002 & - & - & - \\
$W-\gamma_u$ & & & -0.0011 &  0.0116 & -0.0006 &  0.0011 & - & - & - \\
\textbf{Ganymede} &&&&&&&&& \\
$\alpha_S$   & & &  0.0010 & - & -0.0010 &  0.0009 & -0.0020 & -0.0007 & - \\
$\delta_S$   & & & -0.0004 & - &  0.0004 & -0.0002 &  0.0004 &  0.0002 & - \\
$W-\gamma_u$ & & & -0.0009 & - &  0.0009 & -0.0009 &  0.0020 &  0.0007 & - \\
\textbf{Callisto} &&&&&&&&& \\
$\alpha_S$   & & & - & - & - & - & - & -0.0021 & -0.0067 \\
$\delta_S$   & & & - & - & - & - & - &  0.0005 &  0.0014 \\
$W-\gamma_u$ & & & - & - & - & - & - &  0.0021 &  0.0067 \\
\hline
\end{tabular}
\caption{IAU WG solution ($\alpha_S$, $\delta_S$ and $W$) of the Galilean satellites considered as entirely rigid and using the NOE orbital theory, with a limited number of terms ($5$, $9$, $8$, and $4$, respectively) and a limited set of arguments ($14$). The amplitudes are given in degrees.
The column header shows the identification to the fundamental arguments, using the reduced set of arguments $\mathcal{J_j}$, whose physical origin is recalled on the first line. Compared to the full solution of Table~\ref{tab_spinadNOE} for the periodic terms, the precision is degraded to about $0.001^\circ$. 
Our model does not include any physical libration ($\gamma_u=0$).}
\label{tab_spinad_IAUNOE}
\end{table}

\subsection{Effect of the interior model}
\label{sec_liquid}

\begin{table}[!htb]
\small
\centering
\begin{tabular}{lccccccc}
\hline
& Period (y) & $f_j$ (rad/y) & $\varphi_j$ ($^\circ$) & Argument & Range $\varepsilon_j$ ($^\circ$) & Range $\alpha^S_j$ ($^\circ$) & Range $\delta^S_j$ ($^\circ$)\\
\hline
\multicolumn{8}{l}{} \\
\multicolumn{2}{l}{{\bf Europa}} &&&&&& \\
 1                  & -30.2008 & -0.20805 & 184.073 & $\Omega_2$  & -0.035, 0.2 & 1.0, 1.5 & -0.7, -0.4 \\
 4 $(\mathcal{O}2)$ & -15.1004 & -0.41609 &   8.146 & $2\Omega_2$ & - & -0.019, -0.008 & 0.002, 0.004 \\
\multicolumn{8}{l}{} \\
\multicolumn{2}{l}{{\bf Ganymede}} &&&&&& \\
 1                  & -137.328 & -0.04575 &  59.659 & $\Omega_3$    & 0.01, 0.07 & 0.5, 0.6 & -0.3, -0.2 \\
 2                  & -30.2009 & -0.20805 & 183.966 & $\Omega_2$    & -0.4, 0.8 & -1.0, 2.0 & -0.8, 0.4 \\
 8 $(\mathcal{O}2)$ & -24.7565 & -0.25380 & 243.626 & $\Omega_2 + \Omega_3$ & - & -0.02, 0.008  & -0.002, 0.003  \\ 
\multicolumn{8}{l}{} \\
\multicolumn{2}{l}{{\bf Callisto}} &&&&&& \\
1                  & -560.826 & -0.0112 & 309.668 & $\Omega_4$   & -0.02, 0.14 & 0.5, 0.9 & -0.4, -0.2 \\
2                  & -137.327 & -0.0458 &  59.182 & $\Omega_3$   & 0.02, 0.08  & -0.02, 0.12 & -0.05, 0.007 \\ 
3 $(\mathcal{O}2)$ & -280.413 & -0.0224 & 259.337 & $2 \Omega_4$ & - & -0.007, -0.002 & 0.005, 0.0014 \\
\hline
\end{tabular}
\caption{Amplitudes $\varepsilon_j$, $\alpha_S$ and $\delta_S$ for Europa, Ganymede and Callisto for a range of interior models with a liquid layer for the JUP387 orbital theory, after removing the most extreme $5\%$ of values from each range. Only ranges whose limits are clearly far from the solid value and of the order of at least about $0.01$ degrees are kept here. 
The ranges in $\mu_j$ can be approximately estimated by scaling the $\alpha^S_j$ column by a factor of $-0.9$.
For Europa and Callisto, the ranges in corrective term $\delta_{S\,0}$ are also affected by the resonances and are equal to the ranges in the amplitudes $\delta^S_4$ and $\delta^S_3$, respectively, of the term with twice the main frequency (order 2 term). }
\label{tab_liquidinterior}
\end{table}

In the previous sections, the satellites were assumed to be entirely solid. Europa, Ganymede, and Callisto, which are the targets of three future missions or missions currently in cruise phase (Juice, Europa Clipper, Tianwen-4), are believed to harbor a global internal ocean \citep[e.g.][]{Khu98,Kiv02}, however. The presence of a fully molten magma ocean inside Io suggested by (\cite{spo97,Khu11}) has been recently ruled out by the measurement of the satellite's tidal response (\cite{Par25}). Although Io most likely has a liquid core (\cite{spo97}), its effect on the satellite's orientation will not be evaluated here, as the first Galilean satellite is not currently the primary target of any approved space mission, and therefore its rotation will not be observed in the reasonably near future. 

We consider a wide range of interior structure models for Europa, Ganymede and Callisto, consistent with known constraints (mass, radius, and MOI), by varying the densities and sizes of their different layers in ranges similar to those of \cite{Bal12} or \cite{Coy26} and use the theoretical angular momentum model of \cite{Bal19} for a synchronous satellite with an internal global fluid layer to compute the ranges of obliquity amplitudes and build the series for $s_{x/y}$. The series for $\alpha, \delta,$ and $\mu$ are then obtained with the transformations described in Sections \ref{sec_transfo} and \ref{sec_rot}. 
As anticipated in Section~\ref{sec_obliquityseries}, the amplitudes of the spin axis motion at a given forcing frequency can differ significantly from the values obtained in the solid case when one of the free frequencies associated with the interior is sufficiently close to the forcing frequency. In such cases, the ordering of the terms may differ from the ordering in the solid case, and the approximation for $\varepsilon(t), \theta(t), \zeta(t)$, and $\psi(t)$ that assume the existence of a dominant amplitude (e.g.~Eqs.~\ref{eq_dominant}) become ineffective when two amplitudes dominate the solution together.
Because of resonant amplification, obliquity amplitudes can reach very large absolute values, which can be considered unrealistic given that tidal damping is not taken into account in the dynamic model. Therefore, we remove the most extreme $5\%$ of values from each range of obliquity amplitudes before calculating the minimum and maximal values from the remaining set. 
The quoted values represent a typical 95\% interval rather than strict limits, giving us an indicative range for the obliquity, right ascension, and declination amplitudes, see Table~\ref{tab_liquidinterior}, which do not necessarily imply that an amplitude cannot be larger or smaller than these limits.

Typical ranges for the periods of the three modes associated with the precession in space for Ganymede and Callisto can be found in \cite{Coy26}, see their Figs.~8-9. For Europa, the free mode period generally varies between 0 and 0.4 years for the retrograde Free Precession (FP, mode associated with the shell for a satellite with an ocean) and between 3 and 4 years for the retrograde Free Interior Nutation (FIN). The Free Ocean Nutation (FON) can be either prograde or retrograde, with periods ranging from -400 to 400 years. Distinguishing between FON and FIN can be tedious, especially when their periods are similar (\cite{Bal19} mistakenly reversed them for Titan). We generally note that FIN is the mode in which the period tends to remain larger but close to the period of the FP of the solid case, while the period of FON tends to vary within a wider range that may even include negative signs (free prograde motion), see also \cite{Bal25} for the Uranian satellites.

In Table~\ref{tab_liquidinterior}, we present the amplitude ranges whose limits are relatively far from the solid value and of the order of at least about $0.01$ degrees. With the exception of the second term of Ganymede, which is amplified by resonance with the FIN, all first-order terms presented are amplified by resonance with the FON.
For Europa, $\varepsilon_1$ can be positive (up to about $0.2^\circ$ when the most extreme values are discarded), but also negative (up to $-0.035^\circ$), or even close to $0$ and therefore deviate considerably from the solid value ($0.0555^\circ$).
The corresponding ranges in $\alpha_1$ and $\delta_1$ are also widened by the resonance with the FON. Fig.~\ref{fig_changeOrb} shows an example of how a change in interior model from solid (brown line) to liquid (orange line) affects the spin position in the $\alpha, \delta$ plane. The ranges in $\alpha_4$ and $\delta_4$ of the $15$-years second-order term, which is a byproduct of the transformation of the first term in the $s_{x/y}$ series, is also affected by the resonance with the FON.
For Ganymede and Callisto, the resonances also significantly widen the ranges of some of the second-order terms (the eighth and third terms, respectively).

\section{Conclusions and discussion}

This paper presents a method to construct an updated orientation and rotation model for the Galilean satellites.
After introducing the different reference frames and set of angles involved in the problem (Section~\ref{sec_planes}), we develop in Section~\ref{sec_transfo}
a theoretical framework to accurately transform the spin axis Cartesian coordinates 
into the declination and right ascension of the spin axis in the ICRF, while minimizing conversion errors. 
When a target precision of the order of the arcsecond is required, as expected from the Juice mission measurements, second-order terms must be included in the analytical transformation. 
We also show how the expressions for the inertial obliquity, node longitude, orbital obliquity, and offset relative to the Cassini plane
can be obtained from both the spin axis Cartesian or equatorial coordinates. 

Section~\ref{sec_rot} describes the transformation equations for the rotation angles. 
$\phi_{Euler}$ and $W$ are measured along the satellite's equator from the Laplace plane and the ICRF equatorial plane, respectively. However, for the purposes of libration modeling and future measurement interpretation, a third angle which we denote $\phi_{Inertial}$ is used, and is measured from an inertial point on the ICRF equator to the meridian on the satellite's equator. The transformation between $W$ and $\phi_{Inertial}$, denoted $\mu$, is described in Section~\ref{sec_defphi}. Although we do not solve them in this study, we provide the dynamical equations for the librations, see Section~\ref{sec_lib}, in the hope of clarifying the different ways in which librations, and thus the periodic parts of the rotation angle $W$, could be calculated.

The proposed analytical second-order transformation method relies solely on spherical trigonometry and
is therefore applicable regardless of the adopted dynamical model for the spin precession. 
If the spin axis Cartesian coordinates are expressed as trigonometric series, the transformed angles preserve this trigonometric form. 
The other advantages of this method are that the geophysically relevant parameters and the amplitudes of the output series can be readily connected, while error tracing is possible, enabling controlled accuracy. 
Additionally, the first order transformation coefficients can be used to propagate errors between the different coordinate sets. 

Based on the dynamical model described in Section~\ref{sec_dyn_mod}, we generate trigonometric series in spin axis Cartesian coordinates. 
As no direct observations of the Galilean satellites’ spin positions are yet available, we adopted as an example a rigid and solid model consistent with the most recent gravity field coefficients, see section~\ref{sec_sxsyseries}. 
The series are provided with a truncation level of about $0.0001^\circ$. 
If one obliquity amplitude in the spin axis Cartesian coordinates series is dominant (e.g., for a solid Europa), approximate trigonometric representations can be derived for the inertial and orbital obliquities, node longitude, and offset relative to the Cassini plane (Section~\ref{sec_obliquityseries}).
Frequency combinations (sums and differences of the frequencies of the terms in the spin axis Cartesian coordinates series) naturally appear, leading to new frequencies in Euler angles trigonometric series. The offset can be relatively important.
For Europa, the multi-frequency spin position deviates by only 0.005$^\circ$ from the Cassini plane for the solid case, whereas Callisto exhibits the largest offset (up to $0.1^\circ$). 
We converted the spin axis Cartesian coordinates into declination, right ascension, and $W$ series (Section~\ref{sec_admseries}). Second order terms naturally arise from combinations of either twice the dominant term or two distinct terms, corresponding to the $J_7$ or $J_8$ terms in the IAU WG solution.

A solution similar to that of \cite{Arc18} can be recovered by reducing the number of frequencies and assuming certain close phases and frequencies to be equal. Our proposed model now includes 14 frequencies shared among the four Galilean satellites, compared to only 6 in the IAU WG solution (Section~\ref{sec_IAUsolution}). 
In our solution, frequencies have a sign assigned based on the original frequency in the forcing series (corresponding to prograde or retrograde motions), which is crucial for comparison with the free retrograde mode and for evaluating possible amplifications.
The discrepancies between our model and the \cite{Lie79} model incorporated in \cite{Arc18} can reach up to $0.4^\circ$ after $100$ years. 
These differences arise from the use of an updated orbital theory and an non-zero obliquity model (whereas the previous IAU WG solution assumed zero obliquity, our model incorporates a theoretical obliquity consistent with a solid, rigid satellite).
As in \cite{Arc18}, our model contains no forced librations in the $W$ series; the periodic variations in $W$ originate solely from the periodic terms of the $\mu$ angle.  
Because the series in $\alpha$, $\delta$, and $\mu$ have mostly long term periodicities, we do not expect them to be well constrained by a single spacecraft mission, which is typically of limited duration. In such cases, a local approximation is likely more appropriate.

Our nominal series are based on JUP387, but we also provide equivalent series derived from the NOE ephemerides (Section~\ref{sec_orbcomp}). Although the frequencies, phases, and amplitudes differ slightly between the two series, the largest deviations in the spin position are due to the trend differences in the orbital precession.
A SPICE kernel for the case of entirely solid satellites has been generated for the $\alpha, \delta$ and $W$ angles, using both the JUP and NOE ephemerides. 
These 2 kernels are currently available on lara.oma.be/GalileanSat/. Another localization on the SPICE kernel architecture will be available soon.
If Europa, Ganymede, or Callisto possess a liquid global ocean beneath the icy shell, certain amplitudes may be amplified, as summarized in Table~\ref{tab_liquidinterior}.

We validated our transformation of Sections~\ref{sec_transfo} and \ref{sec_rot} by comparing our series of Section~\ref{sec_numericalvalues} with numerical integrations of the full differential equations of motion (Eqs.~\ref{eq_dadd}), finding differences smaller than the intrinsic precision of the series. 
Additional validation was obtained by comparing our Callisto series with those of \cite{Noy09}, showing a good agreement in amplitudes. However, we note that the phases reported by \cite{Noy09} appear inconsistent with both \cite{Arc18} and our results. 

In our computations, the Laplace plane was chosen as the reference plane and consistently with the adopted orbital theory. Each orbital theory defines a slightly different Laplace plane, and various numerical methods and temporal coverages can be employed to determine it. Nevertheless, to first order in small parameters, our multi-frequency theory allows any Laplace plane to be adopted as a reference for frequency decomposition.
If we consider Jupiter's equatorial plane instead of the Laplace plane as the reference plane, 
the resulting differences in spin axis right ascension and declination tend to increase with the tilt of the Laplace plane with respect to Jupiter's equator, and are therefore the largest for Callisto (tilt of $0.43^\circ$). This is because the obliquity relative to this plane becomes larger, making the second-order approximation less appropriate and leading to larger differences in $\alpha_S$ and $\delta_S$. For Io, Europa and Ganymede, the discrepancies reach a few tens of arcseconds $100$ years after J2000. For Callisto, considering the equatorial plane of Jupiter instead of the Laplace plane as a reference plane is definitely not recommended (see also \cite {Noy09}).

This study provides a robust foundation for interpretation of future observations of the Galilean satellites’ orientation and rotation angles, whether by the Juice or Europa Clipper spacecraft, or through Earth-based radar measurements. 
The method can readily be applied to any synchronously rotating satellite whose orbit remains close to its Laplace plane.

\section*{Acknowledgments}
This work was financially supported by the Belgian PRODEX program managed by the European Space Agency in collaboration with the Belgian Federal Science Policy Office. \\
We thank S\'ebastien Le Maistre, Tim Van Hoolst, Valery Lainey, Jean-Luc Margot, and Alexander Stark for their helpful comments and discussions.
We thank Alfonso Caldiero for testing the SPICE kernel and 2 anonymous reviewers for their valuable suggestions.

\bibliography{orientation_references}

@article{And01,
    title = {Io's gravity field and interior structure},
    author = {J.D. {Anderson} and R.A. {Jacobson} and {Lau}, E.L. and W.B. {Moore} and G. {Schubert}},
    journal = {JGR},
    volume = {106},
    number = {E12},
    pages = {32963-32969},
    year = 2001}

@ARTICLE{Arc18,
author = {{Archinal}, B.~A. and {Acton}, C.~H. and {A'Hearn}, M.~F. and {Conrad}, A. and {Consolmagno}, G.~J. and {Duxbury}, T. and {Hestroffer}, D. and {Hilton}, J.~L. and {Kirk}, R.~L. and {Klioner}, S.~A. and {McCarthy}, D. and {Meech}, K. and {Oberst}, J. and {Ping}, J. and {Seidelmann}, P.~K. and {Tholen}, D.~J. and {Thomas}, P.~C. and {Williams}, I.~P.},
        title = "{Report of the IAU Working Group on Cartographic Coordinates and Rotational Elements: 2015}",
      journal = {Celestial Mechanics and Dynamical Astronomy},
         year = 2018,
        month = feb,
       volume = {130},
       number = {3},
          eid = {22},
        pages = {22},
          doi = {10.1007/s10569-017-9805-5}}

@article{Bal11,
title={{T}itan's obliquity as evidence of a subsurface ocean?},
author={Baland, R.~M. and Van Hoolst, T. and Yseboodt, M. and Karatekin, {\"O} },
year={2011},
journal={\aap},
volume={530},
pages={A141}}

@article{Bal12,
title={Obliquity of the {G}alilean satellites: {T}he influence of a global internal liquid layer},
author={Baland, R.-M. and Yseboodt, M. and Van Hoolst, T.},
year={2012},
journal={Icarus},
volume={220},
pages={435-448}}

@article{Bal16,
title={{The obliquity of Enceladus}},
author={Baland, R.-M. and Yseboodt, M. and Van Hoolst, T.},
year={2016},
journal={Icarus},
volume={268},
pages={12-31}}

@ARTICLE{Bal17,
       author = {{Baland}, Rose-Marie and {Yseboodt}, Marie and {Rivoldini}, Attilio and {Van Hoolst}, Tim},
        title = "{Obliquity of Mercury: Influence of the precession of the pericenter and of tides}",
      journal = {\icarus},
         year = 2017,
        month = jul,
       volume = {291},
        pages = {136-159},
          doi = {10.1016/j.icarus.2017.03.020}}

@ARTICLE{Bal19,
       author = {{Baland}, Rose-Marie and {Coyette}, Alexis and {Van Hoolst}, Tim},
        title = "{Coupling between the spin precession and polar motion of a synchronously rotating satellite: application to Titan}",
      journal = {Celestial Mechanics and Dynamical Astronomy},
         year = 2019,
        month = feb,
       volume = {131},
       number = {2},
          eid = {11},
        pages = {11},
          doi = {10.1007/s10569-019-9888-2}}

@ARTICLE{Bal25,
       author = {{Baland}, Rose-Marie and {Filice}, Valerio and {Le Maistre}, S{\'e}bastien and {Trinh}, Antony and {Yseboodt}, Marie and {Van Hoolst}, Tim},
        title = "{Librations and obliquity of the largest moons of Uranus}",
      journal = {\icarus},
         year = 2025,
        month = jan,
       volume = {426},
          eid = {116371},
        pages = {116371},
          doi = {10.1016/j.icarus.2024.116371}}

@ARTICLE{Bil05,
       author = {{Bills}, Bruce G.},
        title = "{Free and forced obliquities of the Galilean satellites of Jupiter}",
      journal = {\icarus},
         year = 2005,
        month = may,
       volume = {175},
       number = {1},
        pages = {233-247},
          doi = {10.1016/j.icarus.2004.10.028}}

@ARTICLE{Bil11,
       author = {{Bills}, Bruce G. and {Nimmo}, Francis},
        title = "{Rotational dynamics and internal structure of Titan}",
      journal = {\icarus},
         year = 2011,
        month = jul,
       volume = {214},
       number = {1},
        pages = {351-355},
          doi = {10.1016/j.icarus.2011.04.028}}

@ARTICLE{Bil22,
       author = {{Bills}, Bruce G. and {Scott}, Bryan R.},
        title = "{Rotation models for the Galilean satellites}",
      journal = {\planss},
         year = 2022,
        month = sep,
       volume = {219},
          eid = {105474},
        pages = {105474},
          doi = {10.1016/j.pss.2022.105474}}

@ARTICLE{Bou20,
       author = {{Bou{\'e}}, Gwena{\"e}l},
        title = "{Cassini states of a rigid body with a liquid core}",
      journal = {Celestial Mechanics and Dynamical Astronomy},
         year = 2020,
        month = apr,
       volume = {132},
       number = {3},
          eid = {21},
        pages = {21},
          doi = {10.1007/s10569-020-09961-9}}

@ARTICLE{Cam43,
       author = {{Camichel}, H. and {Gentili}, M. and {Lyot}, B.},
        title = "{Observations Planetaires au Pic du Midi, en 1941.}",
      journal = {L'Astronomie},
         year = 1943,
        month = jan,
       volume = {57},
        pages = {49-60}}

@ARTICLE{Cap20,
author = {{Cappuccio}, P. and {Hickey}, A. and {Durante}, D. and {Di Benedetto}, M. and {Iess}, L. and {De Marchi}, F. and {Plainaki}, C. and {Milillo}, A. and {Mura}, A.},
        title = "{Ganymede's gravity, tides and rotational state from JUICE's 3GM experiment simulation}",
      journal = {\planss},
         year = 2020,
        month = aug,
       volume = {187},
          eid = {104902},
        pages = {104902},
          doi = {10.1016/j.pss.2020.104902}}

@ARTICLE{Cap22,
       author = {{Cappuccio}, Paolo and {Di Benedetto}, Mauro and {Durante}, Daniele and {Iess}, Luciano},
        title = "{Callisto and Europa Gravity Measurements from JUICE 3GM Experiment Simulation}",
      journal = {\psj},
         year = 2022,
        month = aug,
       volume = {3},
       number = {8},
          eid = {199},
        pages = {199},
          doi = {10.3847/psj/ac83c4}}

@article{Col66,
title={Cassini's {S}econd and {T}hird {L}aws},
author={Colombo, G.},
year={1966},
journal={AJ},
volume={71},
pages={891-896}}

@article{Coy16,
    author = "Alexis {Coyette} and Tim {Van Hoolst} and Rose-Marie {Baland} and Tetsuya {Tokano}",
    title = {Modeling the polar motion of {T}itan },
    journal = "Icarus ",
    volume = "265",
    number = "",
    pages = "1 - 28",
    year = "2016",
    note = "",
    issn = "0019-1035",
    doi = "http://dx.doi.org/10.1016/j.icarus.2015.10.015"}

@ARTICLE{Coy18,
    author = {{Coyette}, Alexis and {Baland}, Rose-Marie and {Van Hoolst}, Tim},
    title = "{Variations in rotation rate and polar motion of a non-hydrostatic Titan}",
    journal = {\icarus},
    year = 2018,
    month = jun,
    volume = {307},
    pages = {83-105},
    doi = {10.1016/j.icarus.2018.02.003}}

@article{Coy26,
  author       = {Alexis Coyette and Rose-Marie Baland and Tim Van Hoolst},
  title        = {Second-order modeling of the Cassini states of large satellites: I—influence of triaxiality and a subsurface ocean},
  journal      = {Celestial Mechanics and Dynamical Astronomy},
  year         = {2026},
  volume       = {138},
  number       = {1},
  pages        = {3},
  doi          = {10.1007/s10569-025-10269-9}
}

@ARTICLE{Dav80,
       author = {{Davies}, M.~E. and {Abalakin}, V.~K. and {Cross}, C.~A. and {Duncombe}, R.~L. and {Masursky}, H. and {Morando}, B. and {Owen}, T.~C. and {Seidelmann}, P.~K. and {Sinclair}, A.~T. and {Wilkins}, G.~A. and {Tjuflin}, Y.~S.},
        title = "{Report of the IAU Working Group on Cartographic Coordinates and Rotational Elements of the Planets and Satellites}",
      journal = {Celestial Mechanics},
         year = 1980,
        month = oct,
       volume = {22},
       number = {3},
        pages = {205-230},
          doi = {10.1007/BF01229508}}

@ARTICLE{Dav81,
       author = {{Davies}, M.~E. and {Katayama}, F.~Y.},
        title = "{Coordinates of features on the Galilean satellites}",
      journal = {\jgr},
         year = 1981,
        month = sep,
       volume = {86},
       number = {A10},
        pages = {8635-8657},
          doi = {10.1029/JA086iA10p08635}}

@ARTICLE{Dav98,
       author = {{Davies}, M.~E. and {Colvin}, T.~R. and {Oberst}, J. and {Zeitler}, W. and {Schuster}, P. and {Neukum}, G. and {McEwen}, A.~S. and {Phillips}, C.~B. and {Thomas}, P.~C. and {Veverka}, J. and {Belton}, M.~J.~S. and {Schubert}, G.},
        title = "{The Control Networks of the Galilean Satellites and Implications for Global Shape}",
      journal = {\icarus},
         year = 1998,
        month = sep,
       volume = {135},
       number = {1},
        pages = {372-376},
          doi = {10.1006/icar.1998.5982}}

@INPROCEEDINGS{Dol74,
       author = {{Dollfus}, A. and {Murray}, J.~B.},
        title = "{La rotation, la cartographie et la photom{\'e}trie des satellites de Jupiter.}",
    booktitle = {Exploration of the Planetary System},
         year = 1974,
       editor = {{Woszczyk}, A. and {Iwaniszewska}, C.},
       series = {IAU Symposium},
       volume = {65},
        month = jan,
        pages = {513-525}}

@article{Gom20,
author = {Luis {Gomez Casajus} and Marco Zannoni and Dario Modenini and Paolo Tortora and Francis Nimmo and Tim {Van Hoolst} and Dustin Buccino and Kamal Oudrhiri},
title = {Updated {E}uropa gravity field and interior structure from a reanalysis of {G}alileo tracking data},
journal = {Icarus},
volume = {358},
pages = {114187},
year = {2021},
issn = {0019-1035},
doi = {https://doi.org/10.1016/j.icarus.2020.114187}}

@ARTICLE{Hen04,
       author = {{Henrard}, Jacques and {Schwanen}, Gabriel},
        title = "{Rotation of Synchronous Satellites: Application to the Galilean Satellites}",
      journal = {Celestial Mechanics and Dynamical Astronomy},
         year = 2004,
        month = mar,
       volume = {89},
       number = {2},
        pages = {181-200},
          doi = {10.1023/B:CELE.0000034515.57763.33}}

@ARTICLE{Hen05,
       author = {{Henrard}, Jacques},
        title = "{The Rotation of Europa}",
      journal = {Celestial Mechanics and Dynamical Astronomy},
         year = 2005,
        month = jan,
       volume = {91},
       number = {1-2},
        pages = {131-149},
          doi = {10.1007/s10569-005-3833-2}}

@ARTICLE{Kiv02,
       author = {{Kivelson}, M.~G. and {Khurana}, K.~K. and {Volwerk}, M.},
        title = "{The Permanent and Inductive Magnetic Moments of Ganymede}",
      journal = {\icarus},
         year = 2002,
        month = jun,
       volume = {157},
       number = {2},
        pages = {507-522},
          doi = {10.1006/icar.2002.6834}
}

@ARTICLE{Khu11,
       author = {{Khurana}, Krishan K. and {Jia}, Xianzhe and {Kivelson}, Margaret G. and {Nimmo}, Francis and {Schubert}, Gerald and {Russell}, Christopher T.},
        title = "{Evidence of a Global Magma Ocean in Io{\textquoteright}s Interior}",
      journal = {Science},
         year = 2011,
        month = jun,
       volume = {332},
       number = {6034},
        pages = {1186},
          doi = {10.1126/science.1201425}
}

@ARTICLE{Khu98,
       author = {{Khurana}, K.~K. and {Kivelson}, M.~G. and {Stevenson}, D.~J. and {Schubert}, G. and {Russell}, C.~T. and {Walker}, R.~J. and {Polanskey}, C.},
        title = "{Induced magnetic fields as evidence for subsurface oceans in Europa and Callisto}",
      journal = {\nat},
         year = 1998,
        month = oct,
       volume = {395},
       number = {6704},
        pages = {777-780},
          doi = {10.1038/27394}
}

@ARTICLE{Lai06,
       author = {{Lainey}, V. and {Duriez}, L. and {Vienne}, A.},
        title = "{Synthetic representation of the Galilean satellites' orbital motions from L1 ephemerides}",
      journal = {\aap},
         year = 2006,
        month = sep,
       volume = {456},
       number = {2},
        pages = {783-788},
          doi = {10.1051/0004-6361:20064941}}

@ARTICLE{Lai09,
       author = {{Lainey}, Val{\'e}ry and {Arlot}, Jean-Eudes and {Karatekin}, {\"O}zg{\"u}r and {van Hoolst}, Tim},
        title = "{Strong tidal dissipation in Io and Jupiter from astrometric observations}",
      journal = {\nat},
         year = 2009,
        month = jun,
       volume = {459},
       number = {7249},
        pages = {957-959},
          doi = {10.1038/nature08108}}

@article{Lem16,
  author    = {Le Maistre, S. and Folkner, W. M. and Jacobson, R. A. and Serra, D.},
  title     = {Jupiter spin-pole precession rate and moment of inertia from Juno radio-science observations},
  journal   = {Planetary and Space Science},
  volume    = {126},
  pages     = {78-92},
  year      = {2016},
  doi       = {10.1016/j.pss.2016.03.006}}

@ARTICLE{Lie79,
       author = {{Lieske}, J.~H.},
        title = "{Poles of the Galilean satellites.}",
      journal = {\aap},
         year = 1979,
        month = may,
       volume = {75},
       number = {1-2},
        pages = {158-163}}

@ARTICLE{Lie93,
       author = {{Lieske}, J.~H.},
        title = "{Algorithm for I.A.U. North Poles and Rotation Parameters - which way is up?}",
      journal = {Celestial Mechanics and Dynamical Astronomy},
         year = 1993,
        month = nov,
       volume = {57},
       number = {3},
        pages = {473-491},
          doi = {10.1007/BF00695716}}

@INPROCEEDINGS{Lie93b,
       author = {{Lieske}, J.~H.},
        title = "{IAU North Poles and Rotation Parameters for Natural Satellites}",
    booktitle = {Developments in Astrometry and their Impact on Astrophysics and Geodynamics},
         year = 1993,
       editor = {{Mueller}, Ivan Istvan and {Kolaczek}, Barbara},
       series = {IAU Symposium},
       volume = {156},
        month = jan,
        pages = {351}}

@INPROCEEDINGS{Mar25,
       author = {{Margot}, Jean-Luc},
        title = "{Spin states of Europa and Ganymede}",
    booktitle = {EGU Meeting Abstracts \#221},
         year = 2025,
       series = {EGU General Assembly 2025, Vienna, Austria, 27 Apr–2 May 2025},
       volume = {EGU25-14788},
          eid = {EGU25-14788},
       }

@ARTICLE{Maz23,
       author = {{Mazarico}, Erwan and {Buccino}, Dustin and {Castillo-Rogez}, Julie and {Dombard}, Andrew J. and {Genova}, Antonio and {Hussmann}, Hauke and {Kiefer}, Walter S. and {Lunine}, Jonathan I. and {McKinnon}, William B. and {Nimmo}, Francis and {Park}, Ryan S. and {Roberts}, James H. and {Srinivasan}, Dipak K. and {Steinbr{\"u}gge}, Gregor and {Tortora}, Paolo and {Withers}, Paul},
        title = "{The Europa Clipper Gravity and Radio Science Investigation}",
      journal = {\ssr},
         year = 2023,
        month = jun,
       volume = {219},
       number = {4},
          eid = {30},
        pages = {30},
          doi = {10.1007/s11214-023-00972-0}}

@BOOK{Mur00,
       author = {{Murray}, Carl D. and {Dermott}, Stanley F.},
        title = "{Solar System Dynamics}",
         year = 2000,
          doi = {10.1017/CBO9781139174817},
    publisher = {Cambridge University Press}}

@ARTICLE{Noy09,
       author = {{Noyelles}, Beno{\^\i}t},
        title = "{Expression of Cassini's third law for Callisto, and theory of its rotation}",
      journal = {\icarus},
         year = 2009,
        month = jul,
       volume = {202},
       number = {1},
        pages = {225-239},
          doi = {10.1016/j.icarus.2008.12.015}}

@ARTICLE{Noy10,
    author = {{Noyelles}, Beno{\^\i}t},
    title = "{Theory of the rotation of Janus and Epimetheus}",
    journal = {\icarus},
    year = 2010,
    month = jun,
    volume = {207},
    number = {2},
    pages = {887-902},
    doi = {10.1016/j.icarus.2009.12.034},
    eprint = {0912.4830}}

@ARTICLE{Par25,
       author = {{Park}, R.~S. and {Jacobson}, R.~A. and {Gomez Casajus}, L. and {Nimmo}, F. and {Ermakov}, A.~I. and {Keane}, J.~T. and {McKinnon}, W.~B. and {Stevenson}, D.~J. and {Akiba}, R. and {Idini}, B. and {Buccino}, D.~R. and {Magnanini}, A. and {Parisi}, M. and {Tortora}, P. and {Zannoni}, M. and {Mura}, A. and {Durante}, D. and {Iess}, L. and {Connerney}, J.~E.~P. and {Levin}, S.~M. and {Bolton}, S.~J.},
        title = "{Io's tidal response precludes a shallow magma ocean}",
      journal = {\nat},
         year = 2025,
        month = feb,
       volume = {638},
       number = {8049},
        pages = {69-73},
          doi = {10.1038/s41586-024-08442-5}
}

@ARTICLE{Pea69,
       author = {{Peale}, Stanton J.},
        title = "{Generalized Cassini's Laws}",
      journal = {\aj},
         year = 1969,
        month = apr,
       volume = {74},
        pages = {483},
          doi = {10.1086/110825}}

@ARTICLE{Ram11,
       author = {{Rambaux}, N. and {van Hoolst}, T. and {Karatekin}, {\"O}.},
        title = "{Librational response of Europa, Ganymede, and Callisto with an ocean for a non-Keplerian orbit}",
      journal = {\aap},
         year = 2011,
        month = mar,
       volume = {527},
          eid = {A118},
        pages = {A118},
          doi = {10.1051/0004-6361/201015304}}

@INBOOK{Sch04,
   author = {{Schubert}, G. and {Anderson}, J.~D. and {Spohn}, T. and {McKinnon}, W.~B.},
    title = "{Interior composition, structure and dynamics of the Galilean satellites}",
booktitle = {Jupiter.~The Planet, Satellites and Magnetosphere},
     year = 2004,
   volume = {1},
publisher = {Cambridge University Press},
    pages = {281-306}}

@ARTICLE{Sei07,
       author = {{Seidelmann}, P. Kenneth and {Archinal}, B.~A. and {A'hearn}, M.~F. and {Conrad}, A. and {Consolmagno}, G.~J. and {Hestroffer}, D. and {Hilton}, J.~L. and {Krasinsky}, G.~A. and {Neumann}, G. and {Oberst}, J. and {Stooke}, P. and {Tedesco}, E.~F. and {Tholen}, D.~J. and {Thomas}, P.~C. and {Williams}, I.~P.},
        title = "{Report of the IAU/IAG Working Group on cartographic coordinates and rotational elements: 2006}",
      journal = {Celestial Mechanics and Dynamical Astronomy},
         year = 2007,
        month = jul,
       volume = {98},
       number = {3},
        pages = {155-180},
          doi = {10.1007/s10569-007-9072-y}}

@INCOLLECTION{Spo97,
       author = {{Spohn}, Tilman},
        title = "{Tides of io}",
    booktitle = {Lecture Notes in Earth Sciences, Berlin Springer Verlag},
         year = 1997,
       editor = {{Wilhelm}, Helmut and {Z{\"u}rn}, Walter and {Wenzel}, Hans-Georg},
       volume = {66},
        pages = {345-377},
          doi = {10.1007/BFb0011459}
}

@INPROCEEDINGS{Sta18,
       author = {{Stark}, Alexander and {Hussmann}, Hauke and {Steinbr{\"u}gge}, Gregor and {Oberst}, J{\"u}rgen and {Roatsch}, Thomas},
        title = "{Resonant Rotation parameters of the Galilean satellites}",
    booktitle = {EGU General Assembly Conference Abstracts},
         year = 2018,
       series = {EGU General Assembly Conference Abstracts},
        month = apr,
        pages = {10488}}

@ARTICLE{Ste19,
       author = {{Steinbr{\"u}gge}, Gregor and {Steinke}, Teresa and {Thor}, Robin and {Stark}, Alexander and {Hussmann}, Hauke},
        title = "{Measuring Ganymede's Librations with Laser Altimetry}",
      journal = {Geosciences},
         year = 2019,
        month = jul,
       volume = {9},
       number = {7},
        pages = {320},
          doi = {10.3390/geosciences9070320}}

@ARTICLE{Ste26,
author={Steinbr{\"u}gge, Gregor and Park, R. S. and Roberts, J. H.
 and Bland, M. and Brooks, S. and Castillo-Rogez, J. and Cascioli, G. and Genova, A. and Greathouse, T. and Hussmann, H. and Kirk, R. and Magnanini, A. and Mazarico, E. and Nimmo, F. and Park, M. S. and Petricca, F. and Retherford, K. and Schroeder, D. M. and Soderlund, K. and Tortora, P. and Zannoni, M.},
  title     = {Geodetic Investigations of the Europa Clipper Mission},
  journal   = {Space Science Reviews},
  volume    = {222},
  number    = {17},
  year      = {2026},
  doi       = {10.1007/s11214-025-01250-x}
}

@ARTICLE{Tur24,
       author = {{Turtle}, E.~P. and {McEwen}, A.~S. and {Patterson}, G.~W. and {Ernst}, C.~M. and {Elder}, C.~M. and {Slack}, K.~A. and {Hawkins}, S.~E. and {McDermott}, J. and {Meyer}, H. and {DeMajistre}, R. and etal},
        title = "{The Europa Imaging System (EIS) Investigation}",
      journal = {\ssr},
         year = 2024,
        month = dec,
       volume = {220},
       number = {8},
          eid = {91},
        pages = {91},
          doi = {10.1007/s11214-024-01115-9}}

@article{VanH08,
author = {T. {Van Hoolst} and N. Rambaux and \"O. Karatekin and V. Dehant and A. Rivoldini},
title = {The librations, shape, and icy shell of Europa},
journal = {Icarus},
volume = {195},
number = {1},
pages = {386-399},
year = {2008},
issn = {0019-1035},
doi = {https://doi.org/10.1016/j.icarus.2007.12.011}}

@ARTICLE{VanH13,
       author = {{Van Hoolst}, Tim and {Baland}, Rose-Marie and {Trinh}, Antony},
        title = "{On the librations and tides of large icy satellites}",
      journal = {\icarus},
         year = 2013,
        month = sep,
       volume = {226},
       number = {1},
        pages = {299-315},
          doi = {10.1016/j.icarus.2013.05.036}}

@ARTICLE{Yse06,
       author = {{Yseboodt}, Marie and {Margot}, Jean-Luc},
        title = "{Evolution of Mercury's obliquity}",
      journal = {\icarus},
         year = 2006,
        month = apr,
       volume = {181},
       number = {2},
        pages = {327-337},
          doi = {10.1016/j.icarus.2005.11.024}}

@INPROCEEDINGS{Yse14,
       author = {{Yseboodt}, Marie and {Van Hoolst}, Tim},
        title = "{The long-period forced librations of Titan}",
    booktitle = {Complex Planetary Systems, Proceedings of the International Astronomical Union},
         year = 2014,
       series = {IAU Symposium},
       volume = {310},
        month = jul,
        pages = {25-28},
          doi = {10.1017/S1743921314007741}}

@ARTICLE{Yse23,
       author = {{Yseboodt}, Marie and {Baland}, Rose-Marie and {Le Maistre}, S{\'e}bastien},
        title = "{Mars orientation and rotation angles}",
      journal = {Celestial Mechanics and Dynamical Astronomy},
         year = 2023,
        month = oct,
       volume = {135},
       number = {5},
          eid = {50},
        pages = {50},
          doi = {10.1007/s10569-023-10159-y}}

\end{document}